\documentclass[%
 reprint,
 amsmath,
 amssymb,
 aps,
 prb
]{revtex4-2}

\usepackage{graphicx}
\usepackage{dcolumn}% Align table columns on decimal point
\usepackage{bm}% bold math
\usepackage{physics}
\usepackage{amsthm}
\usepackage{mathtools}
\usepackage{algpseudocode}
\usepackage{algorithm}
\usepackage{color}
\usepackage{ulem} % to have \sout
\usepackage[caption=false]{subfig}
\usepackage[colorlinks=true, citecolor=blue, urlcolor=blue, linkcolor=blue, pdfencoding=auto, psdextra]{hyperref}

\newcommand*{\Scale}[2][4]{\scalebox{#1}{\ensuremath{#2}}}%
\newcommand{\terml}{\rotatebox[origin=c]{90}{$\blacktriangle$}}
\newcommand{\termr}{\rotatebox[origin=c]{-90}{$\blacktriangle$}}

\newcommand\dual[1]{\overline{#1}}
\newcommand\sma[2]{
\left(\hspace{-3pt}\begin{array}{c}
  #1  \\ \hline
  #2
\end{array}\hspace{-3pt}\right)}

\newcommand\mps[2]{#1:\Scale[0.85]{\setlength{\arraycolsep}{2.5pt}#2}}
\DeclareMathOperator{\ttr}{{\mathcal{T}\!\!\raisebox{-0.4pt}{\it tr}}\hspace{-0.6pt}}

\newcommand{\tmA}{\widetilde{\mathcal{A}}}
\newcommand{\tmC}{\widetilde{\mathcal{C}}}

\makeatletter
\newsavebox{\@brx}
\newcommand{\llangle}[1][]{\savebox{\@brx}{\(\m@th{#1\langle}\)}%
  \mathopen{\copy\@brx\kern-0.5\wd\@brx\usebox{\@brx}}}
\newcommand{\rrangle}[1][]{\savebox{\@brx}{\(\m@th{#1\rangle}\)}%
  \mathclose{\copy\@brx\kern-0.5\wd\@brx\usebox{\@brx}}}
\makeatother
\newcommand{\lgen}{\llangle}
\newcommand{\rgen}{\rrangle}

\newtheorem{theorem}{Theorem}
\newtheorem{prop}[theorem]{Proposition}
\newtheorem{lemma}[theorem]{Lemma}
\newtheorem{corollary}{Corollary}[theorem]
\newtheorem{gentheorem}{Theorem}

\DeclareMathOperator{\diag}{diag}
\DeclareMathOperator{\spn}{span}
\DeclareMathOperator{\svd}{SVD}
\DeclareMathOperator{\smasub}{\mathcal{V}}
\DeclareMathOperator{\orth}{orth}

\begin{document}

\title{Many exact area-law scar eigenstates in the nonintegrable PXP and related models}
\author{Andrew N. Ivanov and Olexei I. Motrunich}
\affiliation{Department of Physics and Institute for Quantum Information and Matter,
  California Institute of Technology, Pasadena, California 91125, USA}

\date{\today}

\begin{abstract} 
  In this work, we present new, highly non-trivial area-law exact zero-energy eigenstates of the one-dimensional (1D) PXP and related models.
  We formulate sufficient conditions for a matrix product state to represent an exact zero-energy eigenstate of a given 1D kinetically constrained model and use them to prove our new states.
  We also demonstrate that all previously known exact eigenstates of PXP-type models satisfy these conditions, and, in fact, can be directly deduced from them.
  We discuss and demonstrate a remarkably effective general numerical technique for discovering finite-bond-dimension eigenstates residing in degenerate subspaces of a broad class of Hamiltonians.
  Our results highlight a previously unrecognized structure characteristic of the exponentially large nullspaces in kinetically constrained models, suggesting the possibly of extensively many increasingly complex area-law zero-energy eigenstates in the thermodynamic limit.
  The important implications of these emergent exact eigenstates for the general thermalization phenomenology are exemplified by one of the states introduced in this work, which we propose is a member of the primary $\mathbb{Z}_2$ quantum many-body scar tower responsible for long-lived revivals in the Rydberg atom chain experiment.
    \end{abstract}

\maketitle

\section{Introduction}
Questions of integrability and thermalization of models with local kinetic constraints, which restrict the allowed state space, have been at the forefront of research for several decades.
More recently, models characterized by the Rydberg blockade constraint, naturally arising in experiments with trapped neutral cold atoms, have attracted significant attention.
This surge of interest followed the observation of unexpected many-body revivals of the N\'{e}el-type $\ket{\mathbb{Z}_2}$ state in the quench experiment~\cite{Bernien_2017}, and the subsequent attribution of that dynamics to the eigenstate thermalization hypothesis (ETH) violations in the one-dimensional (1D) PXP Hamiltonian, which encapsulates the essential physics of the Rydberg atomic system~\cite{Turner_2018weak, Turner_2018quantum}.
Specifically, a quantum many-body scar (QMBS) tower of approximately equally spaced states featuring sub-extensive scaling of entanglement entropy and high overlaps with $\ket{\mathbb{Z}_2}$ (in comparison to the nearby states) was discovered in the spectrum of the PXP Hamiltonian and shown to play a key role in the observable non-ergodic phenomenology.
A very recent discovery of similar approximate scar towers featuring analogous oscillatory dynamical signatures in a broader family of PXP-type models with longer-range constraints~\cite{kerschbaumer2024quantummanybodyscarspxp} further highlighted the ubiquity of scarring in kinetically constrained models.

Although the absence of local conserved quantities in the PXP model was demonstrated recently, suggesting its nonintegrability~\cite{Park_2024}, several exact eigenstates were nonetheless identified within the model's exponentially large nullspace~\cite{lin2019exact, Ivanov_2025}.
Over the years, numerous efforts have been made to uncover additional exact scars in PXP-type models, employing a wide range of techniques.
These include identifying states with anomalously low Schmidt ranks through ED~\cite{Moudgalya_2018, lin2019exact}, entanglement minimization within the nullspace~\cite{Karle_2021}, machine learning~\cite{Szo_dra_2022, feng2024uncoveringquantummanybodyscars}, correlation matrix analysis~\cite{yao2024quantummanybodyscarslens}, analytic continuation of the partition function and Fisher zeros~\cite{meng2025detectingmanybodyscarsfisher}, and DMRG methods~\cite{Zhang_2023, yuan2023exactquantummanybodyscars}.
However, the absence of new exact results beyond the states reported in Refs.~\cite{lin2019exact} and~\cite{Surace_2021} raises a critical question: Do other exact scars exist, and if so, how can they be detected?

We address this question by seeking to uncover an underlying structure that allows the existence of exact scars within the (symmetry-protected) exponentially degenerate nullspaces~\cite{Schecter_2018} of the PXP-type as well as more general kinetically constrained models.
Our pursuit of new exact analytical results stems from the aspiration to gain fresh insights into existing examples of ergodicity-breaking phenomena in these models (in particular, in the thermodynamic limit), and to uncover previously hidden features.
We also aim to enhance our understanding of the limits of exact solvability and integrability within this important family of Hamiltonians.
Our new numerical method for finding exact scars in such highly degenerate nullspaces, while remarkably successful in uncovering the new states presented in this work, also highlights the inherent hardness of this endeavor, which perhaps carries deeper implications by itself.
As a by-product, building on top of one of our new scars in the PXP chain, we find the most accurate theoretical description to date of the primary $\mathbb{Z}_2$ scars with non-zero energies framing the nullspace.
This allows us to also provide a new perspective on the long-lived oscillating revivals in quenches from the $\mathbb{Z}_2$ charge density wave (CDW) states.

This paper is organized as follows.
In Sec.~\ref{sec:models} we review the family of 1D models with Rydberg blockade constrains, which includes the PXP and PPXPP models that are of main focus in this work.
Then in Sec.~\ref{sec:mps} we define several distinct types of translationally-invariant (TI) matrix product state (MPS) representations, which, in the following Sec.~\ref{sec:eigenstates}, are used to express and characterize all presently known exact TI eigenstates of the two models.
Note that our new scar eigenstates will be initially introduced and discussed in some detail without proofs.
We will establish the proofs in Sec~\ref{sec:proofs} in the form of conditions sufficient for an MPS to represent an exact zero energy eigenstate of a Hamiltonian with a local kinetic constraint.
There, we will show that all the states listed in Sec.~\ref{sec:eigenstates} (both new and previously known) satisfy a particular set of nonlinear matrix equations, which can be deduced from the type of the state's MPS representation and the properties of the Hamiltonian.
In Sec.~\ref{sec:extra}, to further highlight the generality of the formalism developed in Sec.~\ref{sec:proofs}, we will provide several additional examples involving models and situations not considered earlier.
In Sec.~\ref{sec:detection}, we present a systematic numerical technique for detecting area-law-entangled states in exponentially degenerate subspaces and provide its demonstration using the PXP Hamiltonian as an example.
Finally, in Sec.~\ref{sec:dynamics}, we provide a new perspective on the extensively studied non-ergodic dynamics associated with the primary $\mathbb{Z}_2$ scars in the PXP model by demonstrating its connection to some of our newly introduced states.

\section{Models}
\label{sec:models}
The PXP model, which is an idealized description of Rydberg atomic systems in the nearest-neighbor blockade regime, is defined for a spin-1/2 chain of length $L$ by the following Hamiltonian:
\begin{equation}
  H_\text{PXP} = \sum_{j=2}^{L-1} P_{j-1} X_j P_{j+1} + H_\text{left} + H_\text{right},
\end{equation}
where $P_i=\ketbra{0}{0}_i$, $X_i = \ketbra{1}{0}_i + \ketbra{0}{1}_i$; for PBC $H_\text{left} = P_L X_1 P_2$ and $H_\text{right} = P_{L-1} X_L P_1$, whereas for open boundary conditions (OBC) $H_\text{left} = X_1 P_2$ and $H_\text{right} = P_{L-1} X_L$.
A natural generalization of the PXP Hamiltonian (assuming PBC) with the blockade radius parametrized by an integer $\alpha$ is the  following:
\begin{equation}
  \label{eq:halpha}
  H^{\alpha} = \sum_{j=1}^L P_{j-\alpha} \dots P_{j-1} X_j P_{j+1} \dots P_{j+\alpha}.
\end{equation}
Clearly, $H_\text{PXP}=H^1$, whereas $H^2$ corresponds to the so-called PPXPP model \cite{Surace_2021, Karle_2021}.
Any $H^\alpha$ has a dynamically decoupled subspace
\begin{equation}
  \mathcal{H}^\alpha = \spn\big\{\ket{s_1 s_2 \cdots s_L}:
  \begin{aligned}[t]
    & s_i s_{i+k} \neq 11,\\
    & \text{for any } i,k=1..\alpha \big\},
  \end{aligned}
\end{equation}
which, due its experimental relevance, will be our focus.

Hamiltonians $H^\alpha$ possess three conventional symmetries: translational symmetry for the case of PBC, which we will denote by $T_x$; inversion symmetry $\mathcal{I}$ about any axis cutting the chain into two equal parts; and spectral reflection symmetry defined by the operator $\mathcal{C} = \prod_jZ_j$, where $Z = \ketbra{0}-\ketbra{1}$.
Since $\left\{\mathcal{C}, H^\alpha\right\} = 0$, the spectrum of $H^\alpha$ is symmetric about $E=0$.
Moreover, all Hamiltonians $H^\alpha$ possess exponentially degenerate nullspaces as a result of the interplay between the $\mathcal{I}$ and $\mathcal{C}$ symmetries \cite{Turner_2018quantum,Buijsman_2022}.

Some parts of our discussion will be more natural in the blocked basis where, following the notation of Ref.~\cite{lin2019exact}, composite two-spin states $\ket{00}$, $\ket{01}$, and $\ket{10}$ are denoted by $\ket{O}$, $\ket{R}$, and $\ket{L}$.
For example, in this blocked basis the PXP Hamiltonian for a PBC chain of size $L=2L_b$ can be written as the following sum of one-body and two-body terms:
\begin{equation}
H_\mathrm{PXP} = H_1 + H_2,
\end{equation}
where (assuming the block-site index $j$ refers to two-atom blocks)
\begin{subequations}
  \begin{align}
    \label{eq:h1pxp}
    H_1 &= \sum_{j=1}^{L_b} \left\{ \left(\ket{O}\left[\bra{L} + \bra{R}\right]\right)_j + \mathrm{h.c.} \right\},\\
    \label{eq:h2pxp}
    H_2 &= -\sum_{j=1}^{L_b} \left\{\left(\ket{RL}\left[\bra{OL} + \bra{RO}\right]\right)_{j,j+1} + \mathrm{h.c.} \right\}.
  \end{align}
\end{subequations}
With summation in Eq.~(\ref{eq:h2pxp}) truncated at $L_b - 1$, this representation is also valid for OBC.

The blocked basis is natural for representing the exact translational-symmetry-breaking eigenstates PBC states $\ket{\Phi_1}$ and $\ket{\Phi_2}$ from Ref.~\cite{lin2019exact}, as well as some of the new eigenstates of both the PXP and PPXPP models to be introduced later.
Such states are intimately connected to the $H_1$ term in Eq.~(\ref{eq:h1pxp}), which, in contrast to the $H_\text{PXP}$ and the $H_2$ term, is an integrable Hamiltonian.
We will elaborate on the details of this connection later.
In what follows, we will often denote various TI Hamiltonians with terms acting on single sites in a particular basis by $H_1$.

\section{Types of MPS representations}
\label{sec:mps}
\subsection{TI MPS}
It is known that any translationally-invariant in a particular local basis $\{\ket{s}\}$ state $\ket{\psi}$ can be expressed as a TI MPS with site-independent matrices $M^s$ as follows~\cite{perez2007matrix}:
\begin{equation}
  \label{eq:timps}
  \ket{\psi} = \sum_{\{s_i\}}\Tr{M^{s_1}M^{s_2}\cdots M^{s_{L_b}}}\ket{s_1s_2\dots s_{L_b}}.
\end{equation}
When defining states via Eq.~(\ref{eq:timps}) we will use either the single-site basis $\{\ket{0}, \ket{1}\}$, or, as in Ref.~\cite{lin2019exact}, the blocked basis $\{\ket{O}, \ket{L}, \ket{R}\} \equiv \{\ket{00}, \ket{10}, \ket{01}\}$ representing blocks consisting of adjacent spin-1/2 sites subject to the Rydberg blockade~\footnote{The corresponding unconstrained (``spin-$1$'') Hilbert space has dimension $3^{L/2}$, which is considerably smaller than the full Hilbert space spanned by all possible spin-1/2 product states since there are no $\ket{11}$ blocks.}.
\subsection{Blocked-antipodal (BA) TI MPS}
Volume-entangled states with well-defined antipodal structure like the ones discussed in Refs.~\cite{Ivanov_2025,Chiba_2024,yoneta2024thermalpurestatessystems} can be expressed in what we will call the BA basis as follows:
\begin{equation}
  \label{eq:abmps}
  \begin{aligned}
    \ket{\psi} = \sum_{\{s_i,s_{\dual i}\}}&\Tr{M^{s_1s_{\dual 1}}M^{s_2s_{\dual 2}}\cdots M^{s_{L_b}s_{\dual{L_b}}}}\\
    &\times\ket{s_1s_2\dots s_{L_b}}\otimes\ket{s_{\dual 1}s_{\dual 2}\dots s_{\dual{L_b}}},
  \end{aligned}
\end{equation}
where $s_i,s_{\dual i} \in \{0, 1\}$ and the sites in the second line are adjacent to each other on a chain with PBCs (see Fig. 1 in Ref. \cite{Ivanov_2025}). Note that the BA representation generates TI states only when $M^{s_is_{\dual i}} = M^{s_{\dual i}s_i}$.

\subsection{Twisted translationally-invariant (TTI) MPS}
\label{subsec:ttimps}
We stated earlier that any TI state has a TI MPS representation.
It is, however, not always possible to write such states, even those with finite Schmidt index in the thermodynamic limit, using system-size-independent tensors.
This striking deficiency of the standard TI MPS representation is famously exemplified by the so-called $W$-state given by
\begin{equation}
  \label{eq:wstate}
  \ket{W} = \left(\frac{1}{\sqrt{L}}\sum_{i=1}^{L}X_i\right)\ket{00\dots 0}.
\end{equation}
It is known that the TI MPS representation of the $W$-state, whose Schmidt index is 2, requires system-size-dependent tensors with bond dimension  $\chi \succeq \mathcal{O}(L^{1/3})$~\cite{perez2007matrix}; there is also more recent evidence suggesting that in the best case $\chi = \lfloor L/2 \rfloor + 1$ \cite{klimov2023translationinvariantmatrixproductstates}.
Leaving the search for the optimal TI MPS representation of the $W$-state to dedicated studies, we pose the following pragmatic question: is there a more natural manifestly TI tensor network representation of the $W$-state in terms of site- and system-size-independent tensors, and can it be generalized to express other states suffering similar limitation of the TI MPS form?

\begin{figure}
  \subfloat{\label{fig:h1genmpo}}
  \subfloat{\label{fig:contr}}
  \subfloat{\label{fig:tensor}}
  \includegraphics[width=\columnwidth]{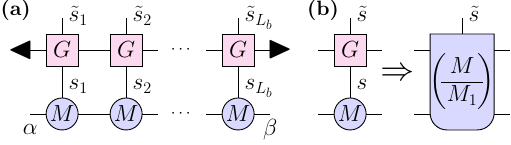}
  \caption{Action of MPO representation of $H_1$ on MPS $\ket{\psi}$.
    (a) Initial tensor network.
    (b) Contraction of inner physical indices.
  }
\end{figure}

Structurally, the $W$-state in Eq.~(\ref{eq:wstate}) can be viewed as the result of the action of a TI operator
\begin{equation}
  \label{eq:h1gen}
  Q = \sum_{i=1}^{L_b}q_i,
\end{equation}
whose identical strictly local terms $q$ have support on individual sites of a vacuum state representable as a TI MPS with site- and system-size-independent tensors. Specifically, for the $W$-state, the vacuum state is $\ket{00\dots 0}$ and $h = X/\sqrt{L}$. In general, the operator $Q$ in Eq.~(\ref{eq:h1gen}) acting on an MPS $\ket{\psi}$ generated by the tensor $M^s$ can be expressed as a matrix product operator (MPO) \cite{McCulloch_2007,mcculloch2008infinitesizedensitymatrix} defined by tensors
    \begin{equation}
      \label{eq:h1genmpo}
      \terml = \begin{pmatrix}
        0 & 1
      \end{pmatrix},\quad
      G = \begin{pmatrix}
        \mathbf{1} & \mathbf{0} \\
        \mathbf{q} & \mathbf{1}
      \end{pmatrix},\quad
      \termr = \begin{pmatrix}
        1 & 0
      \end{pmatrix}^T,
    \end{equation}
where the elements of $G$ act on the physical degrees of freedom and $\mathbf{q}$ represents terms of $Q$ with support on individual sites.
The state $Q\ket{\psi}$ as a tensor network is shown in Fig.~\ref{fig:h1genmpo}.

Contracting inner physical indices as shown in Fig.~\ref{fig:contr}, we obtain tensors of the form
\begin{equation}
  \label{eq:ttitensor}
  \sma{M}{M_1}^{\tilde s} = \begin{pmatrix}
    M^{\tilde s} & \mathbf{0} \\
    M_1^{\tilde s} & M^{\tilde s}
  \end{pmatrix},
\end{equation}
where $M^{\tilde s}$ is the tensor generating $\ket{\psi}$ and $M_1^{\tilde s} = \sum_{s} \mathbf{q}_s^{\tilde s}M^s$. Then, the contraction of the tensor network in Fig.~\ref{fig:h1genmpo} is given by
\begin{equation}
  \label{eq:ttimps}
  \begin{aligned}
    H_1\ket{\psi} = \sum_{\{s_i\}}&\Tr{\mathcal{T} \sma{M}{M_1}^{s_1}\sma{M}{M_1}^{s_2}\cdots \sma{M}{M_1}^{s_{L_b}}}\\
    &\times \ket{s_1s_2\dots s_{L_b}},
  \end{aligned}
\end{equation}
where
\begin{equation}
  \label{eq:ttimpstau}
  \mathcal{T} = (\termr\cdot \terml)\otimes B_{\alpha\beta} = \begin{pmatrix}
    \mathbf{0} & B_{\alpha\beta} \\
    \mathbf{0} & \mathbf{0}
  \end{pmatrix}
\end{equation}
is the ``twist'' matrix resulting from the terminations of the MPO and the boundary conditions of $\ket{\psi}$ encoded by the tensor $B_{\alpha\beta}$.

For the $W$-state, we have $M^0 = M_1^1 = \begin{pmatrix}1\end{pmatrix}$, $M^1 = M_1^0 = \begin{pmatrix}0\end{pmatrix}$ and $B_{\alpha\beta}=\begin{pmatrix}1\end{pmatrix}$ giving
\begin{equation}
  \sma{M}{M_1}^0 = \begin{pmatrix}
    1 & 0 \\
    0 & 1
  \end{pmatrix},
  \sma{M}{M_1}^1 = \begin{pmatrix}
    0 & 0 \\
    1 & 0
  \end{pmatrix},
  \mathcal{T} = \begin{pmatrix}
    0 & 1 \\
    0 & 0
  \end{pmatrix},
\end{equation}
which recovers the state's bulk MPS commonly used with OBC terminations to compensate for the absence of an adequate TI MPS representation.

When $\ket{\psi}$ is a state with PBCs, $B_{\alpha\beta}=\delta_{\alpha\beta}$ and the construction of Eq.~(\ref{eq:ttimps}) is manifestly TI since it represents the action of a TI Hamiltonian on a TI state.
This means that the position of $\mathcal{T}$ is irrelevant or,
\begin{equation}
  \label{eq:cyclic}
  \begin{aligned}
    &\Tr{\mathcal{T} \sma{M}{M_1}^{s_1}\sma{M}{M_1}^{s_2}\cdots \sma{M}{M_1}^{s_{L_b}}}\\
    &= \Tr{\mathcal{T} \sma{M}{M_1}^{s_{L_b}}\sma{M}{M_1}^{s_1}\cdots \sma{M}{M_1}^{s_{L_b-1}}},
  \end{aligned}
\end{equation}
and so on.
In fact, if we drop the requirement that the tensor $M_1$ represents the action of some operator on the physical index of tensor $M$ and assume the former to be completely arbitrary, or even assume inhomogeneous site-dependent tensors of the form given in Eq.~(\ref{eq:ttitensor}), the cyclic property like that in Eq.~(\ref{eq:cyclic}) will still hold, which can readily be verified by a direct calculation using the identity
\begin{equation}
  \label{eq:smaidentity}
  \sma{A}{A_1}^{s_1}\sma{B}{B_1}^{s_2} = \sma{AB}{A_1B+AB_1}^{s_1s_2}.
\end{equation}
With this in mind, to emphasize the translational invariance of the construction, let us define the (homogeneous) TTI MPS representation of a state $\ket{\psi'}$ in terms of the ``twisted'' trace operator $\ttr\left\{[\cdots]\right\} \equiv \Tr{\mathcal{T}[\cdots]}$ and two arbitrary tensors $M$ and $M_1$ of the same shape as follows:
\begin{equation}
  \begin{aligned}
    \ket{\psi'} = \sum_{\{s_i\}}&\ttr\left\{\sma{M}{M_1}^{s_1}\sma{M}{M_1}^{s_2}\cdots \sma{M}{M_1}^{s_{L_b}}\right\} \\
    &\times \ket{s_1s_2\dots s_{L_b}}.
  \end{aligned}
\end{equation}

Our TTI MPS framework is, effectively, a more analytically convenient variant of the MPS ``single-mode approximation'' (SMA) used as a variational ansatz in Refs.~\cite{Haegeman_2012,Haegeman_2013,lin2019exact}.
The primary advantage of the TTI MPS formulation over the traditional SMA is its close resemblance to the standard TI MPS form.
This similarity allows us to apply the extensive toolset developed for TI MPS to TTI MPSs with only slight modifications.
Clearly, any state that has a finite-bond-dimension  TI MPS representation can trivially be expressed as a TTI MPS; the converse, however, is generically not true (e.g., the $W$-state).
In this sense, the TTI MPS can be seen as an extension of the TI MPS.
In App.~\ref{app:ground}, we present several demonstrations of the analytical merits of the TTI MPS form, along with additional insights into the quasi-particle models discussed in Refs.~\cite{Omiya_2023quantum, Chandran_2023}.

\section{New exact \texorpdfstring{$E=0$}{E=0} MPS scars in the PXP and PPXPP models}
\label{sec:eigenstates}
\begin{figure}
  \subfloat{\label{fig:eepxp}}
  \subfloat{\label{fig:eeppxpp}}
  \includegraphics[width=\columnwidth]{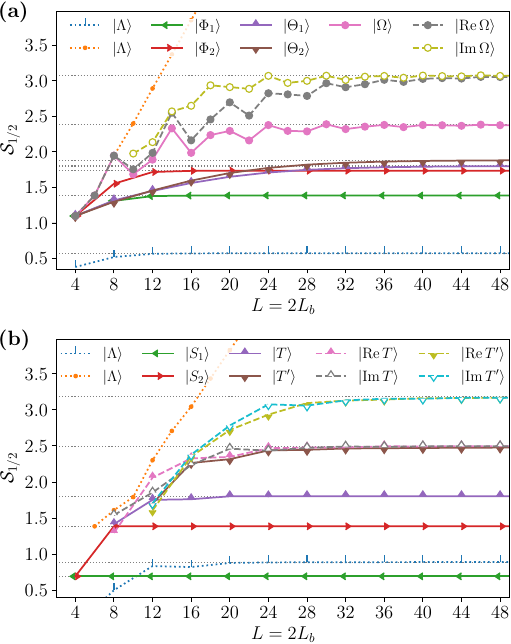}
  \caption{Dependence of bipartite entanglement entropy on the system size in (a) the PXP and (b) the PPXPP models.
    For each state, the system sizes are chosen such that both half-systems contain an integer number of local basis blocks.
    We plot two characteristic entanglement entropies of the maximally entangled (with respect to its standard bipartition) volume-law zero energy eigenstate $\ket{\Lambda}$ found in Ref.~\cite{Ivanov_2025}, which for the two systems is defined on the appropriate constrained Hilbert space.
    The entanglement entropy of $\ket{\Lambda}$ with respect to the standard bipartition, essentially, continues growing linearly at the rate shown, whereas the entropy with respect to the entanglement minimizing bipartition quickly saturates at a low value.
    The increasing saturation values of the entanglement entropy of the consecutively introduced area-law states reflect a certain ``complexity hierarchy'' among them.
    Note also that saturation of the entanglement entropy for the new states introduced in this work typically occurs in significantly larger systems than that for the previously known states.}
  \label{fig:ee}
\end{figure}
In this section, for easy reference, we will list the MPS representations of all currently known eigenstates of the PXP and PPXPP Hamiltonians with PBC.
We will mark the states that have never appeared in the literature before as ``\textbf{new}.''
For the states first introduced in other works (not necessarily as MPSs) we will provide a reference to the original paper.

All our new states will exhibit a more complex structure compared to the previously known ones.
This can be seen in Fig.~\ref{fig:ee} where we show the dependence of the bipartite entanglement entropy on the system size for all the states included in this section: in both the PXP and PPXPP models, the previously known eigenstates are the three states with the lowest values of the saturation entanglement entropy.

Our naming convention for the new eigenstates will roughly adhere to that used in previous works.
We will use subscripts, as in $\ket{\Phi_2} = T_x\ket{\Phi_1}$, to denote pairs of translational symmetry breaking states related by the single-site translation operator $T_x$.
The volume-entangled states from Ref.~\cite{Ivanov_2025} (expressed using the BA TI MPS form) as $\ket{\Lambda}$.
For the states in the TTI MPS form, we will use the ``prime'' symbol; for example $\ket{T'}$ is a TTI MPS on top of the TI MPS representation of the state $\ket{T}$.
Some of our states will be complex-valued, which means, given the real-valuedness of $H_\text{PXP/PPXPP}$, that their separate real and imaginary parts will themselves be linearly independent eigenstates.
\subsection{PXP model}
The $E=0$ exact eigenstates of the PXP model with PBC are the following:
\begin{subequations}
  \begin{flalign}
    \label{eq:lambdapxpmps}
    &\mps{
      \begin{aligned}[t]&\ket{\Lambda} \\ &\text{Ref.~\cite{Ivanov_2025}} \end{aligned}
      }{
      \underbrace{
      \frac{1}{2}\begin{pmatrix}
        1 & 1 \\ 1 & 1 
      \end{pmatrix}}_{M^{00}},\,
      \underbrace{
      -2\begin{pmatrix}
        0 & 0  \\ 1 & 0
      \end{pmatrix}}_{M^{11}},\,
      \underbrace{
      \begin{pmatrix}
        0 & 0 \\ 0 & 0
      \end{pmatrix}}_{M^{01/10}}};&&\\
    \label{eq:phi1mps}
    &\mps{
      \begin{aligned}[t]&\ket{\Phi_1} \\ &\text{Ref.~\cite{lin2019exact}} \end{aligned}
      }{
      \underbrace{
      \begin{pmatrix}
        0 & -1 \\ 1 & 0 
      \end{pmatrix}}_{M^O},\,
      \underbrace{
      \begin{pmatrix}
        0 & 0  \\ 0 & -\sqrt{2}
      \end{pmatrix}}_{M^L},\,
      \underbrace{
      \begin{pmatrix}
        \sqrt{2} & 0 \\ 0 & 0
      \end{pmatrix}}_{M^R};}&&\\
    \label{eq:phi2mps}
    &\mps{
      \begin{aligned}[t]&\ket{\Phi_2} \\ &\text{Ref.~\cite{lin2019exact}} \end{aligned}
      }{
      \underbrace{
      \begin{pmatrix}
        0 & -1 & 0 \\
        1 & 0 & 0 \\
        0 & 0 & 0
      \end{pmatrix}}_{M^O},\,
      \underbrace{
      \begin{pmatrix}
        \sqrt{2} & 0 & 0 \\
        0 & 0 & 0 \\
        -\sqrt{2} & 0 & 0
      \end{pmatrix}}_{M^L},\,
      \underbrace{
      \begin{pmatrix}
        -\sqrt{2} & 0 & -\sqrt{2} \\
        0 & 0 & 0 \\
        0 & 0 &  0
      \end{pmatrix}}_{M^R};}&&\\
    \label{eq:theta1mps}
    &\mps{
      \begin{aligned}[t]&\bf\ket{\Theta_1} \\ &\textbf{new} \end{aligned}
      }{
      \underbrace{
      \begin{pmatrix}
        0 & 3 & -1 & 0 \\ -1 & 0 & 0 & 1 \\ 2 & 0 & 0 & -6 \\ 0 & 0 & 0 & 0
      \end{pmatrix}}_{M^O},\,
      \underbrace{
      \begin{pmatrix}
        0 & 0 & 0 & 0 \\ 0 & 0 & 0 & 0 \\ 0 & 3 & -1 & 0 \\ 1 & 0 & 0 & -3
      \end{pmatrix}}_{M^L},\,
      \underbrace{
      \begin{pmatrix}
        1 & 0 & 0 & 0 \\ 0 & 3 & 0 & 0 \\ 0 & 0 & 0 & 0 \\ 0 & 0 & 0 & 0
      \end{pmatrix}}_{M^R};\hspace{-2em}}&&\\
    \label{eq:theta2mps}
    &\mps{
      \begin{aligned}[t]&\bf\ket{\Theta_2} \\ &\textbf{new} \end{aligned}
      }{
      \underbrace{
      \begin{pmatrix}
        0 & 3 & -1 & 0 \\ -\frac{5}{3} & 0 & 0 & 3 \\ 0 & 0 & 0 & 0 \\ 0 & 0 & 0 & 0
      \end{pmatrix}}_{M^O},\,
      \underbrace{
      \begin{pmatrix}
        \frac{9}{2} & 0 & 0 & -\frac{9}{2} \\ 0 & 0 & 0 & 0 \\ 0 & -3 & 1 & 0 \\ \frac{3}{2} & 0 & 0 & -\frac{3}{2}
      \end{pmatrix}}_{M^L},\,
      \underbrace{
      \begin{pmatrix}
        0 & 0 & 0 & 0 \\ 0 & -1 & 0 & 0 \\ 0 & 0 & 0 & 0 \\ 1 & 0 & 0 & -3
      \end{pmatrix}}_{M^R};\hspace{-2em}}&&\\
    \label{eq:omegamps}
    &\mps{
      \begin{aligned}[t]&\bf\ket{\Omega} \\ &\textbf{new} \end{aligned}
      }{
    \underbrace{
    \begin{pmatrix}
      0 & \frac{1}{8} & 2 & \frac{5}{6} \\
      -8 & -2 & 0 & -2 \\
      -\frac{1}{2} & -\frac{1}{8} & 0 & -\frac{1}{8} \\
      0 & \frac{1}{2} & 8 & -2 \\
    \end{pmatrix}}_{M^0},\,
    \underbrace{
    \frac{i}{\sqrt{3}}\begin{pmatrix}
      0 & 1 & 0 & 0 \\
      0 & 0 & 0 & 0 \\
      0 & 0 & 0 & 1 \\
      0 & 0 & 0 & 0 \\
    \end{pmatrix}}_{M^1}}.
  \end{flalign}
\end{subequations}
Note that $\ket{\Omega}$ is the first reported exact eigenstate (comprised of two orthogonal eigenstates $\ket{\Re\Omega}$ and $\ket{\Im\Omega}$) that is defined for both even- and odd-length PBC chains.

Some of the basic properties of the states in Eqs.~(\ref{eq:lambdapxpmps})--(\ref{eq:omegamps}) --- such as $T_x$, $\mathcal{C}$, and $\mathcal{I}$ symmetry quantum numbers, the saturation values of the bipartite entanglement entropy $\lim_{L\to\infty}\mathcal{S}_{1/2}$ [cf.~Fig.~\ref{fig:eepxp}], and correlation lengths $\xi$ --- are given in Table~\ref{tab:zmpxp}.
The instances where the correlation length is defined correspond to injective MPSs.
Derivations of these and some additional properties (such as norms and various overlaps) are available in Apps.~\ref{app:propspxp} and \ref{app:entpxp}.

\begin{table}[h]
\caption{\label{tab:zmpxp}Properties of the $E=0$ eigenstates of $H_\text{PXP}$.}
\begin{ruledtabular}
\begin{tabular}{lllllll}
  State & $T_x$ & $\mathcal{C}$, $\mathcal{I}$ & $\lim_{L\to\infty} \mathcal{S}_{1/2}$ & $\xi$ \\
  $\ket{\Lambda}$ & $+1$ & $+1$ & $\sim 0.5895$ & $\sim 1.0390 $ \\
  $\ket{\Phi_1}$ & --- & $(-1)^{L_b}$ & $ 2\log 2$ & $\sim 1.8205$ \\
  $\ket{\Phi_2}$ & --- & $(-1)^{L_b}$ & $ 2\log(3/2^{1/3})$ & $\sim 1.8205$ \\
  $\ket{\Theta_1}$ & --- & $(-1)^{L_b}$ & $\sim 1.8010$ & $\sim 6.9834$ \\
  $\ket{\Theta_2}$ & --- & $(-1)^{L_b}$ & $\sim 1.8839$ & $\sim 6.9834$ \\
  $\ket{\Omega}$ & $+1$ & --- & $\sim 2.3811$ & $\sim 7.9294$ \\
  $\ket{\Re\Omega}$ & $+1$ & $+1$ & $\mathcal{S}^{\ket{\Omega}}_{1/2}+\log2$ & --- \\
  $\ket{\Im\Omega}$ & $+1$ & $-1$ & $\mathcal{S}^{\ket{\Omega}}_{1/2}+\log2$ & --- 
\end{tabular}
\end{ruledtabular}
\end{table}

\subsection{PPXPP model}
The $E=0$ exact eigenstates of the PPXPP model with PBC that are representable as TI MPSs are the following:
\begin{subequations}
  \begin{flalign}
    \label{eq:lambdappxpp}
    &\mps{
      \begin{aligned}[t]&\ket{\Lambda} \\ &\text{Ref.~\cite{Ivanov_2025}} \end{aligned}
      }{
      \underbrace{
      \begin{pmatrix}
        0 & 0 & 1 \\ 0 & 0 & -1 \\ 1 & 1 & 1
      \end{pmatrix}}_{M^{00}},\,
      \underbrace{
      \begin{pmatrix}
        0 & 0 & 0  \\ -1 & 0 & 0 \\ 0 & 0 & 0
      \end{pmatrix}}_{M^{11}},\,
      \underbrace{
      \begin{pmatrix}
        0 & 0 & 0 \\ 0 & 0 & 0 \\ 0 & 0 & 0
      \end{pmatrix}}_{M^{01/10}};}&&\\
    \label{eq:s1mps}
    &\mps{
      \begin{aligned}[t]&\ket{S_1} \\ &\text{Ref.~\cite{Surace_2021}} \end{aligned}
      }{
      \underbrace{
      \begin{pmatrix}
        0 & 1 \\ 0 & 0
      \end{pmatrix}}_{M^O},\,
      \underbrace{
      \begin{pmatrix}
        0 & 0  \\ -1 & 0
      \end{pmatrix}}_{M^L},\,
      \underbrace{
      \begin{pmatrix}
        0 & 0 \\ 1 & 0
      \end{pmatrix}}_{M^R}};&&\\
    \label{eq:s2mps}
    &\mps{
      \begin{aligned}[t]&\ket{S_2} \\ &\text{Ref.~\cite{Surace_2021}} \end{aligned}
      }{
      \underbrace{
      \begin{pmatrix}
        0 & 1 & 0 \\ 0 & 0 & 1 \\ 0 & 0 & 0
      \end{pmatrix}}_{M^O},\,
      \underbrace{
      \begin{pmatrix}
        0 & 0 & 0  \\ 0 & 0 & 0 \\ 0 & 1 & 0
      \end{pmatrix}}_{M^L},\,
      \underbrace{
      \begin{pmatrix}
        0 & 0 & 0 \\ -1 & 0 & 0 \\ 0 & 0 & 0
      \end{pmatrix}}_{M^R}};&&\\
    \label{eq:tmps}
    &\mps{
      \begin{aligned}[t]&\bf\ket{T} \\ &\textbf{new} \end{aligned}
      }{
      \underbrace{
      \begin{pmatrix}
        1 & 0 & \gamma^* - 1 \\
        0 & \gamma & 0 \\
        1 & 0 & \gamma^* - 1
      \end{pmatrix}}_{M^O},\,
      \underbrace{
      \begin{pmatrix}
        0 & 1 & 0 \\
        0 & 0 & 0 \\
        0 & 0 & 0
      \end{pmatrix}}_{M^L},\,
      \underbrace{
      \begin{pmatrix}\
        0 & 0 & 0 \\
        0 & 0 & 1 \\
        0 & 0 & 0
      \end{pmatrix}}_{M^R}};&&\\
    \label{eq:s1tti}
    &\mps{
      \begin{aligned}[t]&\bf\ket{S'_1} \\ &\textbf{new} \end{aligned}
      }{
      \underbrace{
      \begin{pmatrix}
        1 & 1 \\ 0 & 0
      \end{pmatrix}}_{M_1^O},\,
      \underbrace{
      \begin{pmatrix}
        0 & 0  \\ 0 & 0
      \end{pmatrix}}_{M_1^L},\,
      \underbrace{
      \begin{pmatrix}
        0 & 0 \\ 0 & 0
      \end{pmatrix}}_{M_1^R}};&&\\
    \label{eq:s2tti}
    &\mps{
      \begin{aligned}[t]&\bf\ket{S'_2} \\ &\textbf{new} \end{aligned}
      }{
      \underbrace{
      \begin{pmatrix}
        0 & 1 & 0 \\ 0 & 1 & 1 \\ 0 & 0 & 0
      \end{pmatrix}}_{M_1^O},\,
      \underbrace{
      \begin{pmatrix}
        0 & 0 & 0  \\ 0 & 0 & 0 \\ 0 & 0 & 0
      \end{pmatrix}}_{M_1^L},\,
      \underbrace{
      \begin{pmatrix}
        0 & 0 & 0 \\ 0 & 0 & 0 \\ 0 & 0 & 0
      \end{pmatrix}}_{M_1^R}};&&\\
    \label{eq:ttti}
    &\mps{
      \begin{aligned}[t]&\bf\ket{T'} \\ &\textbf{new} \end{aligned}
      }{
      \underbrace{
      \begin{pmatrix}
        0 & \gamma^* - 1 & 0 \\
        1 & 0 & -1 \\
        0 & -1 & 0
      \end{pmatrix}}_{M_1^O},\,
      \underbrace{
      \begin{pmatrix}
        0 & 0 & 1 \\
        0 & 0 & 0 \\
        0 & 0 & 0
      \end{pmatrix}}_{M_1^L},\,
      \underbrace{
      \begin{pmatrix}
        0 & 0 & -1 \\
        0 & 0 & 0 \\
        0 & 0 & 0
      \end{pmatrix}}_{M_1^R}}.
  \end{flalign}
\end{subequations}
Here $\gamma = 2^{-1/6}e^{\pm i\pi/4}$; the two choices for $\gamma$ generate two linearly independent, but not orthogonal, states related by complex conjugation.

While all the MPSs representing the known eigenstates of the PXP model listed in Eqs.~(\ref{eq:lambdapxpmps})--(\ref{eq:omegamps}) are injective, which suggests that they are native to the bases they are expressed in, this is not the case for some of the MPSs in Eqs.~(\ref{eq:lambdappxpp})--(\ref{eq:tmps}) --- here, we do not consider the TTI MPSs, whose bulk tensors are non-injective by construction.
Specifically, only the representations of $\ket{\Lambda}$ in Eq.~(\ref{eq:lambdappxpp}) and $\ket{T}$ in Eq.~(\ref{eq:tmps}) are injective MPSs, whereas those of $\ket{S_1}$ and $\ket{S_2}$ in Eqs.~(\ref{eq:s1mps}) and (\ref{eq:s2mps}) are not.
It is easy to see that the state $\ket{S_1}$ is a $T_x^2$-invariant version of the state
\begin{equation}
  \label{eq:phi2ppxpp}
  \ket{\phi_2} = \bigotimes_{i_b=1}^{L_b/2}[\ket{R}-\ket{L}]_{2i_b-1}\otimes\ket{O}_{2i_b},
\end{equation}
first reported in Ref.~\cite{Surace_2021}, which means that when $L_b$ is even there exists a $p$-periodic (with $p = 2$ blocks) decomposition of $\ket{S_1}$ into two injective MPSs with bond dimension 1, whereas when $L_b$ is odd $\ket{S_1}$ vanishes~\cite{perez2007matrix}.
Similarly, $\ket{S_2}$ is decomposable into injective MPSs with bond dimensions $1$ and $2$ when $L_b$ is even, and when $L_b$ is odd $\ket{S_2} = 0$.
In Sec.~\ref{sec:extrap_pxp_p}, we will re-examine states such as $\ket{\phi_2}$ and other states from Ref.~\cite{Surace_2021} utilizing the framework to be introduced in Sec.~\ref{sec:proofs}.

Surprisingly, TTI MPSs on top of the states $\ket{S_1}$ and $\ket{S_2}$ with the $M_1$ tensors given in Eqs.~(\ref{eq:s1tti}) and (\ref{eq:s2tti}), respectively, generate previously unknown exact eigenstates $\ket{S'_1}$ and $\ket{S'_2}$ that do not vanish when $L_b$ is odd; when $L_b$ is even, $\ket{S'_1}$ and $\ket{S'_2}$ are identical to, respectively, $\ket{S_1}$ and $\ket{S_2}$.

On the other hand, a TTI MPS $\ket{T'}$ on top of the new state $\ket{T}$ generates a new complex-valued state which is different from $\ket{T}$ in all system sizes.
This means that $\ket{T}$ and $\ket{T'}$ together generate four new linearly independent eigenstates of $H_\text{PPXPP}$.

Table~\ref{tab:zmppxpp} summarizes the same set of basic properties for the eigenstates of the PPXPP in Eqs.~(\ref{eq:s1mps})--(\ref{eq:ttti}) as those given in Table~\ref{tab:zmpxp} for the eigenstates of the PXP model.
As before, the instances where the correlation length is defined correspond to injective MPSs.
Derivations of these and some additional properties (such as norms and various overlaps) were performed analogously to those in App.~\ref{app:propspxp}.
\begin{table}[h]
\caption{\label{tab:zmppxpp}Properties of the $E=0$ eigenstates of $H_\text{PPXPP}$.}
\begin{ruledtabular}
\begin{tabular}{lllllll}
  State & $T_x$ & $\mathcal{C}$, $\mathcal{I}$ & $\lim_{L\to\infty}\mathcal{S}_{1/2}$ & $\xi$ \\
  $\ket{S_1}$, $\ket{S'_1}$ & --- & $(-1)^{\left\lfloor \frac{L_b}{2} \right\rfloor}$ & $\log 2$ & --- \\
  $\ket{S_2}$, $\ket{S'_2}$ & --- & $(-1)^{\left\lfloor \frac{L_b}{2} \right\rfloor}$ & $2\log 2$ & --- \\
  $\ket{\Lambda}$ & $+1$ & $+1$ & $\sim 0.8900$ & $\sim 1.7441$ \\
  $\ket{T}$ & --- & $+1$ & $\sim 1.8035$ & $\sim 3.4882$ \\
  $\ket{T'}$ & --- & $-1$ & $\mathcal{S}^{\ket{T}}_{1/2}+\log2$ & --- \\
  $\ket{\Re T}$ & $+1$ & $+1$ & $\mathcal{S}^{\ket{T}}_{1/2}+\log2$ & --- \\
  $\ket{\Im T}$ & $-1$ & $+1$ & $\mathcal{S}^{\ket{T}}_{1/2}+\log2$ & --- \\
  $\ket{\Re T'}$ & $-1$ & $-1$ & $\mathcal{S}^{\ket{T'}}_{1/2}+\log2$ & --- \\
  $\ket{\Im T'}$ & $+1$ & $-1$ & $\mathcal{S}^{\ket{T'}}_{1/2}+\log2$ & ---  
\end{tabular}
\end{ruledtabular}
\end{table}

\section{Proofs}
\label{sec:proofs}
In the previous section we introduced several new $E=0$ eigenstates of the PXP and PPXPP chains with PBC without providing any proofs.
Now, our objective is to develop a framework that allows for direct proofs of such $E=0$ eigenstates in kineticaly constrained models.
This framework will provide a procedure for generating conditions based on the specific properties of a given Hamiltonian that are sufficient for an MPS to represent its exact zero energy eigenstate.
Our approach is applicable to both the newly introduced and previously reported eigenstates of PXP-type models.
This suggests that it should also be applicable to $E=0$ eigenstates of these models yet to be discovered.

For concreteness, let us start by revisiting the technique used in Ref.~\cite{lin2019exact} to prove the eigenstate $\ket{\Phi_1}$ of the PXP model in the context of the following two technical results (applicable for both PBC and OBC cases assuming an appropriate definition of the Fibonacci subspace $\mathcal{H}^1$):
\begin{lemma}
  \label{lemma:fibh1}
  Any $\ket{\psi} \in \mathcal{H}^1$ such that $H_1\ket{\psi} = \lambda\ket{\psi}$ is an exact eigenstate of $H_\mathrm{PXP}$ with energy $\lambda$.
  \begin{proof}
    Let $\mathcal{P}_f$ be a projector onto the Fibonacci subspace.
    The condition that $\ket{\psi} \in \mathcal{H}^1$ is equivalent to $\mathcal{P}_f\ket{\psi} = \ket{\psi}$.
    Therefore, acting with $\mathcal{P}_f$ on both sides of the eigenvalue equation $H_1\ket{\psi} = \lambda\ket{\psi}$, we get $\mathcal{P}_fH_1\ket{\psi} = \lambda\ket{\psi}$.
    On the subspace $\mathcal{H}^1$, $H_\text{PXP} = \mathcal{P}_f H_\text{PXP}\mathcal{P}_f = \mathcal{P}_f H_1\mathcal{P}_f$, where we have used $\mathcal{P}_f H_2 \mathcal{P}_f = 0$ easily seen by inspection of individual terms in $H_2$. Therefore, $H_\text{PXP}\ket{\psi} = \mathcal{P}_f H_1 \ket{\psi} = \lambda\ket{\psi}$.
  \end{proof}
\end{lemma}

\begin{corollary}
  \label{corr:fibh1}
  Any $\ket{\psi} \in \mathcal{H}^1$ such that
  $H_1\ket{\psi} = \lambda\ket{\psi}$ is annihilated by $H_2$ \footnote{We can make an even stronger claim: any such $\ket{\psi}$ is \emph{locally} annihilated by individual terms of $H_2$. This implies that any such $\ket{\psi}$ is expected to be short-range-entangled and have an MPS form.}.
\end{corollary}

The proofs in Ref. \cite{lin2019exact} establishing their state $\ket{\Phi_1}$ as an exact eigenstate of $H_\text{PXP}$ first explicitly checked that the state (residing in $\mathcal{H}^1$ by construction) was annihilated by $H_2$ and then showed that it was an exact eigenstate of $H_1$ with zero energy. Per Corollary~\ref{corr:fibh1}, however, it would suffice to show only the latter.
We state this not for the sake of pointing out that certain calculations were unnecessary, but rather to appreciate the fact that the approach used in Ref. \cite{lin2019exact} to prove that an MPS defined on $\mathcal{H}^1$ is an exact eigenstate of $H_1$ can be generalized into a framework for proving (and potentially discovering) new exact eigenstates of $H_\text{PXP}$ satisfying the conditions of Lemma~\ref{lemma:fibh1}.
We summarize such generalization in the following:
\begin{theorem}
  \label{thrm:h1eig}
  Suppose matrices $M^s$, $s\in \{O, L, R\}$ are an MPS representation of $\ket{\psi} \in \mathcal{H}^1$, satisfying
  \begin{equation}
    \label{eq:fibconstr}
    M^RM^L = \mathbf{0}.
  \end{equation}
  In the case of PBC, where $M^s$ define a TI MPS, if there exists a matrix $X$ such that
  \begin{subequations}
    \begin{align}
      \label{eq:h1eigcond_fo}
      &[X, M^O] = F^O,\\
      \label{eq:h1eigcond_fl}
      &[X, M^L] = F^L,\\
      \label{eq:h1eigcond_fr}
      &[X, M^R] = F^R,
    \end{align}
  \end{subequations}
  where
  \begin{equation}
    \label{eq:fs}
    F^O = M^{L} + M^{R},\quad F^L = F^R = M^{O} ~,
  \end{equation}
  then $\ket{\psi}$ satisfies the conditions of Lemma~\ref{lemma:fibh1} with $\lambda = 0$ (i.e., is an exact zero energy eigenstate of $H_\text{PXP}$).
  In the case of OBC, if the terminations are chosen to be left and right eigenvectors $v^T$ and $w$ of $X$ with eigenvalues $\lambda_v$ and $\lambda_w$, then $\ket{\psi}$ satisfies the conditions of Lemma~\ref{lemma:fibh1} with $\lambda = \lambda_v - \lambda_w$.
  \begin{proof}
    The action of $H_1$ on $\ket{\psi}$ has the following TTI MPS representation:
    \begin{equation}
      \mps{H_1\ket{\psi}}{\left\{\sma{M}{F}^s\right\}}.
    \end{equation}
    To prove the Theorem for the case of PBC, we want to show that
    \begin{equation}
      \ttr\left\{\sma{M}{F}^{s_1}\sma{M}{F}^{s_2}\cdots \sma{M}{F}^{s_{L_b}}\right\} = 0
    \end{equation}
    for any choice of the physical indices.
    The identity in Eq.~(\ref{eq:smaidentity}) gives
    \begin{equation}
      \sma{M}{F}^{s_1}\sma{M}{F}^{s_2} = \sma{MM}{MF + FM}^{s_1s_2}.
    \end{equation}
    Invoking Eqs.~(\ref{eq:h1eigcond_fo})--(\ref{eq:h1eigcond_fr}),
    \begin{equation}
      \label{eq:fmmf}
      \begin{aligned}
        &F^{s_1}M^{s_2} + M^{s_1}F^{s_2} \\
        &= [X,M^{s_1}]M^{s_2} + M^{s_1}[X,M^{s_2}]\\
        &=[X, M^{s_1}M^{s_2}].
      \end{aligned}
    \end{equation}
    where in the last line we applied the commutator identity $[X, A]B + A[X, B] = [X, AB]$.
    Thus,
    \begin{equation}
      \sma{M}{F}^{s_1}\sma{M}{F}^{s_2} = \sma{MM}{[X,MM]}^{s_1s_2}.
    \end{equation}

    Suppose, up to some $k$ we have
    \begin{equation}
      \label{eq:indm}
      \sma{M}{F}^{s_1}\cdots\sma{M}{F}^{s_k} = \sma{M\cdots M}{[X,M\cdots M]}^{s_1\cdots s_k}.
    \end{equation}
    Then
    \begin{equation}
      \label{eq:q}
      \sma{M}{F}^{s_1}\cdots\sma{M}{F}^{s_k}\sma{M}{F}^{s_{k+1}} = \sma{M\cdots M}{Q}^{s_1\cdots s_{k+1}}
      \end{equation}
    where [again invoking Eqs.~(\ref{eq:h1eigcond_fo})--(\ref{eq:h1eigcond_fr}) and the same commutator identity]
    \begin{equation}
      \begin{aligned}
        Q&= [X, M^{s_1}M^{s_2}\cdots M^{s_k}]M^{s_{k+1}} \\
         &\qquad + M^{s_1}M^{s_2}\cdots M^{s_k}[X, M^{s_{k+1}}] \\
         &= [X, M^{s_1}M^{s_2}\cdots M^{s_{k+1}}].
      \end{aligned}
    \end{equation}
    Clearly, by induction, Eq.~(\ref{eq:indm}) holds for any $k\in [1..L_b]$, which means that for the case of PBC
    \begin{equation}
      \label{eq:h1psimps}
      \begin{aligned}
        &\ttr\left\{\sma{M}{F}^{s_1}\sma{M}{F}^{s_2}\cdots \sma{M}{F}^{s_{L_b}}\right\}\\
        &= \Tr{[X, M^{s_1}M^{s_2}\cdots M^{s_{L_b}}]} = 0,
      \end{aligned}
    \end{equation}
    where in the last line we used the tracelessness of the commutator.
    
    In the case of OBC, the amplitudes depend on the boundary of $\ket{\psi}$.
    Specifically, using Eqs.~(\ref{eq:ttimps}) and (\ref{eq:ttimpstau}) with $B_{\alpha\beta} = wv^T$,
    \begin{equation}
      \begin{aligned}
        &\mel{s_1s_2\cdots s_{L_b}}{H_1}{\psi}\\
        &= \Tr{wv^T[X, M^{s_1}M^{s_2}\cdots M^{s_{L_b}}]}\\
        &=v^T [X, M^{s_1}M^{s_2}\cdots M^{s_{L_b}}] w \\
        &= (\lambda_v - \lambda_w) v^T M^{s_1}M^{s_2}\cdots M^{s_{L_b}}  w,
      \end{aligned}
    \end{equation}
    which implies that $H_1\ket{\psi} = (\lambda_v - \lambda_w)\ket{\psi}$.
    
  \end{proof}
\end{theorem}

For an alternative proof of Theorem~\ref{thrm:h1eig} without the TTI MPS formalism (and a self-contained review of the main results and proof techniques of Ref.~\cite{lin2019exact}) see App.~\ref{app:proofh1eig}.

Per Theorem~\ref{thrm:h1eig}, any four matrices, $M^{O,L,R}$ and $X$, satisfying Eqs.~(\ref{eq:fibconstr}) and (\ref{eq:h1eigcond_fo})--(\ref{eq:h1eigcond_fr}) give an MPS representation of an exact eigenstate of $H_\text{PXP}$ with (and with OBC with the additional conditions in the Theorem). 
We can immediately rule out the prospect of finding any new finite energy eigenstates that would satisfy these equations: such states, being eigenstates of $H_1$, would be expected to have energies $m\sqrt{2}, m\in \mathbb{Z}^{\neq 0}$, which, based on numerics, is known to occur only in systems with OBC for two already known states $\ket{\Gamma_{1,2}}$ and $\ket{\Gamma_{2,1}}$ from Ref.~\cite{lin2019exact}.
Nevertheless, it may seem plausible that new eigenstates residing in the exponentially large nullspace of $H_\mathrm{PXP}$ could be found.
Unfortunately, however, that possibility is also ruled out by the following:

\begin{theorem}
\label{thrm:fibh1dims}
The subspaces spanned by states satisfying the requirements of Lemma~\ref{lemma:fibh1} are at most 1- and 4-dimensional in PBC and OBC systems, respectively.
Further, the unique state in PBC systems must have zero energy, be translationally invariant in the blocked basis, and possess definite $\mathcal{I}$ and $C$ quantum numbers equal to $(-1)^{L_b}$.
\begin{proof}
See App.~\ref{app:fibh1dims}.
\end{proof}
\end{theorem}

Thus, per Theorem~\ref{thrm:fibh1dims}, the states found in \cite{lin2019exact} exhaust all the possibilities given by Lemma~\ref{lemma:fibh1} and no other simultaneous eigenstates of $H_\text{PXP}$ and $H_1$ (with or without MPS representations) exist. It also follows from the pidgeonhole principle that all (infinitely many) non-trivial solutions of Eqs.~(\ref{eq:fibconstr}) and (\ref{eq:h1eigcond_fo})--(\ref{eq:h1eigcond_fr}), regardless of the size of the matrices, must yield representations of the same state $\ket{\Phi_1}$.

Although Theorem~\ref{thrm:h1eig} is not directly applicable to our quest for proving or discovering new exact eigenstates of $H_\text{PXP}$ [indeed, it is easy to verify by solving a linear system of equations for components of matrix $X$ that among the states in Eqs.~(\ref{eq:phi1mps})--(\ref{eq:omegamps}), assuming the state $\ket{\Omega}$ is expressed in the blocked basis, only $\ket{\Phi_1}$ satisfies its conditions], the intuition we have developed above informs the following generalization:
\begin{theorem}
  \label{thrm:genzm}
  Suppose matrices $M^s$, $s\in \{O, L, R\}$ are a TI MPS representation of $\ket{\psi} \in \mathcal{H}^1$.
  If there exists a matrix $X$ such that
  \begin{subequations}
    \begin{align}
      \label{eq:padding_fo}
      & [X, M^O] = F^O,\\
      \label{eq:padding_mofl}
      & M^O [X, M^L] = M^O F^L,\\
      \label{eq:padding_mlfl}
      & M^L [X, M^L] = M^L F^L, \\
      \label{eq:padding_frmo}
      & [X, M^R] M^O = F^R M^O,\\
      \label{eq:padding_frmr}
      & [X, M^R] M^R = F^R M^R,
    \end{align}
  \end{subequations}
  then $\ket{\psi}$ is an exact $E=0$ eigenstate of $H_\text{PXP}$ with PBC.
  
  \begin{proof}
    In general, for any $\ket{\psi} \in \mathcal{H}^1$, $H_1: \mathcal{H}^1 \to \mathcal{H}^1\oplus\mathcal{\overline H}^1$, where $\mathcal{\overline H}^1$ is a subspace orthogonal to $\mathcal{H}^1$ spanned by blocked computational basis state vectors with exactly one Rydberg blockade violation of the form $\ket{RL}$.

    We have $H_\text{PXP} \ket{\psi} = \mathcal{P}_f H_1 \ket{\psi}$.
    From the computational point of view, this means that the action of $H_\text{PXP}$ on $\ket{\psi}$ can be realized by first acting with $H_1$ and then projecting from $\mathcal{H}^1\oplus\mathcal{\overline H}^1$ onto $\mathcal{H}^1$, which is equivalent to simply discarding all the components of $H_1\ket{\psi}$ that belong to $\mathcal{\overline H}^1$.

    In terms of the formalism and notation introduced in the proof of Theorem~\ref{thrm:h1eig}, we want to find conditions on matrices $M^s$ that are less restrictive than Eqs.~(\ref{eq:h1eigcond_fo})--(\ref{eq:h1eigcond_fr}) and yet sufficient for Eq.~(\ref{eq:h1psimps}) to be true for any $\ket{s_1s_2\cdots s_{L_b}} \in \mathcal{H}^1$.

    Note that each time we invoked Eqs.~(\ref{eq:h1eigcond_fo})--(\ref{eq:h1eigcond_fr}) --- in Eq.~(\ref{eq:fmmf}) and Eq.~(\ref{eq:q}) --- $F^{s}$ entered as part of a matrix product $F^{s_i}M^{s_{i+1}}$ or $M^{s_i}F^{s_{i+1}}$.  
    Therefore, if $\exists X$ such that
    \begin{subequations}
      \begin{align}
        \label{eq:pfm}
        [X, M^{s_1}]M^{s_2} = F^{s_1}M^{s_2},\\
        \label{eq:pmf}
        M^{s_1}[X, M^{s_2}] = M^{s_1}F^{s_2}
      \end{align}
    \end{subequations}
    for any $s_1s_2 \neq RL$, the entire argument of Theorem~\ref{thrm:h1eig} proceeds without change for any $\ket{s_1s_2\cdots s_{L_b}} \in \mathcal{H}^1$, from which it follows that in the case of PBC $H_\text{PXP}\ket{\psi} = \mathcal{P}_fH_1\ket{\psi} = 0$, whereas in the case of OBC, $H_\text{PXP}\ket{\psi} = (\lambda_v - \lambda_w)\ket{\psi}$ for $\lambda_v$ and $\lambda_w$ as defined in Theorem~\ref{thrm:h1eig}.
    Note that Eqs.~(\ref{eq:pfm}) and (\ref{eq:pmf}) can be less restrictive than Eqs.~(\ref{eq:h1eigcond_fo})--(\ref{eq:h1eigcond_fr}) when some of the matrices $M^s$ are not full rank. Generally, matrices $M^L$ and $M^R$ which belong to an MPS corresponding to a state in $\mathcal{H}^1$ will be singular, so padding with them would tend to alleviate the inconsistencies which prevent any $X$ from satisfying Eqs.~(\ref{eq:h1eigcond_fo})--(\ref{eq:h1eigcond_fr}); we also expect states with singular $M^O$.
    
    Note that satisfying both (\ref{eq:pfm}) and (\ref{eq:pmf}) guarantees exact eigenstates in both OBC and PBC, and we chose these conditions for our first quick illustration.
    However, these conditions are not satisfied in most of our states that are exact in PBC but do not have exact OBC counterparts.
    Our PBC states instead satisfy somewhat weaker conditions stated in the Theorem, which we turn to next.

    Let us now restrict our attention to the case of PBCs.
    Proceeding with the proof of Theorem~\ref{thrm:h1eig} up to Eq.~(\ref{eq:h1psimps}) without applying Eqs.~(\ref{eq:h1eigcond_fo})--(\ref{eq:h1eigcond_fr}) one would obtain [via repeatedly applying Eq.~(\ref{eq:smaidentity}), or simply expressing the TTI MPS in the equivalent SMA form]
    \begin{equation}
      \label{eq:h1psimpsgen}
      \begin{aligned}
        &\ttr\left\{\sma{M}{F}^{s_1}\sma{M}{F}^{s_2}\cdots \sma{M}{F}^{s_{L_b}}\right\}\\
        &=  \Tr
          \begin{aligned}[t]
            &\big\{F^{s_1}M^{s_2}\cdots M^{s_{L_b}} \\
            &+ M^{s_1}F^{s_2}\cdots M^{s_{L_b}} + \cdots \\
            &+ M^{s_1}M^{s_2}\cdots F^{s_{L_b}}\big\}.
          \end{aligned}
      \end{aligned}
    \end{equation}
    Clearly, arbitrary cyclic permutations of the order of matrices in individual terms of the right-hand side of Eq.~(\ref{eq:h1psimpsgen}) have no effect on the result.
    This means that in contrast with the OBC case, one has freedom in choosing how matrices $F^s$ are paired with matrices $M^s$ when producing a less restrictive substitute of Eqs.~(\ref{eq:h1eigcond_fo})--(\ref{eq:h1eigcond_fr}), given that a full set of possibilities is exhausted. For example, it is sufficient for any given $F^s$ to satisfy either of Eqs.~(\ref{eq:pfm}) and (\ref{eq:pmf}), and not necessarily both of them, in order for the substitution $F^s \to [X, M^s]$ to be allowed in Eq.~(\ref{eq:h1psimps}) restricted to $\mathcal{H}^1$.

    Clearly, Eqs.~(\ref{eq:padding_fo})--(\ref{eq:padding_frmr}) exhaust all possibilities in the sense described above and, if satisfied, guarantee that Eq.~(\ref{eq:h1psimps}) holds for any $\ket{s_1s_2\cdots s_{L_b}} \in \mathcal{H}^1$, which means $H_\text{PXP}\ket{\psi} = \mathcal{P}_fH_1\ket{\psi} = 0$.
  \end{proof}
\end{theorem}

Per Theorem~\ref{thrm:genzm}, checking whether a specific MPS constrained to $\mathcal{H}^1$ represents an exact zero energy eigenstate of $H_\text{PXP}$ with PBCs amounts to the straighforward task of solving a system of linear equations for the matrix elements of $X$. Hence, we have the following:
\begin{corollary}
  \label{corr:proofx}
  All MPS representations given in Eqs.~(\ref{eq:phi1mps})--(\ref{eq:omegamps}) correspond to exact zero energy eigenstates of $H_\text{PXP}$ with PBCs and an even number of sites.
  \begin{proof}
    Matrices $X$ satisfying the conditions of Theorem~\ref{thrm:genzm} for each of the states exist and are given in App.~\ref{app:proofxpxp}.
    Here we assume the blocked basis representation of state $\ket{\Omega}$ (i.e., $M^{O}=M^0M^0$, $M^L = M^1M^0$, $M^R = M^0M^1$).

    Note that Ref.~\cite{lin2019exact} only provided an indirect proof for $\ket{\Phi_2}$ by showing that it is related to $\ket{\Phi_1}$ by translation. Here, the proof for $\ket{\Phi_2}$ is direct and does not rely on any knowledge about its relation to $\ket{\Phi_1}$.
  \end{proof}
\end{corollary}

In App.~\ref{app:obctheta} we investigate whether the states other than $\ket{\Phi_1}$ also have simple OBC counterparts and rule out that possibility; having OBC states with the same bulk MPS as $\ket{\Theta_{1,2}}$ or $\ket{\Omega}$ deep inside the chain is likely possible only in the iDMRG sense.

Since Theorem~\ref{thrm:genzm} and Corollary~\ref{corr:proofx} only address PXP chains of even length, they are not sufficient to support our claim that the state $\ket{\Omega}$ is also an eigenstate for odd length chains.
Our general proof of the state $\ket{\Omega}$ as well as of all the other eigenstates from Sec.~\ref{sec:eigenstates} (including those of the PPXPP chain) will follow from a further generalization of our results so far.
Consider the following theorem where we effectively recap the structure of the proofs given earlier in a slightly more abstract way.
\begin{theorem}
  \label{thrm:gengenzm}
  Suppose $\mathcal{H}$ is a dynamically decoupled subspace of Hamiltonian $\widetilde H$ defined by a local kinetic constraint and let $\ket{\psi}\in \mathcal{H}$.
  Suppose $\widetilde H$ can be expressed as
  \begin{equation}
    \widetilde H = \widetilde H_1 + \widetilde H_2,
  \end{equation}
  and, when acting on $\ket{\psi}$, $\widetilde H_1: \mathcal{H} \to \mathcal{H}\oplus\mathcal{\overline H}$ and $\widetilde H_2: \mathcal{H} \to \mathcal{\overline H}$ for some subspace $\mathcal{\overline H}$ orthogonal to $\mathcal{H}$. If $\widetilde H_1 \ket{\psi} = \lambda\ket{\psi} + \ket*{\overline\psi}$ for some $\ket*{\overline\psi}\in\mathcal{\overline H}$, then $\ket{\psi}$ is an exact eigenstate of $\widetilde H$ with energy $\lambda$.

  Let $\ket{\psi}$ be a TI MPS defined in some local basis $\{\ket{s}\}$ in terms of matrices $M^s$, and suppose $\widetilde H_1$ is a sum of single-site operators $\mathbf{h}: \{\ket{s}\}\to\{\ket{s}\}$, formally like the MPO in Eq.~(\ref{eq:h1genmpo}) and Fig.~\ref{fig:h1genmpo}. Then, if there exist a matrix $X$ and two integers $P\geq 0$ and $Q \geq 0$ such that equation
  \begin{equation}
    \label{eq:padding}
    M^{l_1}\cdots M^{l_P}\left([X, M^s]- F^s\right)M^{r_1}\cdots M^{r_Q} = 0
  \end{equation}  
  holds for any $\ket{l_1\cdots l_Ps r_1\cdots r_Q}$ that does not violate the local kinetic constraint, with generalized matrices $F^s$ defined in analogy with Eq.~(\ref{eq:fs}) according to the action of $\mathbf{h}$ on the local computational basis, $\ket{\psi}$ is a zero energy eigenstate of $\widetilde H$ with PBCs for any system size $N > P+Q+1$ (in units of the local basis).
  \begin{proof}
    The general argument regarding the action of $\widetilde H_1$ on $\ket{\psi}$ is identical to that used in Lemma~\ref{lemma:fibh1}.
    For the TI MPS case,  completely analogously to the argument used in Theorem~\ref{thrm:genzm} for PBCs, one can show that Eq.~(\ref{eq:padding}) guarantees that Eq.~(\ref{eq:h1psimps}) (assuming its physical indices are expressed in the appropriate local basis) holds for any $\ket{s_1s_2\cdots s_{L_b}} \in \mathcal{H}$.
  \end{proof}
\end{theorem}

\begin{corollary}
  \label{corr:rydbproof}
  Specializing to the PXP chain, suppose matrices $M^s$, $s\in \{0, 1\}$ are a TI MPS representation of $\ket{\psi} \in \mathcal{H}^1$ (i.e., $M^1M^1=0$).
  Then $\ket{\psi}$ is an exact zero mode of $H_\text{PXP}$ with PBC of size $L > 3$ (in the spin-1/2 basis) if there exists a matrix $X$ such that
  \begin{subequations}
    \begin{align}
      \label{eq:f0}
      &[X, M^0] = F^0 = M^1, \\
      \label{eq:f1}
      &M^0[X, M^1]M^0 = M^0F^1M^0 = [M^0]^3.
    \end{align}
  \end{subequations}
  \begin{proof}
    Rewriting the Hamiltonian as $H_\text{PXP} = \widetilde H_1 + \widetilde H_2$, where $\widetilde H_1 = \sum_{i=1}^L X_i$, one can easily confirm that $\widetilde H_1: \mathcal{H}^1 \to \mathcal{H}^1\oplus\mathcal{\overline H}^1$ and $\widetilde H_2: \mathcal{H}^1 \to \mathcal{\overline H}^1$ (here, $\mathcal{\overline H}^1$ is a subspace orthogonal to $\mathcal{H}^1$ and spanned by computational basis states with one or two Fibonacci constraint violations). Applying Theorem~\ref{thrm:gengenzm} with $\mathbf{h}=\ketbra{0}{1}+\ketbra{1}{0}$, which defines matrices $F^s$, we arrive at a valid set of sufficient conditions given in Eqs.~(\ref{eq:f0}) and (\ref{eq:f1}).    
  \end{proof}
\end{corollary}

Corollary~\ref{corr:rydbproof} gives a direct means for validating states expressed in the spin-1/2 basis, like $\ket{\Omega}$ in Eq.~(\ref{eq:omegamps}).
One can easily verify that the spin-1/2 MPS representation of $\ket{\Omega}$ along with the same matrix $X$ given in App.~\ref{app:proofxpxp} used earlier when proving Corollary~\ref{corr:proofx} also satisfies Eqs.~(\ref{eq:f0}) and (\ref{eq:f1}). Thus, per Corollary~\ref{corr:rydbproof}, $\ket{\Omega}$ is defined on PXP chains of both even and odd lengths and is an exact zero-energy eigenstate, as we claimed.

Note that the matrix $M^1$ of the MPS representation of $\ket{\Omega}$ is a canonical nilpotent with index 2 (i.e., $M^1M^1=0$).
Any MPS representation in the spin-1/2 basis constrained to $\mathcal{H}^1$ must have a nilpotent $M^1$ matrix, which can always be brought into a canonical form via a gauge similarity transformation.
In App.~\ref{app:rydbti} we demonstrate this by providing spin-1/2 TI variants of the states $\ket{\Phi_{1,2}}$ and $\ket{\Theta_{1,2}}$ with the corresponding $M^1$ matrices in the canonical nilpotent form.
Although such TI variants are not injective MPSs unlike their $T_x$ symmetry-breaking constituents (which also means that the states vanish when the system size is odd), they also satisfy the requirements of Corollary~\ref{corr:rydbproof}.

Let us now prove the TI MPS $E=0$ eigenstates of the PPXPP model given in Eqs.~(\ref{eq:s1mps})--(\ref{eq:tmps}).
In the same local blocked basis we used for the PXP model, $\mathcal{H}^2$ can be defined as a subspace spanned by computational basis states without $\ket{RL}$, $\ket{LL}$, and $\ket{RR}$.
Then, using Theorem~\ref{thrm:gengenzm}, we can construct the following:
\begin{corollary}
  \label{corr:ppxpp}
  Suppose matrices $M^s$, $s\in \{O, L, R\}$ are a TI MPS representation of $\ket{\psi} \in \mathcal{H}^2$ (i.e., $M^R M^L=0$, $M^L M^L=0$, and $M^R M^R=0$).
  If there exists a matrix $X$ such that
  \begin{subequations}
    \begin{align}
      \label{eq:ppxpp_padding_fo}
      & [X, M^O] = F^O,\\
      \label{eq:ppxpp_padding_moflmo}
      & M^O [X, M^L] M^O = M^O F^L M^O,\\
      \label{eq:ppxpp_padding_moflmr}
      & M^O [X, M^L] M^R = M^O F^L M^R,\\
      \label{eq:ppxpp_padding_mofrmo}
      & M^O[X, M^R] M^O = M^O F^R M^O,\\
      \label{eq:ppxpp_padding_mlfrmo}
      & M^L[X, M^R] M^O = M^L F^R M^O,
    \end{align}
  \end{subequations}
  for matrices $F^s$ given in Eq.~(\ref{eq:fs}), then $\ket{\psi}$ is an exact zero mode of $H_\text{PPXPP}$ with PBC (of size greater than 3 blocks).
\begin{proof}
  Rewriting the Hamiltonian as $H_\text{PPXPP} = \widetilde H_1 + \widetilde H_2$, where $\widetilde H_1 = H_1$ in Eq.~(\ref{eq:h1pxp}), one can easily confirm that $\widetilde H_1: \mathcal{H}^2 \to \mathcal{H}^2\oplus\mathcal{\overline H}^2$ and $\widetilde H_2: \mathcal{H}^2 \to \mathcal{\overline H}^2$ (where $\mathcal{\overline H}^2 \perp \mathcal{H}^2$). Thus, per Theorem~\ref{thrm:gengenzm}, Eqs.~(\ref{eq:ppxpp_padding_fo})--(\ref{eq:ppxpp_padding_mlfrmo}) constitute a valid set of sufficient conditions.
\end{proof}
\end{corollary}

The requirements of Corollary~\ref{corr:ppxpp} are satisfied by the states $\ket{S_1}$, $\ket{S_2}$ and $\ket{T}$ in Eqs.~(\ref{eq:s1mps})--(\ref{eq:tmps}) with the corresponding matrices $X$ given in App.~\ref{app:proofxppxpp}.
Therefore $\ket{S_1}$, $\ket{S_2}$, and $\ket{T}$ are exact zero energy eigenstates of $H_\text{PPXPP}$.
\begin{gentheorem}
  \label{thrm:gengenzmtti}
  (Generalization of Theorem~\ref{thrm:gengenzm} to TTI MPS.)
  If $\ket{\psi}$ is a TTI MPS defined in some local basis $\{\ket{s}\}$ in terms of matrices $\sma{M}{M_1}^s$, Theorem~\ref{thrm:gengenzm} holds under the substitutions $M^s\to \sma{M}{M_1}^s$ and $F^s \to \sma{F}{F_1}$, where $F_1^s$ are derived from $M_1^s$ completely analogously to how $F^s$ are derived from $M^s$.
  \begin{proof}
    The formalism developed in Sec.~\ref{subsec:ttimps} can be used to express the action of the Hamiltonian $\widetilde H_1$ on a TTI MPS.
    To that end, we interpret the twist matrix
    \begin{equation}
      \mathcal{T} = \begin{pmatrix}
        \mathbf{0} & \mathbf{1} \\
        \mathbf{0} & \mathbf{0}
      \end{pmatrix}
    \end{equation}
    of $\ket{\psi}$ as the boundary conditions $B_{\alpha\beta}$ in Eq.~(\ref{eq:ttimpstau}), which means that the bulk tensors and the twist matrix for $\widetilde H_1\ket{\psi}$ have the following form:
    \begin{equation}
      \mps{\widetilde H_1\ket{\psi}}
      {
        \left\{
          \sma{\sma{M}{M_1}}{\sma{F}{F_1}}^s
        \right\},
        \mathcal{T}' = \begin{pmatrix}
          \mathbf{0} & \mathcal{T} \\
          \mathbf{0} & \mathbf{0}
        \end{pmatrix}
      }.
    \end{equation}
    Thus, the amplitudes of the components of $\widetilde H_1\ket{\psi}$ [cf.~Eq.~(\ref{eq:h1psimpsgen})] are given by
    \begin{equation}
      \label{eq:h1psittimpsgen}
      \begin{aligned}
        &\mel{s_1s_2\cdots s_{L_b}}{\widetilde H_1}{\psi} \\
        &=  \ttr
          \begin{aligned}[t]
            &\Bigg\{\sma{F}{F_1}^{s_1}\sma{M}{M_1}^{s_2}\cdots \sma{M}{M_1}^{s_{L_b}} \\
            &+ \sma{M}{M_1}^{s_1}\sma{F}{F_1}^{s_2}\cdots \sma{M}{M_1}^{s_{L_b}} + \cdots \\
            &+ \sma{M}{M_1}^{s_1}\sma{M}{M_1}^{s_2}\cdots \sma{F}{F_1}^{s_{L_b}}\Bigg\}.
          \end{aligned}
      \end{aligned}
    \end{equation}
    The cyclicity of $\ttr$ even in the case of inhomogeneous bulk tensors [discussed in Sec.~\ref{subsec:ttimps}] allows applying Eq.~(\ref{eq:padding}), with the substitutions stated earlier, to all the individual terms on the right-hand side of Eq.~(\ref{eq:h1psittimpsgen}) with no regard to the twist matrix $\mathcal{T}$ (this can be thought of as exercising the freedom of placing $\mathcal{T}$ maximally far from the defect $\sma{F}{F_1}^{s_j}$ in each term when evaluating $\ttr$ via the standard trace $\Tr$).
    Hence, assuming Eq.~(\ref{eq:padding}) with the substitutions holds for the TTI MPS tensors generating $\ket{\psi}$,
    \begin{equation}
      \label{eq:h1psittimpsgentele}
      \begin{aligned}
        &\mel{s_1s_2\cdots s_{L_b}}{\widetilde H_1}{\psi} \\
        &=  \ttr
          \begin{aligned}[t]
            &\Bigg\{\left[X, \sma{M}{M_1}^{s_1}\right]\sma{M}{M_1}^{s_2}\cdots \sma{M}{M_1}^{s_{L_b}} \\
            &+ \sma{M}{M_1}^{s_1}\left[X,\sma{M}{M_1}^{s_2}\right]\cdots \sma{M}{M_1}^{s_{L_b}} + \cdots \\
            &+ \sma{M}{M_1}^{s_1}\sma{M}{M_1}^{s_2}\cdots \left[X,\sma{M}{M_1}^{s_{L_b}}\right]\Bigg\},
          \end{aligned}
      \end{aligned}
    \end{equation}
    where we assume that $\ket{s_1s_2\cdots s_{L_b}} \in \mathcal{H}$.
    Clearly, the right-hand side of Eq.~(\ref{eq:h1psittimpsgentele}) has the form of telescoping series, which gives $\mel{s_1s_2\cdots s_{L_b}}{\widetilde H_1}{\psi} = 0$ for any $\ket{s_1s_2\cdots s_{L_b}} \in \mathcal{H}$.
  \end{proof}
\end{gentheorem}
Per Theorem~\ref{thrm:gengenzmtti}, Corollary~\ref{corr:ppxpp} with appropriate substitutions for $M^s$ and $F^s$ is suitable for proving the states $\ket{S'_1}$, $\ket{S'_2}$, and $\ket{T'}$ in Eqs.~(\ref{eq:s1tti})--(\ref{eq:ttti}).
Matrices $X$ establishing the proofs of these states via such an extension of Corollary~\ref{corr:ppxpp} are provided in App.~\ref{app:proofxppxpp}.

We have now direct proofs for all the eigenstates of $H_\text{PXP}$ and $H_\text{PPXPP}$ listed in Sec.~\ref{sec:eigenstates} except the volume-entangled with respect to the standard bipartition states $\ket{\Lambda}$ given in Eqs.~(\ref{eq:lambdapxpmps}) and (\ref{eq:lambdappxpp}).
For such states expressed in the BA basis we can formulate the following:
\begin{corollary}
  \label{corr:bacond}
  Suppose matrices $M^{s\dual s}$, $s,\dual s \in \{0, 1\}$ are a BA TI MPS representation of $\ket{\psi} \in \mathcal{H}^\alpha$ (i.e., $M^{s_1\dual s_1}M^{s_2\dual s_2}\cdots M^{s_1 \dual s_\alpha}=0$ for any $\ket{s_1s_2\cdots s_\alpha} \in \mathcal{\overline H}^\alpha$ and any $\ket{\dual s_1\dual s_2\cdots \dual s_\alpha} \in \mathcal{\overline H}^\alpha$).
  Then $\ket{\psi}$ is an exact zero mode of $H^\alpha$ with PBC of size $L > 2\alpha + 1$ (in the spin-1/2 basis) if there exists a matrix $X$ such that
  \begin{equation}
    \label{eq:bacond}
    \begin{aligned}
      & M^{l_1\dual l_1}\cdots M^{l_\alpha\dual l_\alpha}\\
      &\times\left([X, M^{s\dual s}]-F^{s\dual s}\right)M^{r_1\dual r_1}\cdots M^{r_\alpha\dual r_\alpha}= 0,\\
    \end{aligned}
  \end{equation}
  where $\ket{l_1\cdots l_\alpha\,s\,r_1\cdots r_\alpha} \in \mathcal{H}^\alpha$, $\ket{\dual l_1\cdots \dual l_\alpha\,\dual s\,\dual r_1\cdots \dual r_\alpha} \in \mathcal{H}^\alpha$ and
  \begin{equation}
    \begin{aligned}
      &F^{00} = F^{11} = M^{01} + M^{10},\\
      &F^{01} = F^{10} = M^{00} + M^{11}.
    \end{aligned}
  \end{equation}
  \begin{proof}
    Rewriting the Hamiltonian as $H^\alpha = \widetilde H_1 + \widetilde H_2$, where $\widetilde H_1 = \sum_{i=1}^L X_i$, one can easily confirm that $\widetilde H_1: \mathcal{H}^\alpha \to \mathcal{H}^\alpha\oplus\mathcal{\overline H}^\alpha$ and $\widetilde H_2: \mathcal{H}^\alpha \to \mathcal{\overline H}^\alpha$. Applying Theorem~\ref{thrm:gengenzm} with $\mathbf{h}=X\otimes\mathbf{1} + \mathbf{1}\otimes X$, which defines matrices $F^{s\dual s}$, we arrive at a valid set of sufficient conditions expressed by Eq.~(\ref{eq:bacond}).
  \end{proof}
\end{corollary}
If we apply Corollary~\ref{corr:bacond} to states with $M^{01}= M^{10} = 0$, such as $\ket{\Lambda}$ with any blockade radius $\alpha$, the only non-trivial condition that follows from Eq.~(\ref{eq:bacond}) is
\begin{equation}
  \label{eq:lambdacond}
  \left(M^{00}\right)^\alpha\left(M^{00}+M^{11}\right)\left(M^{00}\right)^\alpha = 0,
\end{equation}
since all the equations involving the commutator with $X$ can be satisfied by choosing $X$ proportional to identity.
The condition in Eq.~(\ref{eq:lambdacond}), along with the appropriate Rydberg blockade constraints given in Corollary~\ref{corr:bacond}, is satisfied by the MPSs in Eqs.~(\ref{eq:lambdapxpmps}) and (\ref{eq:lambdappxpp}), which means they both represent exact zero energy eigenstates of, respectively, $H_\text{PXP}$ and $H_\text{PPXPP}$.

\section{Additional examples}
\label{sec:extra}
To demonstrate the broader applicability of Theorem~\ref{thrm:gengenzm}, let us apply it to a range of situations and Hamiltonians that we have not previously explored.
We start by revisiting the main results from Ref.~\cite{Surace_2021}.

\subsection{P\dots PXP\dots P models}
\label{sec:extrap_pxp_p}
All the eigenstates of the Hamiltonian $H^\alpha$ in  Eq.~(\ref{eq:halpha}) found in Ref.~\cite{Surace_2021} feature ``frozen'' sequences of the local ground state $\ket{0}$ on a subset of sites in the chain.
Such frozen sequences of any length are not possible when $\alpha = 1$.
However, for $\alpha>1$ and the case of PBCs, there exist exact eigenstates composed of the motifs $\ket{s} \otimes \ket{0}^{\otimes (\alpha-1)}$, in which the three-site blocks $\ket{s} \in \{\ket{000}, \ket{100}, \ket{010}, \ket{001}\}$ can be controlled in the wavefunction via TI MPS tensors $M^s$.
In what follows, we will demonstrate how applying Theorem~\ref{thrm:gengenzm} in a slightly different way from our earlier examples allows systematically recovering all such states discussed in Ref.~\cite{Surace_2021}.

Suppose the system has PBCs and $L=(\alpha+2)n$ for some integer $n$.
Let us rewrite $H^\alpha$ as $H^\alpha_1 + H^\alpha_2$. Consider terms of the form
\begin{equation}
  \label{eq:pxpexth1term}  
  \begin{aligned}
    &h_j = \underbrace{H^\alpha_\text{OBC}(j\dots j+2)}_{\text{term ``A'', range 3}}\\
    &\quad + \underbrace{H^\alpha_\text{OBC}(j+3\dots j+3 + \alpha-2)}_\text{term ``B'', range $\alpha-1$},
  \end{aligned}
\end{equation}
where $H^\alpha_\text{OBC}(j\dots k)$ is an OBC version of $H^\alpha$ acting on sites $j\dots k$ for $k \geq j$.
Note that both terms in Eq.~(\ref{eq:pxpexth1term}) have ranges shorter than that of the terms in $H^\alpha$, so truncation is implicit.
Given
\begin{equation}
  H^\alpha_1 = \sum_{k=0}^{n-1} h_{(\alpha + 2)k + 1},
\end{equation}
it follows that $H^\alpha_2 \equiv H^\alpha - H^\alpha_1$ (which we will not write explicitly) must take any state $\ket{\psi} \in \mathcal{H}^\alpha$ entirely outside of $\mathcal{H}^\alpha$.

\begin{figure}
  \subfloat{\label{fig:frozen}}
  \subfloat{\label{fig:sansfrozen}}
    \includegraphics[width=\columnwidth]{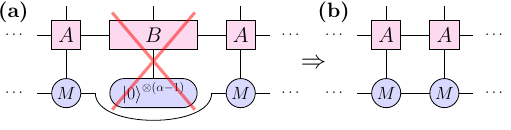}
    \caption{Action of MPO representation of $H_1^\alpha$ on MPS with frozen sequences.
    Frozen sequences and terms of type ``B'' can be eliminated from the initial tensor network (a) if additional constrains (discussed in the text) are imposed upon the tensors that terms of type ``A'' act on; the effective TI tensor network (b) augmented with additional constraints becomes amenable to analysis put forward in Theorem~\ref{thrm:gengenzm}.}
\end{figure}
The action of $H^\alpha_1$ MPO on an MPS whose frozen sequences are aligned with terms of type ``B" in Eq.~(\ref{eq:pxpexth1term}) is shown in Fig.~\ref{fig:frozen}.
In order for the MPS to be a zero mode of $H^\alpha$, individual terms of type ``B'' must take it outside of $\mathcal{H}^\alpha$.
This is only possible if adjacent three-site blocks acted upon by terms of type ``A'' exclude configurations $\ket{000}\ket{000}$, $\ket{000}\ket{001}$, $\ket{100}\ket{000}$, and $\ket{100}\ket{001}$. If this condition is satisfied, terms of type ``B'' become irrelevant and we can discard them, together with the frozen sequences they act on, to get the effective MPO/MPS shown in Fig.~\ref{fig:sansfrozen}.
The effective MPS must also exclude configurations $\ket{001}\ket{100}$ in order to satisfy the original kinetic constraint. With these considerations in mind, we can construct the following:

\begin{corollary}
  \label{corr:extpxp}
  Let matrices $M^{s}$ for $s~\in~\{000, 100, 010, 001\}$ be a TI MPS representation of $\ket{\psi} \in \mathcal{H}^\alpha$ with the assumption that adjacent three-site blocks controlled in the wavefunction by tensors $M^s$ are separated by a frozen sequence $\ket{0}^{\otimes (\alpha - 1)}$, see Fig.~\ref{fig:frozen}.
  Suppose
  \begin{equation}
    \label{eq:constrextpxp}
    \begin{aligned}
      M^{s_1}M^{s_2} = \mathbf{0}, \text{ for any } &(s_1, s_2) \in \{(001, 100), \\
                                                                         &(000, 000), (000, 001),\\
      &(100, 000), (100, 001)\},
    \end{aligned}
  \end{equation}
  and there exists a matrix $X$ such that
\begin{equation}
  \label{eq:extpxp_padding}
  [X, M^{s}] = F^{s}, \text{ for any } s\in\{000,100,010,001\},
\end{equation}
where
\begin{subequations}
  \begin{align}
    &F^{000} = M^{100} + M^{010} + M^{001},\\
    &F^{100}=F^{010}=F^{001}=M^{000}.
  \end{align}
\end{subequations}
Then $\ket{\psi}$ is an exact zero mode of $H^\alpha$ (with PBC).

In the case of OBCs, assuming all the ``frozen'' sequences are inside the chain, if the terminations are chosen to be left and right eigenvectors $v^T$ and $w$ of $X$ with eigenvalues $\lambda_v$ and $\lambda_w$, then $\ket{\psi}$ is an eigenstate of $H^\alpha_\text{OBC}$ of size L=$(\alpha+2)n - (\alpha-1)$ with energy $\lambda = \lambda_v - \lambda_w$.
\begin{proof}
  The rationale behind additional constraints in Eq.~(\ref{eq:constrextpxp}) and justification for the applicability of Theorem~\ref{thrm:gengenzm} were given in the discussion preceding the Corollary, thus proving the case of PBC.
  Since the conditions set by Eq.~(\ref{eq:extpxp_padding}) are unrestricted, arguments similar to those used in Theorem~\ref{thrm:h1eig} prove also the case of OBC.
\end{proof}
\end{corollary}

By inspection, we immediately find that the following two one-dimensional MPS representations related to the generalization of the state $\ket{\phi_2}$ in Eq.~(\ref{eq:phi2ppxpp}) satisfy the requirements of Corollary~\ref{corr:extpxp}:
\begin{subequations}
  \begin{align}
    \label{eq:phi_alpha}
    &\mps{\ket{\phi_\alpha}}{
      \underbrace{
      \begin{pmatrix}
        0
      \end{pmatrix}}_{M^{000}},
      \underbrace{
      \begin{pmatrix}
        1
      \end{pmatrix}}_{M^{100}},
      \underbrace{
      \begin{pmatrix}
        -1
      \end{pmatrix}}_{M^{010}},
      \underbrace{
      \begin{pmatrix}
        0
      \end{pmatrix}}_{M^{001}}};&&\\
    \label{eq:tx_phi_alpha}
    \mps{T_x&\ket{\phi_\alpha}}{
      \underbrace{
      \begin{pmatrix}
        0
      \end{pmatrix}}_{M^{000}},
      \underbrace{
      \begin{pmatrix}
        0
      \end{pmatrix}}_{M^{100}},
      \underbrace{
      \begin{pmatrix}
        1
      \end{pmatrix}}_{M^{010}},
      \underbrace{
      \begin{pmatrix}
        -1
      \end{pmatrix}}_{M^{001}}}.
  \end{align}
\end{subequations}

We also find that all non-trivial two-dimensional representations satisfying the requirements of Corollary~\ref{corr:extpxp} with non-zero $M^{000}$ are gauge-equivalent to
\begin{equation}
  \label{eq:psialpha}
  \mps{\ket{\psi^\pm_\alpha}}{
    \underbrace{
      \begin{pmatrix}
        0 & \pm\sqrt{3} \\
        0 & 0
      \end{pmatrix}}_{M^{000}_{\pm}},
    \underbrace{
      \begin{pmatrix}
        0 & 1 \\
        0 & 1
      \end{pmatrix}}_{M^{100}},
    \underbrace{
      \begin{pmatrix}
        1 & 1 \\
        0 & -1
      \end{pmatrix}}_{M^{010}},
    \underbrace{
      \begin{pmatrix}
        -1 & 1 \\
        0 & 0
      \end{pmatrix}}_{M^{001}}},
\end{equation}
i.e., the state from Ref.~\cite{Surace_2021} when $M^{000}_+$ is chosen.
For this state, we can take the matrix $X$ as:
\begin{equation}
  \label{eq:psialphax}
  \Scale[0.88]{\setlength{\arraycolsep}{2pt}
    X_{\psi^\pm_\alpha} = 
    \begin{pmatrix}
      0 &  0\\
      0 & \mp\sqrt{3}
    \end{pmatrix}.
  }
\end{equation}

Several comments on the discussion in Ref.~\cite{Surace_2021} are now in order.
First, we point out that in the case of PBC, the MPS in Eq.~(\ref{eq:psialpha}) is completely identical to the following block-diagonal superposition of the states given in Eqs.~(\ref{eq:phi_alpha}) and (\ref{eq:tx_phi_alpha}):
\begin{equation}
  \label{eq:phi_alpha_ti}
  \mps{\ket{\phi_\alpha}+T_x\ket{\phi_\alpha}}{
    \underbrace{
      \begin{pmatrix}
        0 & 0 \\
        0 & 0
      \end{pmatrix}}_{M^{000}},
    \underbrace{
      \begin{pmatrix}
        0 & 0 \\
        0 & 1
      \end{pmatrix}}_{M^{100}},
    \underbrace{
      \begin{pmatrix}
        1 & 0 \\
        0 & -1
      \end{pmatrix}}_{M^{010}},
    \underbrace{
      \begin{pmatrix}
        -1 & 0 \\
        0 & 0
      \end{pmatrix}}_{M^{001}}}.
\end{equation}
Indeed, it is easy to see that when trace is taken, the off-diagonal elements in the matrices for $\ket{\psi^\pm_\alpha}$ cannot have any contribution (in particular, the two MPSs $\ket{\psi^\pm_\alpha}$ give the same state in chains with PBCs).
Thus, the expression of the TI state generated by the MPS in Eq.~(\ref{eq:psialpha}) as a cat state whose two components were not eigenstates of $H^\alpha$ [see Eq.~(21) in Ref.~\cite{Surace_2021}] was not the simplest and most natural.

Second, in the case of OBCs, per Corollary~\ref{corr:extpxp}, given that the matrix in Eq.~(\ref{eq:psialphax}) has eigenvalues $0$ and $\mp\sqrt{3}$, the MPS in Eq.~(\ref{eq:psialpha}) is expected to generate 4 states (only 3 of which were explicitly identified in Ref.~\cite{Surace_2021}) with energies $-\sqrt{3}$, $0$, $0$,  and $\sqrt{3}$ for appropriate choices of terminations.
Indeed, the MPS in Eq.~(\ref{eq:psialpha}) generates a four-dimensional manifold similar to that generated by the MPS in Eq.~(\ref{eq:phi1mps}) corresponding to the state $\ket{\Phi_1}$; however, the choice of terminations $v^T=(0, 1)$ and $w^T=(1, 0)$ always generates a zero state vector as a consequence of non-injectivity of the MPS.
As a result, $\ket{\psi_\alpha^+}$ and $\ket{\psi_\alpha^-}$ MPSs each generate three linearly independent OBC states, overlapping in two of them.
Clearly, eigenstates with energies $\pm\sqrt{3}$ can be generated choosing $v^T=(1, 0)$ and $w^T=(0, 1)$ along with the matching $M^{000}_{\pm}$. Two zero energy eigenstates identical to $\ket{\phi_\alpha}$ and $T_x\ket{\phi_\alpha}$ in Eqs.~(\ref{eq:phi_alpha}) and (\ref{eq:tx_phi_alpha}) are generated when choosing, respectively, $v^T=(0, 1)$, $w^T=(0, 1)$, and $v^T=(1, 0)$, $w^T=(1, 0)$ --- thus, taking the trace in the case of OBCs, as suggested in Ref.~\cite{Surace_2021}, gives their equally weighted superposition and conceals the existence of two linearly independent zero modes.

In some of our examples from this section, and also in the case of the state $\ket{S_1}$ in Eq.~(\ref{eq:s1mps}), it was easy to obtain non-trivial MPSs satisfying the requirements of Theorem~\ref{thrm:gengenzm} by inspection.
Typically, this was because Eq.~(\ref{eq:padding}) admitted solutions with vanishing $[X, M^s]$.
The vanishing of the commutator means that the individual terms of $\widetilde H_1$ (and not $\widetilde H_1$ as a whole like in the earlier PXP model examples) take the MPS entirely outside of the constrained Hilbert space $\mathcal{H}$.
For example, individual terms of $H_1$ in Eq.~(\ref{eq:h1pxp}) annihilate $\ket{\phi_2}$ in Eq.~(\ref{eq:phi2ppxpp}) when they act on blocked sites with $\ket{R}-\ket{L}$, whereas acting on sites with $\ket{O}$ they produce $\ket{R}+\ket{L}$, which results in a state entirely in $\mathcal{\overline H}^2$.
Since states given in Eqs.~(\ref{eq:s1mps}) and (\ref{eq:phi_alpha_ti}) are closely related to $\ket{\phi_2}$, it is not surprising that they satisfy the requirements of Corollaries \ref{corr:ppxpp} and \ref{corr:extpxp}, respectively, with $X = \mathbf{1}$.

\subsection{PSP and related models}
Another example of the special case described in the paragraph above is the MPS eigenstate of the higher spin generalization of the PXP model~\cite{Ho_2019} discovered numerically via the sophisticated DMRG-S algorithm in Ref.~\cite{Zhang_2023}.
The higher spin generalization is defined by the ``PSP'' Hamiltonian
\begin{equation}
  H_\text{PSP}=\sum_{j=1}^L \bar P_{j-1}S_j^x \bar P_{j+1},
\end{equation}
where $S^\mu
$ is the spin-$s$ generator of rotations around the axis $\mu \in \{x, y, z\}$, the local Hilbert space is spanned by $2s + 1$ states $\{\ket{-s},\ket{-s+1},\cdots\ket{s-1},\ket{s}\}$ in the $S^z$ basis, and $\bar P_j=\ketbra{-s}_j$.
The kinetic constraint, therefore, requires that for any pair of adjacent sites at least one must be in the local state $\ket{-s}$ (analogous to the Rydberg atom ground state $\ket{0}$ in the PXP model).

Let us demonstrate how the zero energy MPS eigenstate of the PSP model, first reported in Ref.~\cite{Zhang_2023}, is recovered through the following application of Theorem~\ref{thrm:gengenzm}:
\begin{corollary}
  \label{corr:psp}
  Suppose matrices $M^q$, $q\in \{-s,-s+1,\dots,s-1,s\}$ are a TI MPS representation of $\ket{\psi} \in \mathcal{H}_\text{PSP}$ (i.e., $M^{q}M^{r}=0$ if $q\neq -s$ and $r \neq -s$).
  Then $\ket{\psi}$ is an exact zero mode of $H_\text{PSP}$ with PBC if there exists a matrix $X$ such that
  \begin{subequations}
    \begin{align}
      \label{eq:psp_padding}
      & [X, M^{-s}] = F^{-s},\\
      \label{eq:psp_padding_k}
      & M^{-s} [X, M^k] M^{-s} = M^{-s} F^{k} M^{-s},
    \end{align}
  \end{subequations}
  for any $k \in [-s+1,\dots,s]$ and matrices $F^q$ given by (with appropriate spin ladder terminations)
  \begin{equation}
    \label{eq:f_q}
    \begin{aligned}
      F^q = &\mel{s,m_z=q}{S^x}{s,m_z=q+1} M^{q+1} \\
            &+ \mel{s,m_z=q}{S^x}{s,m_z=q-1} M^{q-1}.
    \end{aligned}
  \end{equation}
  \begin{proof}
    The equations in the statement of the Corollary follow from writing $H_\text{PSP} = \widetilde H_1 + \widetilde H_2$, where $\widetilde H_1 = \sum_{j=1}^LS_j^x$.
  \end{proof}
\end{corollary}

Similarly to the simplest PPXPP model examples, let us attempt to guess an MPS satisfying the requirements of Corollary~\ref{corr:psp} for which $X$ is the identity matrix (i.e., the MPS is annihilated by the individual PSP terms, or taken outside of the constrained Hilbert space by individual terms of $\widetilde H_1$).
This makes the left-hand side of Eqs.~(\ref{eq:psp_padding}) and (\ref{eq:psp_padding_k}) vanish, and we are left with
\begin{subequations}
  \begin{align}
    \label{eq:fspsp}
    &0 = F^{-s} \nonumber \\
    &\quad = \mel{s,m_z=-s}{S^x}{s,m_z=-s+1} M^{-s+1},\\
    &0 = M^{-s}F^kM^{-s} \nonumber \\
    \label{eq:fkpsp}
    &\quad = M^{-s}\big(\mel{s,m_z=k}{S^x}{s,m_z=k+1} M^{k+1} \nonumber \\
             &\quad + \mel{s,m_z=k}{S^x}{s,m_z=k-1} M^{k-1}\big)M^{-s},
  \end{align}
\end{subequations}
where $k \in [-s+1,\dots, s]$.
Note that the case $k=s$ at the top of the ladder gives $M^{-s} M^{s-1} M^{-s} = 0$.
This, combined with Eq.~(\ref{eq:fspsp}), suggests that the simplest way to satisfy the recurrence of Eq.~(\ref{eq:fkpsp}) is by stipulating that $s$ is an integer and setting
\begin{equation}
  \label{eq:pspodd}
  M^{-s+2n-1} = 0
\end{equation}
for any $n\in [1, \dots, s]$.
Then we still need to satisfy
\begin{equation}
  \label{eq:psprec}
  \begin{aligned}
    &M^{-s} M^{-s+2n} M^{-s} \\
    &=-\frac{\mel{s,m_z=-s+2n-1}{S^x}{s,m_z=-s+2n-2}}{\mel{s,m_z=-s+2n-1}{S^x}{s,m_z=-s+2n}} \\
    &\quad\times M^{-s} M^{-s+2(n-1)} M^{-s} \\
  \end{aligned}
\end{equation}
for any $n \in [1,\dots,s]$.
Using $\mel{s,m_z}{S^x}{s,m_z\pm 1} = \sqrt{(s\pm m_z+1)(s\mp m_z)}/2$, Eq.~(\ref{eq:psprec}) gives
\begin{equation}
  \label{eq:psprec_clean}
  \begin{aligned}
    &M^{-s} M^{-s+2n} M^{-s} \\
    &=-\sqrt{\frac{(2n-1)(s-n+1)}{n(2s-2n+1)}} M^{-s} M^{-s+2(n-1)} M^{-s} \\
    &=a_n M^{-s}M^{-s}M^{-s},
  \end{aligned}  
\end{equation}
where in the last line we used $a_n$ to denote the analytical solution to the recurrence obtained in Ref.~\cite{Zhang_2023}.

The final step is finding any set of matrices $\{M^{-s+2n}: n \in [1,\dots,s]\}$ together with $M^{-s}$ satisfying Eq.~(\ref{eq:psprec_clean}).
Restricting ourselves to bond dimension $\chi=2$, the kinetic constraint requires that any $M^{-s+2n}$ for any $n \in [1,\dots,s]$ is proportional to the same nilpotent matrix, which without loss of generality can be gauge-fixed as $\left(\begin{smallmatrix}0 & 1 \\ 0 & 0\end{smallmatrix}\right)$. It is then easy to see that the following state satisfies the requirements of Corollary~\ref{corr:psp}:
\begin{equation}
  \label{eq:psis}
  \mps{\ket{\Psi}_s}{
    \underbrace{
      \begin{pmatrix}
        0 & 0 \\ 1 & 1
      \end{pmatrix}}_{M^{-s}}, \;
    \underbrace{
      \begin{pmatrix}
        0 & 0  \\ 0 & 0
      \end{pmatrix}}_{M^{-s+2n-1}}, \;
    \underbrace{
      a_n\begin{pmatrix}
        0 & 1 \\ 0 & 0
      \end{pmatrix}}_{M^{-s+2n}}},
\end{equation}
where $n \in [1,\dots,s]$.
The state $\ket{\Psi}_s$ in Eq.~(\ref{eq:psis}) is gauge-equivalent to the state reported in Ref.~\cite{Zhang_2023} for the integer spin model.
We note that Ref.~\cite{Zhang_2023} arrived at an equation identical to our Eq.~(\ref{eq:psprec_clean}) when proving their specific eigenstate; here, in contrast, the equation emerges as a special case of a more general construction, and the eigenstate is recovered via simple analytical means.

Finally, let us revisit the additional eigenstates of the spin-1 and spin-3/2 PSP model that were introduced in Ref.~\cite{yuan2023exactquantummanybodyscars}, see App.~B in this reference.
Upon closer inspection, we recognize structural similarly of these states and those discussed in Sec.~\ref{sec:extrap_pxp_p}.
In both $s=1$ and $s=3/2$ cases, the states feature ``frozen'' sequences of the local $\ket{-s}$ state.
The states with PBCs found in Ref.~\cite{yuan2023exactquantummanybodyscars} are composed of the motifs $\ket{k}\otimes\ket{-s}$, where $\ket{-s}$ is the local frozen state and $\ket{k}$ is a single-site state controlled through TI MPS tensors $M^k$.
We demonstrate below that these states can, in fact, be extended to a family $H_\text{PSP}^\alpha$ of PSP-type Hamiltonians with generalized blockade radius $\alpha$, defined analogously to Eq.~(\ref{eq:halpha}).
With this generalization, the motifs take on the form $\ket{k}\otimes\ket{-s}^{\otimes\alpha}$.

To provide a general set of conditions defining the MPS representations of the states from Ref.~\cite{Zhang_2023}, we need a straightforward extension of Theorem~\ref{thrm:gengenzm} to MPSs with finite local energy density.
Assuming for simplicity that $P = Q  = 0$ in Theorem~\ref{thrm:gengenzm}, this can be achieved by replacing Eq.~(\ref{eq:padding}) with
\begin{equation}
  \label{eq:finite_e}
  [X, M^k]- F^k + \epsilon M^k = 0,
\end{equation}
where $\epsilon$ is the local energy density.
This generalization allows formulating the following (cf.~Corollary~\ref{corr:extpxp}):
\begin{corollary}
  \label{corr:pspext}
  Let matrices $M^q$, $q\in \{-s,-s+1,\dots,s-1,s\}$ be a TI MPS representation of $\ket{\psi} \in \mathcal{H}^\alpha_\text{PSP}$ with the assumption that adjacent blocks controlled by tensors $M^q$ are separated by a frozen sequence $\ket{-s}^{\otimes \alpha}$.
  Suppose $M^{-s}M^{-s} = 0$ (which ensures that the $\ket{-s}^{\otimes \alpha}$ sequences in between remain ``frozen'' --- see the discussion in Sec.~\ref{sec:extrap_pxp_p} on the additional constraints for context) and there exists a matrix $X$ such that Eq.~(\ref{eq:finite_e}) is satisfied  for any $k$ and matrices $F^q$ given by Eq.~(\ref{eq:f_q}).
  Then $\ket{\psi}$ is an exact eigenstate of $H^\alpha_\text{PSP}$ with energy $E_\text{PBC} = \epsilon N$, where $N$ is the number of $M^q$ tensors in the PBC chain.

  Further, if there exist multiple tensors $M^q_\beta$ satisfying the above conditions with the same matrix $X$ as well as $M^{-s}_\beta M^{-s}_\gamma = 0$ for any $\beta \text{ and } \gamma$, then $\ket{\psi}$ formed with position-dependent $M^q_\beta$ is an exact eigenstate of $H^\alpha_\text{PSP}$ with energy $E_\text{PBC} = \sum_\beta \epsilon_\beta N_\beta$, where $N_\beta$ is the number of $M_\beta^q$ tensors in the PBC chain.

  Since the conditions in Eq.~(\ref{eq:finite_e}) are unconstrained (i.e., not padded), the states satisfying the requirements of the Corollary for PBC are also guaranteed to have OBC counterparts defined analogously to those in Corollary \ref{corr:extpxp}.
  In the finite-energy case, the OBC counterparts can be shown to have energies $E_\text{OBC} = E_\text{PBC} + \lambda_v - \lambda_w$ with $\lambda_v$ and $\lambda_w$ defined as in Corollary \ref{corr:extpxp}.
\end{corollary}

Note that the smallest possible bond dimension for the states satisfying the requirements of Corollary~\ref{corr:pspext} is $\chi=2$ since $M^{-s}$ (assuming it is non-vanishing) must be a nilpotent matrix.
Taking this into account, by solving the equations that follow from Corollary~\ref{corr:pspext} in each case, it is easy to recover the exact zero energy eigenstate of the spin-1 PSP model and the family of eigenstates of the spin-3/2 PSP model with local energy densities $\epsilon=\pm 1/2$  found in Ref.~\cite{yuan2023exactquantummanybodyscars}.
The MPS representations of these states are as follows:
\begin{subequations}
  \begin{align}
  \label{eq:mpsspin1}
  &\mps{\ket{\Psi}^{\epsilon=0}_{s=1}}{
    \underbrace{
      \begin{pmatrix}
        0 & 0 \\
        -\sqrt{2} & 0
      \end{pmatrix}}_{M^{-1}},
    \underbrace{
      \begin{pmatrix}
        1 & 0 \\
        0 & -1
      \end{pmatrix}}_{M^0},
    \underbrace{
      \begin{pmatrix}
        0 & \sqrt{2} \\
        0 & 0
      \end{pmatrix}}_{M^{1}}};&&\\
  \label{eq:mpsspin32plus}
  &\mps{\ket{\Psi}^{\epsilon = +\frac{1}{2}}_{s=\frac{3}{2}}}{
    \underbrace{
      \begin{pmatrix}
        0 & 0 \\
        -\sqrt{3} & 0
      \end{pmatrix}}_{M^{-\frac{3}{2}}},
    \underbrace{
      \begin{pmatrix}
        -1 & 0 \\
         -1 & 1
      \end{pmatrix}}_{M^{-\frac{1}{2}}},
    \underbrace{
      \begin{pmatrix}
        -1 & 1 \\
         0 & 1
      \end{pmatrix}}_{M^{\frac{1}{2}}},
    \underbrace{
      \begin{pmatrix}
        0 & \sqrt{3} \\
        0 & 0
      \end{pmatrix}}_{M^{\frac{3}{2}}}};&&\\
  \label{eq:mpsspin32minus}
  &\mps{\ket{\Psi}^{\epsilon = -\frac{1}{2}}_{s=\frac{3}{2}}}{
    \underbrace{
      \begin{pmatrix}
        0 & 0 \\
        \sqrt{3} & 0
      \end{pmatrix}}_{M^{-\frac{3}{2}}},
    \underbrace{
      \begin{pmatrix}
        1 & 0 \\
        -1 & -1
      \end{pmatrix}}_{M^{-\frac{1}{2}}},
    \underbrace{
      \begin{pmatrix}
        -1 & -1 \\
         0 &  1
      \end{pmatrix}}_{M^{\frac{1}{2}}},
    \underbrace{
      \begin{pmatrix}
        0 & \sqrt{3} \\
        0 & 0
      \end{pmatrix}}_{M^{\frac{3}{2}}}}.
  \end{align}
\end{subequations}
All of the above MPSs satisfy the requirements of Corollary~\ref{corr:pspext} with $X=\frac{1}{2}\left(\begin{smallmatrix}0 & 1 \\ 1 & 0\end{smallmatrix}\right)$.
Hence, the tensors in Eqs.~(\ref{eq:mpsspin32plus}) and (\ref{eq:mpsspin32minus}) can be used jointly to generate states with total energies $E_\text{PBC} = (N_+-N_-)/2$, where $N_+$ and $N_-$ denote the number of tensors with the corresponding sign of the local energy density.
All MPSs in Eqs.~(\ref{eq:mpsspin1})--(\ref{eq:mpsspin32minus}) are injective, which means that in the case of OBCs every unique eigenstate of the PBC chain will split into four linearly independent eigenstates defined by appropriate terminations.

Although we will not pursuse this here, similar techniques can be applied to formulate the necessary conditions on the MPSs hosted by the kinetically constrained clock model~\cite{Bull_2019, Zhang_2023} and the effective kinetically constrained Hamiltonian corresponding to the tilted Fermi-Hubbard model~\cite{Desaules_2021,yuan2023exactquantummanybodyscars}.

\section{Numerical distillation of exact finite-bond-dimension eigenstates in exponentially degenerate subspaces of systems with PBC}
\label{sec:detection}
Apart from the simple special cases, which effectively correspond to all the previously known eigenstates of the models discussed earlier, guessing MPSs satisfying a set of requirements derived from Theorem~\ref{thrm:gengenzm} and its extensions amounts to the formidable task of solving nonlinear tensor equations.
Generically, when formulated as a decision or optimization problem, this task is expected to be NP-hard~\cite{hillar2013tensorproblemsnphard}.
On the other hand, certifying that a specific MPS is an exact eigenstate for all system sizes greater than some fixed integer reduces to the computationally simple polynomial-time task of solving a finite number of linear matrix equations for $X$ (and, in general, local energy density).

None of the new MPSs introduced in this work were discovered as direct solutions to specific nonlinear tensor equations.
Finding such solutions seems to be beyond the capabilities of even the most advanced software.
Instead, the new eigenstates were obtained through indirect methods.
While these methods, relying on solving the rank minimization problem (RMP) over a convex set of matrices~\cite{fazel2004}, do not circumvent the NP-hardness characteristic of the direct approach, they enable considerable progress toward finding interesting solutions corresponding to provable eigenstates.

In this section, in hopes to motivate future similar searches and also provide deeper insights into the emergence of QMBSs in nonintegrable models, we outline a systematic numerical algorithm for detecting and characterizing well-defined area-law eigenstates in exponentially large subspaces of TI Hamiltonians.
All the new $E=0$ eigenstates of the PXP and PPXPP models listed in Sec.~\ref{sec:eigenstates} were discovered via this algorithm (some via early and more heuristic versions) and afterwards proven as exact eigenstates of the respective models using Theorem~\ref{thrm:gengenzm}.
Importantly, our algorithm is general and applicable to a wide range of Hamiltonians.

\begin{figure}
  \subfloat{\label{fig:mapping}}
  \subfloat{\label{fig:altproj}}
  \includegraphics[width=\columnwidth]{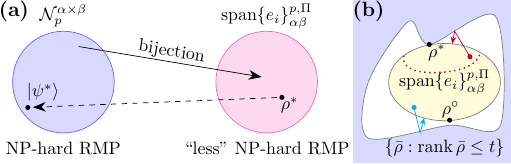}
  \caption{
    General approach for the detection of exact area-law scar states in exponentially degenerate subspaces.
    (a)~Bijective mapping to a more tractable problem.
    Let $\mathcal{N}_p^{\alpha\times\beta}$ (notation to be explained later in the text) represent the exponentially degenerate subspace under investigation obtained via ED.
    Identifying anomalous area-law states in $\mathcal{N}_p^{\alpha\times\beta}$ can be formulated as an RMP.
    However, solving such an RMP over $\mathcal{N}_p^{\alpha\times\beta}$ is computationally challenging.
    Suppose there exists a bijective function that maps $\mathcal{N}_p^{\alpha\times\beta}$ to another linear vector space $\spn\{e_j\}_{\alpha\beta}^{p,\Pi}$ spanned by $\dim\mathcal{N}_p^{\alpha\times\beta}$ matrices $\rho_j$.
    If RMP over $\spn\{e_j\}_{\alpha\beta}^{p,\Pi}$ is computationally easier than that over $\mathcal{N}_p^{\alpha\times\beta}$, we can solve the former and map the solution $\rho^*$ back to the original linear vector space $\mathcal{N}_p^{\alpha\times\beta}$.
    Due to the bijective nature of our mapping, an anomalous $\rho^*$ is effectively guaranteed to correspond to an anomalous $\ket{\psi^*}$.
    (b)~Heuristic solution for the RMP. The objective is finding the intersection between $\spn\{e_j\}_{\alpha\beta}^{p,\Pi}$ and the set of dimension-compatible matrices with rank upper-bounded by an integer $t$ (the shaded region extending beyond the margins).
    Starting from a random initial $\rho_\text{init}\in \spn\{e_j\}_{\alpha\beta}^{p,\Pi}$, we can perform alternating orthogonal projections (two iterations of which are denoted by zigzagging arrows) between the two sets.
    A sequence of such projections is guaranteed to converge to a point $\rho^*$ in the intersection of the two sets (assuming the intersection is nonempty) for a subset of initial points in $\spn\{e_j\}_{\alpha\beta}^{p,\Pi}$ (schematically, the set of such initial points, the basin of attraction of $\rho^*$, is above the dotted line).
    Since the set $\{\bar\rho: \rank\bar\rho \leq t\}$ is not convex, convergence is guaranteed only probabilistically, and some initial points yield null-results like $\rho^\circ$.
  }
  \label{fig:approach}
\end{figure}

Many previous studies have performed numerical detection of both exact and approximate QMBSs in various nonintegrable models.
The most direct approach, used in Refs.~\cite{Moudgalya_2018, lin2019exact}, involves finding states with anomalously low Schmidt ranks among those obtained via exact diagonalization (ED).
Less direct approaches include entanglement minimization \cite{Karle_2021}, machine learning \cite{Szo_dra_2022,feng2024uncoveringquantummanybodyscars}, correlation matrix analysis \cite{yao2024quantummanybodyscarslens}, analytic continuation of the partition function \cite{meng2025detectingmanybodyscarsfisher}, and density matrix renormalization group (DMRG) techniques~\cite{Zhang_2023, yuan2023exactquantummanybodyscars}.
To the best of our knowledge, no existing and documented numerical method has been successful in discovering previously unknown eigenstates in the PXP-type models; additionally, apart from the entanglement minimization approach of Ref.~\cite{Karle_2021}, no systematic exploration of exponentially degenerate nullspaces has been attempted.
Hence, our approach, outlined below and illustrated schematically in Fig.~\ref{fig:approach}, sets a new benchmark for scar detection.

\begin{figure}[!ht]
  \subfloat{\label{fig:ndfull}}
  \subfloat{\label{fig:ndproj}}
  \subfloat{\label{fig:ndmatrix}}
  \subfloat{\label{fig:ndmatrixtrans}}
  \includegraphics[width=\columnwidth]{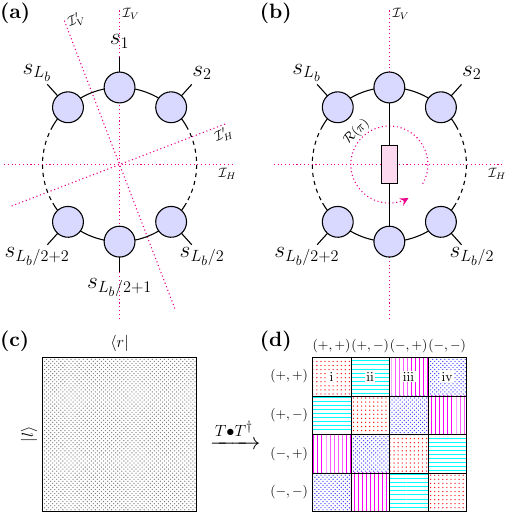}
  \caption{
    (a) TI MPS representation of a $p$-periodic state $\ket{\psi}$ on a PBC chain of $L_b$ $p$-blocks.
    Due to translational invariance, low entanglement entropy (or low Schmidt index) of $\ket{\psi}$ associated with a bipartition made along any symmetry axis (e.g., $\mathcal{I}'_V$) means that cuts along all other symmetry axes exhibit identical (for translationally equivalent axes) or qualitatively similar (for translationally nonequivalent ones) properties.
    (b) Problem size reduction via elimination of $p$-blocks $s_1$ and $s_{L_b/2 + 1}$.
    The resulting state $f_\Pi\ket{\psi}$ produced from a sought-after finite-bond-dimensional MPS $\ket{\psi}$ will typically have even fewer non-zero entries in its entanglement spectrum associated with the bipartition along the $\mathcal{I}_V$ axis than $\ket{\psi}$ has in its entanglement spectrum associated with a bipartition along the $\mathcal{I}'_V$ axis shown in (a).
    Thus, instead of minimizing the zeroth-order R\'enyi entropy over the entire subspace $\mathcal{N}_p$, we can achieve the same result on the reduced subspace $f_\Pi\mathcal{N}_p$ (now with a unique natural choice for the bipartition) without losing of any useful information.
    (c) Matricized state $\mathfrak{M}f_\Pi\ket{\psi}$ in the standard computational basis.
    Assuming typical lexicographic ordering of the bitstring states $\ket{l}$ and $\ket{r}$, no obvious structure is apparent.
    (d) Symmetry-resolved blocks of $T\left(\mathfrak{M}f_\Pi\ket{\psi}\right)T^\dagger$.
    States $\ket{(-1)^{|f|}, \pm 1}_f$ are grouped in the rows and columns according to the $\mathfrak{c}$ and $\mathfrak{i}$ symmetry quantum numbers.
    These groups are labeled as $(\pm, \pm)$, where the signs correspond to the $\mathfrak{c}$ and $\mathfrak{i}$ quantum numbers, respectively.
    The ordering within the groups is lexicographic in $f$.
    The correspondence between the symmetries of the state $\ket{\psi}$ (assuming they are well-defined) and the types of the non-vanishing blocks (as labeled by Roman numerals) is as follows:
    i)~$\ev{\mathcal{C}} = +1$, $\ev{\mathcal{I}} = +1$;
    ii)~$\ev{\mathcal{C}} = +1$, $\ev{\mathcal{I}} = -1$;
    iii)~$\ev{\mathcal{C}} = -1$, $\ev{\mathcal{I}} = +1$;
    iv)~$\ev{\mathcal{C}} = -1$, $\ev{\mathcal{I}} = -1$.
  }
\end{figure}

Let us assume that a given real-valued and TI in the spin-1/2 basis Hamiltonian $H$ satisfies $\left\{\mathcal{C}, H\right\} = [\mathcal{I}, H] = 0$ and thus, as discussed in Sec.~\ref{sec:models}, has an exponentially degenerate symmetry-protected nullspace in which we would like to identify eigenstates with exact finite-bond-dimension MPS representations well-defined for arbitrary system sizes.
Suppose we are interested to discover TI MPSs defined in a $d$-dimensional $p$-blocked basis $\{\ket{s}\}$, where each $\ket{s}$ represents an allowed state of $p$ spin-1/2 sites.
This means that we can restrict our attention to the $p$-periodic TI sector of $H$.
For example, when $p=2$ and the system size is chosen such that $L = p L_b = 2L_b$, this sector will contain states with the original momenta $0$ and $\pi$.
If $L_b$ is even, there is a natural bipartition of the system into two subsystems, each containing $L_b/2 \in \mathbb{Z}^+$  $p$-blocks.
Such a bipartition is shown in Fig.~\ref{fig:ndfull}, where the cut is made along the axis labeled $\mathcal{I}'_V$.
For the rest of this section, we will always assume bipartitions into two subsystems of equal sizes, each containing an integer number of $p$-blocks.

Let $\mathcal{N}_p = \spn\left\{\ket{\psi_i}\right\}$ be the complete $p$-periodic in terms of the original spin-1/2's --- or TI in terms of the block-spins --- nullspace of $H$, where $\ket{\psi_i} \in \mathbb{R}^{dL_b}$ are orthonormal wavefunctions computed via ED for a system of size $L=pL_b$.
Note that since our Hamiltonians of interest are real-valued, their $p$-periodic nullspaces can also be assumed real-valued; although this assumption simplifies our discussion to some extent, it is not strictly necessary.
The nullspace $\mathcal{N}_p$ can always be decomposed into four symmetry-resolved mutually orthogonal subspaces $\mathcal{N}_p^{(\pm,\pm)}$, where the superscript contains the signs of, respectively, $\mathcal{C}$ and $\mathcal{I}$ symmetry quantum numbers.
Some of these four subspaces can be empty for certain values of $p$; for instance, $\mathcal{N}_2 = \mathcal{N}_2^{(+,+)} \cup \mathcal{N}_2^{(-,-)}$ for $H = H_\text{PXP/PPXPP}$~\footnote{Although this is easily seen numerically, to the best of our knowledge, no explanation exists for why this phenomenon occurs. Some related discussion for the case of OBCs can be found in Ref.~\cite{Schecter_2018}.}.
Our objective is discovering states $\ket{\psi} \in \mathcal{N}_p$ whose bipartite entanglement spectra have a number of non-zero entries upper-bounded by a fixed number for any $L$.
In systems of numerically accessible sizes, the entanglement spectra of such states will have anomalously few non-zero entries in comparison to generic states in $\mathcal{N}_p$.
Minimization of the number of non-zero entries in the entanglement spectrum --- which is equivalent to the minimization of the zeroth order R\'enyi entropy $\mathcal{S}_0(\rho) = \log\rank \rho$, where $\rho$ is the reduced density matrix (RDM) associated with a bipartition of $\ket{\psi}$ --- is an instance of the RMP.
The obvious jump-discontinuous nature of $\mathcal{S}_0(\rho)$ renders this task highly numerically unstable.
Consequently, it is crucial to simplify the problem as much as possible and identify computationally tractable proxies that can facilitate minimizing the rank of the full RDM.

Our first simplification, shown in Fig.~\ref{fig:ndproj}, allows reducing the size of the problem --- here the number of components in the considered wavefunctions --- by approximately a factor of $d^2$ as well as, typically, substantially decreasing the zeroth order R\'enyi entropy that we hope to effectively optimize for.
Consider acting on $\ket{\psi}$ with a projector $\Pi=\ketbra{s}_{1}\otimes \ketbra{s}_{L_b/2+1}$, where $\ket{s}$ is chosen such that $[\Pi, \mathcal{C}] = 0$ and $[\Pi, \mathcal{I}_H] = [\Pi, \mathcal{I}_V] = 0$, where $\mathcal{I}_H$ and $\mathcal{I}_V$ denote mirror reflections (equivalent to inversions in the 1D case) with respect to axes marked with the  corresponding labels in Fig.~\ref{fig:ndproj}.

Let us define a linear map $f_\Pi: \left[\mathbb{R}^{d}\right]^{L_b} \to \left[\mathbb{R}^{d}\right]^{L_b-2}$ via the following equation:
\begin{equation}
  \Pi\ket{\psi} \equiv \ket{ss}_{1,L_b/2+1}\otimes f_\Pi\ket{\psi}.
\end{equation}
$f_\Pi$ takes a wavefunction defined on $L_b$ block-sites to its image produced by applying the projector $\Pi$, and discarding the sites $1$ and $L_b/2 + 1$ acted on by $\Pi$.
Thus, $f_\Pi\ket{\psi}$ is a wavefunction defined on $L_b-2$ block-sites composed of the right and left halves with block-sites $2, \dots, L_b/2$ and $L_b/2+2, \dots, L_b$ respectively.
To guarantee that this simplification is reversible, we want to choose $\Pi$ such that $f_\Pi$ is injective on $\mathcal{N}_p$.
For models under consideration, given a large enough system size, this will typically be the case for most choices of $\ket{s}$ defining the projector $\Pi$.
The simplest such choice on our $p$-block sites is $\ket{s} = \ket{0}^{\otimes p}$.

Since the zeroth order R\'enyi entropy of $f_\Pi\ket{\psi}$ associated with the bipartition by the $\mathcal{I}_V$ symmetry axis in Fig.~\ref{fig:ndproj} and that of $\ket{\psi}$ associated with the bipartition by the $\mathcal{I}'_V$ symmetry axis in Fig.~\ref{fig:ndfull} are closely related, rank minimization over the images in $f_\Pi\mathcal{N}_p$ is expected to result in similar minimization over the pre-images in $\mathcal{N}_p$.
The relationship between the two R\'enyi entropies is easy to see if we assume that the right subsystem in the initial bipartition includes sites $1$ and $L_b/2+1$.
For a large enough system such a bipartition will have the same zeroth order R\'enyi entropy as that associated with the bipartition by the $\mathcal{I}'_V$ symmetry axis in Fig.~\ref{fig:ndfull}.
Then, assuming Schmidt basis and taking into account that $f_\Pi$ affects the right subsystem only, $f_\Pi\ket{\psi} = f_\Pi\sum_i\alpha_i\ket{\psi_i^L}\otimes\ket{\psi_i^R} = \sum_i\alpha_i\ket{\psi_i^L}\otimes\ket*{\widetilde\psi_i^R}$, where $\ket*{\widetilde \psi_i^R}$ will generically be non-orthonormal. 
Hence, $f_\Pi$ never increases the Schmidt rank, and the zeroth order R\'enyi entropy of $f_\Pi\ket{\psi}$ is upper-bounded by that of $\ket{\psi}$.

Clearly, the state $f_\Pi\ket{\psi}$ retains the $\mathcal{C}$, $\mathcal{I}_H$, and $\mathcal{I}_V$ symmetries of $\ket{\psi}$ (provided they are well-defined in the original state), as well as the invariance (eigenvalue +1) under $\mathcal{R}(\pi)$, which represents a rotation of the system by half a turn (equivalently, translation by $L_b/2-1$ block-sites).
Moreover, these symmetries of $f_\Pi\ket{\psi}$ can be expressed in terms of the symmetry operators $\mathfrak{c}$ and $\mathfrak{i}$ acting on the individual (left and right) subsystems (with $L_b/2-1$ sites each) as follows:
\begin{equation}
  \label{eq:symmetries}
\begin{aligned}
&\mathcal{C} = \mathfrak{c} \otimes \mathfrak{c}, \\
&\mathcal{I}_H = \mathfrak{i} \otimes \mathfrak{i}, \\
&\mathcal{I}_V = \mathcal{R}(\pi) (\mathfrak{i} \otimes \mathfrak{i}) = (\mathfrak{i} \otimes \mathfrak{i}) \mathcal{R}(\pi) = \mathfrak{i} \otimes \mathfrak{i}. \\
\end{aligned}
\end{equation}
The above equalities are short-hands for $f_\Pi (\mathcal{C} \ket{\psi}) = \mathfrak{c} \otimes \mathfrak{c} (f_\Pi \ket{\psi})$, etc., where it is crucial that $\ket{s}$ has definite Rydberg excitation parity number and $\ket{ss}_{1, L_b/2+1}$ is invariant under $\mathcal{I}_H$ and $\mathcal{I}_V$; in the last equation we also used the fact that $\ket{\psi}$ is TI in the block-site basis.
Note that Eq.~(\ref{eq:symmetries}) will be valid if the support of the projector $\Pi$ has an even integer multiple of $p$-blocks that are antipodal on the chain, and if $\Pi$ is invariant under $\mathcal{C}$ and inversions about its two symmetry axes.
Thus, one can try increasing the range of $\Pi$ to even further reduce the size of the problem (making sure $f_\Pi$ remains injective).
For simplicity, in what follows, we will assume that the support of $\Pi$ has two $p$-blocks, as shown in Fig.~\ref{fig:ndproj}; the generalizations discussed above are straightforward.

Using the original bitstring spin-$1/2$ basis, let us now introduce an orthogonal transformation that expresses either of the half-systems in terms of states with fully resolved $\mathfrak{c}$ and $\mathfrak{i}$ symmetries, which we will label as $\ket{(-1)^{|f|}, \pm}_f$ for any $f \in \mathcal{F}$, where $\mathcal{F}$ is the set of length $d(L_b-2)/2$ bitstrings that generates the computational basis of a subsystem, and $\mathfrak{c}\ket{f} = (-1)^{|f|}\ket{f}$.
The correspondence between states $\ket{(-1)^{|f|},\pm 1}_f$ and the states of the conventional computational basis $\ket{f}$ becomes evident through the orthogonal basis transformation
\begin{equation}
  \label{eq:basischange}
  \begin{aligned}
    T = & \sum_{f \in F, f=\mathfrak{i}(f) } \ket{(-1)^{|f|},+1}_f \bra{f} + \\
        & + \sum_{f \in F, f < \mathfrak{i}(f)} \sum_{\zeta = \pm 1}
          \ket{(-1)^{|f|}, \zeta}_f \frac{\bra{f} + \zeta \bra{\mathfrak{i}(f)}}{\sqrt{2}},
  \end{aligned}
\end{equation}
where $\mathfrak{i}(f)$ denotes the inversion of the bitstring $f$ and the first sum in the second line is over pairs of inversion-related configurations, $\{f, \mathfrak{i}(f) \}$, with some fixed ordering within the pair denoted schematically as $f < \mathfrak{i}(f)$.

Our second simplification will lie in the application the matricization operation $\mathfrak{M}$ defined as
\begin{equation}
  f_\Pi\ket{\psi} \equiv \sum_{l,r \in \mathcal{F}} \alpha_{l,r} \ket{l}\otimes\ket{r} \xrightarrow{\mathfrak{M}} \sum_{l,r}\alpha_{l,r} \ketbra{l}{r},
\end{equation}
followed by a basis transformation via $T\bullet T^\dagger$, which --- when applied to the matricized state $\mathfrak{M}f_\Pi\ket{\psi}$ --- accomplishes the change of basis defined by Eq.~(\ref{eq:basischange}) in both subsystems.
The structures of the matrices $\mathfrak{M}f_\Pi\ket{\psi}$ and $T\left(\mathfrak{M}f_\Pi\ket{\psi}\right)T^\dagger$ are illustrated in Figs.~\ref{fig:ndmatrix} and \ref{fig:ndmatrixtrans}, respectively.
Importantly, if $\ket{\psi}$ has fully (or partially) resolved $\mathcal{C}$ and $\mathcal{I}$ symmetries, $T\left(\mathfrak{M}f_\Pi\ket{\psi}\right)T^\dagger$ will have a well-defined block structure.
For example, if $\mathcal{C}\ket{\psi} = \mathcal{I}\ket{\psi} = \ket{\psi}$, all symmetry-resolved blocks of $T\left(\mathfrak{M}f_\Pi\ket{\psi}\right)T^\dagger$, except the four blocks along the main diagonal, will contain zeros; whereas in the case when $\mathcal{C}\ket{\psi} = \mathcal{I}\ket{\psi} = -\ket{\psi}$, the non-zero blocks will be along the anti-diagonal.
Note that singular value decomposition (SVD) of the matrix $\mathfrak{M}f_\Pi\ket{\psi}$, which yields the Schmidt values of $f_\Pi\ket{\psi}$, is identical to that of the matrix $T\left(\mathfrak{M}f_\Pi\ket{\psi}\right)T^\dagger$.
The symmetry-resolved block structure of the latter allows obtaining the full SVD in multiple steps from individual blocks.

For our third simplification, let us introduce the operation $[\bullet]_{\alpha\beta}$ that extracts a symmetry-resolved block specified by $\alpha,\beta \in \{(\pm,\pm)\}$ from the matrix $T\left(\mathfrak{M}f_\Pi\ket{\psi}\right)T^\dagger$.
Only states in $\mathcal{N}_p^{\alpha\times\beta} \subseteq \mathcal{N}_p$ [where $\times$ denotes element-wise multiplication; e.g.,  $(-,-)\times(-,-) = (+,+)$ and so on] can have a non-vanishing $\alpha\beta$ block.
Consider the set of matrices
\begin{equation}
  \label{eq:srbs}
  \{e_j \}^{p,\Pi}_{\alpha\beta} \equiv \orth\{[T (\mathfrak{M}f_\Pi\ket{\psi_i}) T^\dagger]_{\alpha\beta}: \ket{\psi_i} \in \mathcal{N}_p\},
\end{equation}
where $\orth$ is the orthogonalization operation which ensures the vectorized matrices satisfy $\braket{e_i}{e_j} = \delta_{ij}$ for any $i,j$.
If $f_\Pi$ is injective and there exist $\alpha,\beta$ such that
\begin{equation}
  \label{eq:D}
  \left|\{e_j\}^{p,\Pi}_{\alpha\beta}\right| = \dim\mathcal{N}_p^{\alpha\times\beta} \equiv D,
\end{equation}
the map from any $\ket{\psi} \in \mathcal{N}_p^{\alpha\times\beta}$ to $\rho = [T (\mathfrak{M}f_\Pi\ket{\psi}) T^\dagger]_{\alpha\beta}$ is also injective, and hence any $\rho^* \in \spn\{e_j\}_{\alpha\beta}^{p,\Pi}$ can be mapped back to its unique pre-image $\ket{\psi^*}$ [as illustrated in Fig.~\ref{fig:mapping}].
In practice, the inverse mapping can be accomplished by acting on $\ket{\rho^*}$ with the linear transformation
\begin{equation}
  \label{eq:invmap}
  \begin{pmatrix}
    \ket{\psi_1} & \ket{\psi_2} & \cdots & \ket{\psi_D}
  \end{pmatrix}
  \begin{pmatrix}
    \ket{\rho_1} & \ket{\rho_2} & \cdots & \ket{\rho_D}
  \end{pmatrix}^{+},
\end{equation}
where $\{\ket{\psi_i}\}$ is the orthonormal basis of $\mathcal{N}_{p}^{\alpha\times\beta}$, $\rho_i = [T (\mathfrak{M}f_\Pi\ket{\psi_i}) T^\dagger]_{\alpha\beta}$, and ``$+$'' denotes the Moore-Penrose pseudoinverse \cite{Penrose_1955generalized}, which --- under the assumptions above --- is unique and always exists.
The assumption made in Eq.~(\ref{eq:D}) might initially appear quite strong.
However, considering that the Hamiltonians of interest are chaotic, all the blocks in the image of a typical ``thermal'' state are expected to be strongly intertwined and full-rank.
Consequently, finding $\left|\{e_j\}_{\alpha\beta}^{p,\Pi}\right| < \dim\mathcal{N}_p^{\alpha\times\beta}$ for some fixed $\alpha\times\beta$ would be highly anomalous, as that would imply the existence of states with vanishing block $\alpha\beta$ in their image under our mapping.
As we will see in Sec.~\ref{sec:distdemo}, such anomaly would itself indicate the presence of scars.

Since the set $\{e_j \}_{\alpha\beta}^{p,\Pi}$ contains matrices with dimensions approximately four times smaller than those in $\mathfrak{M}f_\Pi\mathcal{N}_p$, we have achived an overall reduction of the original problem by about a factor of $(4d)^2$ (in terms of the total number of elements) without losing any information necessary to execute all our simplification steps in reverse.
Compared to the original problem, performing a rank minimization search over $\spn\{e_j \}_{\alpha\beta}^{p,\Pi}$ has lower computational cost (associated with the SVD operations) by approximately a factor of $(4d)^3$.
More importantly, the heuristics for solving the RMP are effective only when working with reasonably small matrices.
Therefore, our simplifications not only improve the time complexity of the search, but, in fact, make it practically feasible.

Amoung the various approaches to solving the RMP~\cite{fazel2004}, we have chosen to adopt a variant of the alternating projections algorithm due to its simplicity.
Originally studied by von Neumann, this algorithm aims to find a point in the intersection of two arbitrary closed non-empty convex subsets of a Hilbert space~\cite{Bauschke_1993}.
A sequence of alternating orthogonal projections onto these two sets always converges to a point in their intersection, provided that the intersection is non-empty.
We are interested in discovering points in the intersection of $\spn\{e_j \}_{\alpha\beta}^{p,\Pi}$ and the set of dimension-compatible matrices whose rank is upper-bounded by an integer parameter $t$ [see Fig.~\ref{fig:altproj}].
The latter set is non-convex and not closed.
However, assuming the intersection of the sets is non-empty, local convergence of the alternating projections algorithm is still guaranteed for a subset of initial points $\rho_\text{init} \in \spn\{e_j \}_{\alpha\beta}^{p,\Pi}$.
Algorithm~\ref{alg:tsvd}, with the integer parameter $t \ll \max_{\rho \in \spn\{e_j \}_{\alpha\beta}^{p,\Pi}}\{\rank \rho\}$ and a numerical tolerance $\varepsilon_0\ll 1$, applied to $\{e_j\}_{\alpha\beta}^{p,\Pi}$ handles such an ideal ``happy path'' scenario.
For simplicity, we have omitted explicit handling of the case where the algorithm enters a limit cycle, which occurs when $\rho_\text{init}$ is in the basin of attraction of a null-result, such as $\rho^\circ$ in Fig.~\ref{fig:altproj}.
\begin{algorithm}[H]
  \caption{Alternating projections (``happy path'').}\label{alg:tsvd}
  \begin{algorithmic}
    \Require $\left\{e_i: e_i \in \mathbf{R}^{m\times n}\right\}$,  $t > 0$, $\varepsilon_0 \geq 0$
    \Ensure $\braket{e_i}{e_j}=\delta_{ij}$ \Comment{Vectorized matrices}
    \State $c \gets \text{random vector in } \mathbf{R}^{|\{e_i\}|}$
    \State $\varepsilon \gets 1$
    \While{$\varepsilon > \varepsilon_0$}
    \State $\rho \gets \sum_i c_ie_i$
    \State $\mathbf{U}_t, \mathbf{\Sigma}_t, \mathbf{V}^T_t \gets \svd_t(\rho)$ \Comment{Truncated SVD, $\mathbf{\Sigma}_t \in \mathbf{R}^{t\times t}$}
    \State $\bar \rho \gets \mathbf{U}_t\mathbf{\Sigma}_t\mathbf{V}^T_t$
    \State $\bar\rho \gets \tilde\rho / \braket{\bar\rho}$
    \State $c_i \gets \braket{e_i}{\bar\rho}$ for any $i$
    \State $\varepsilon \gets 1 - \lVert c\rVert$
    \EndWhile
  \end{algorithmic}
\end{algorithm}

We perform rank minimization search by repeatedly running Algorithm~\ref{alg:tsvd} with particular input parameters.
Each run either converges to an intersection point $\rho^*$, or gets stuck in a limit cycle.
Given that RMPs are NP-hard, the solutions are heuristic, and, unfortunately, there is no guarantee that all intersection points will be discovered.
The basins of attraction of certain intersection points can be arbitrarily small relative to the size of $\spn\{e_j \}_{\alpha\beta}^{p,\Pi}$.
Therefore, even though this approach has been remarkably successful in discovering new eigenstates of PXP-type models (as well as other nonintegrable models with exponentially degenerate subspaces), no claims regarding its exhaustiveness can be made.
The alternating projections approach to the RMP also suffers from slow convergence, which typically means that the number of iterations in the main loop of Algorithm~\ref{alg:tsvd} can be rather high.

Once an anomalous $\rho^* \in \spn\{e_j \}_{\alpha\beta}^{p,\Pi}$ is obtained using Algorithm~\ref{alg:tsvd} or another RMP heuristic, we can perform the inverse mapping via Eq.~(\ref{eq:invmap}) and acquire the corresponding $\ket{\psi^*}$.
In practice, repeated convergence on the same $\rho^*$ signifies that the corresponding $\ket{\psi^*}$ is likely a finite-system-size instance of a scar state.

The state $\ket{\psi^*}$ can be fit to a variational MPS ansatz of a certain bond dimension (inferred from the entanglement spectrum of $\ket{\psi^*}$) and then proven using Theorem~\ref{thrm:gengenzm}.
Alternatively, an algebra of parent Hamiltonians~\cite{perez2007matrix,Schuch_2010} of $\ket{\psi^*}$ with range $r$ in the $p$-blocked basis can be generated from the basis of the nullspace of the RDM, $\spn\{\ket{v_i}\} \equiv \ker\Tr_{[r+1 \dots L_b]}\ketbra{\psi^*}$, where sites $[r+1 \dots L_b]$ are being traced over.
Here, we assume that for some $r < L_b/2$ the RDM is not full-rank and that state vectors trivially violating the kinetic constraint are excluded from $\{\ket{v_i}\}$.
Generically, as long as the same holds for $\ket{\psi^*}$, we can always assume that each state vector in $\{\ket{v_i}\}$ has definite $\mathcal{C}$ and $\mathcal{I}$ symmetry quantum numbers (with the inversion for $\ket{v_i}$ understood with respect to the middle of the region $[1\dots r]$).

We can generate a family of frustration-free extensive-local parent Hamiltonians from the basis $\{\ket{v_i}\}$ via
\begin{equation}
  \label{eq:tiph}
  V_r(\mathbf q) = \sum_{j=1}^{L_b} \sum_i q_i{P_i}_{[j \dots j+r-1]},
\end{equation}
where $q_i \geq 0$ are some constants and $P_i=\ketbra{v_i}$ are strictly local projectors of system-size-independent range $r$.
If $\ket{\psi^*}$ is an instance of an injective MPS, there are guaranteed to exist $r$ and a choice of $\mathbf{q}$ such that $\ket{\psi^*}$ is the unique ground state of $V_r(\mathbf q)$.
In such a case, $\ket{\psi^*}$ is the unique state in the common kernel of local projectors $P_i$ and, using the notation of Ref.~\cite{Moudgalya_2024}, we have the ``Shirashi-Mori'' bond algebra and commutant pair
\begin{subequations}
  \begin{align}
    \label{eq:bondalg}
    \tmA^\text{MPS}_\text{scar} &= \lgen \{P_{[j\dots j+ r- 1]} h_{[j]} P_{[j\dots j+ r -1]}\} \rgen, \\
    \label{eq:commutant}
    \tmC^\text{MPS}_\text{scar} &= \lgen \ketbra{\psi^*} \rgen,
\end{align}
\end{subequations}
where $h_{[j]}$ are generic local operators aligned with site $j$ whose support may be different from that of $P_{[j\dots j+r-1]}$ (per Ref.~\cite{Moudgalya_2024} App.~A, such $h_{[j]}$ with large enough but finite range exist to exhaustively generate the bond algebra as specified by the above scar commutant).
When $\ket{\psi^*}$ is not an instance of an injective MPS, the common kernel of the local projectors $P_i$ will be spanned by states $\{\ket{\psi^*_\alpha}\}$, which means the commutant will take the form $\tmC^\text{MPS}_\text{scar} = \lgen \ketbra*{\psi^*_\alpha}{\psi^*_\beta} \rgen$.

If range $r$ is not sufficient to produce a parent Hamiltonian $V_r(\mathbf q)$ with a unique ground state (or a unique ground manifold of non-injective MPSs), the generators of the local bond algebra $\tmA^\text{MPS}_\text{scar}$ in Eq.~(\ref{eq:bondalg}) can be augmented with the Hamiltonian $H$ serving as a degeneracy lifting operator.
This will typically be enough to generate the full symmetry algebra of the scar manifold related to the state $\ket{\psi^*}$.
The dimension of this manifold can be expressed as
\begin{equation}
  \label{eq:dscar}
  d_\text{scar}^{\ket*{\psi^*}} \equiv \lim_{L_b\to\infty} \min_{\mathbf q} \dim\ker (H +  V_r(\mathbf q)),
\end{equation}
where the convergence is anticipated when $L_b \gtrsim 2r$.
Hence, in practice, Eq.~(\ref{eq:dscar}) can be evaluated numerically using finite-size systems accessible via ED.
For choices of $\mathbf q$ that minimize $d_\text{scar}^{\ket*{\psi^*}}$, instances of scar states spanning the manifold associated with $\ket{\psi^*}$ can be obtained for all numerically accessible system sizes satisfying $\dim\ker (H +  V_r(\mathbf q)) = d_\text{scar}^{\ket*{\psi^*}}$.
In general, a perturbation $V_r(\mathbf q)$ satisfying $\{V_r(\mathbf q), \mathcal{IC}\} \neq 0$ eliminates the symmetry-protected exponential degeneracy in the nullspace of $H$, which arises from the condition $\{H, \mathcal{IC}\} = 0$~\cite{Schecter_2018, Turner_2018,Buijsman_2022}.
\subsection{Demonstration using the PXP Hamiltonian}
To demonstrate the procedure discussed, we will use the PXP Hamiltonian as an example.
In what follows, we will work in the 2-blocked basis ($p=2$) and use the projector $\Pi = \ketbra{O}_1\otimes\ketbra{O}_{L_b/2 + 1}$, where $\ket{O} = \ket{00}$ in the spin-$1/2$ basis.
Although our algorithm can handle larger systems, a modest system size of $L_b = 10$ (or $L = 20$) is sufficient for this demonstration.

\begin{figure}
  \subfloat{\label{fig:psi_npp}}
  \subfloat{\label{fig:psi_npp_mapped}}
  \subfloat{\label{fig:psi_nmm}}
  \subfloat{\label{fig:psi_nmm_mapped}}
  \includegraphics[width=\columnwidth]{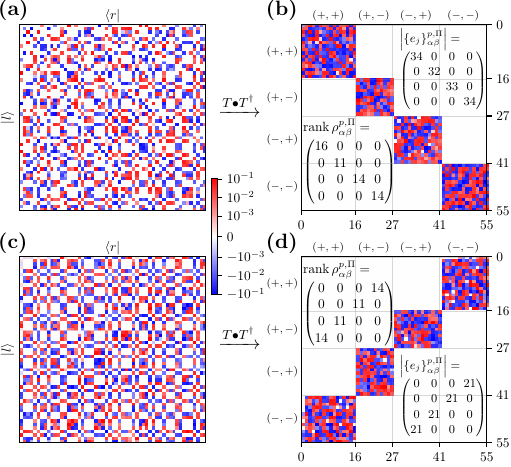}
  \caption{Bijective mapping of $\mathcal{N}_2^{(+,+)}$ and $\mathcal{N}_2^{(-,-)}$ from $H_\text{PXP}$.
    The system size is $L_b=10$, $\dim\mathcal{N}_2^{(+,+)} = 34$ and $\dim\mathcal{N}_2^{(-,-)} = 21$.
    (a)~$\mathfrak{M}f_\Pi\ket{\psi}$ for a state $\ket{\psi} \in \mathcal{N}_2^{(+,+)}$; (b)~its transform $T\left(\mathfrak{M}f_\Pi\ket{\psi}\right)T^{-1}$.
    Note that all non-vanishing blocks $\rho_{\alpha\beta}^{p,\Pi}$ of the transformed $\ket{\psi}$ in (b) are full-rank, which indicates that $\ket{\psi}$ is a rather typical state.
    However, considering full symmetry-resolved block subspaces generated via Eq.~(\ref{eq:srbs}) from $\mathcal{N}_2^{(+,+)}$, there exist blocks for which $\left|\{e_j\}_{\alpha\beta}^{p,\Pi}\right| < \dim\mathcal{N}_p^{\alpha\times\beta}$ (i.e., when $\alpha=\beta=(\pm,-)$).
    This points to the existence of at least two scar states with vanishing corresponding blocks in their images under our mapping.
    (c),(d)~The same transformations as in (a),(b) for $\mathcal{N}_2^{(-,-)}$.
    Note that here $\left|\{e_j\}_{\alpha\beta}^{p,\Pi}\right| = \dim\mathcal{N}_p^{\alpha\times\beta}$ for all non-vanishing blocks.
    \label{fig:pxp_mapping}
  }
\end{figure}

Suppose we have acquired the full nullspace $\mathcal{N}_2$ of $H_\text{PXP}$ consisting of mutually orthogonal subspaces  $\mathcal{N}_2^{(+,+)}$ and $\mathcal{N}_2^{(-,-)}$.
These symmetry-resolved subspaces can be naturally separated by utilizing the semi-momentum basis with definite inversion symmetry (see Ref.~\cite{Sandvik_2010}).
In Fig.~\ref{fig:pxp_mapping} we show the intermediate steps of our bijective mapping process and provide additional insights on the structure of the resultant image subspaces.
Note that since the states in $\mathcal{N}_2^{(+,+)}$ and $\mathcal{N}_2^{(-,-)}$ have definite inversion symmetries, and inversion about the $\mathcal{I}_V$ axis in Fig.~\ref{fig:ndproj} corresponds to the transposition of $T\left(\mathfrak{M}f_\Pi\ket{\psi}\right)T^\dagger$ for any $\ket{\psi}$, the matrices in Figs.~\ref{fig:psi_npp_mapped} and \ref{fig:psi_nmm_mapped} are, respectively, symmetric and antisymmetric.
This means that while there can be as many as four distinct bijective mappings for $\mathcal{N}_2^{(+,+)}$, at most two such mappings are available for $\mathcal{N}_2^{(-,-)}$; e.g., we can only use the block images above the main diagonal in $T\left(\mathfrak{M}f_\Pi\ket{\psi}\right)T^{-1}$ for $\ket{\psi} \in \mathcal{N}_2^{(-,-)}$.

\label{sec:distdemo}

\begin{figure}
  \subfloat{\label{fig:phi1_img}}
  \subfloat{\label{fig:phi2_img}}
  \subfloat{\label{fig:theta1_img}}
  \subfloat{\label{fig:theta2_img}}
  \subfloat{\label{fig:imomega_img}}
  \subfloat{\label{fig:reomega_img}}
  \includegraphics[width=\columnwidth]{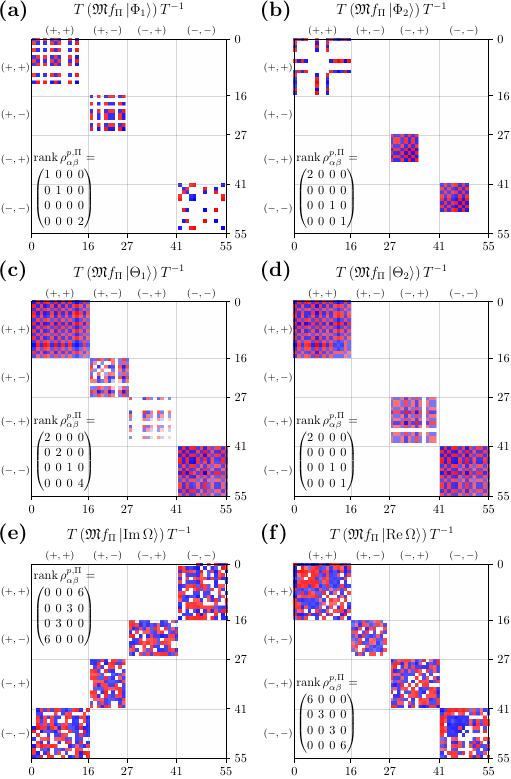}
  \caption{States discovered through our approach.
    (a)--(d)~$t = 2$, mapping of $\mathcal{N}_2^{(+,+)}$ to the first diagonal block ($\alpha=\beta=(+,+)$).
    (e)~$t = 3$, mapping of $\mathcal{N}_2^{(-,-)}$ to the first block above the main diagonal ($\alpha=(+,-)$, $\beta=(-,+)$).
    (f)~State discovered as a non-injective MPS sibling of the state in (e).
    This state is also discoverable with $t = 6$ and the mapping of $\mathcal{N}_2^{(+,+)}$ as in (a)--(d).
  }
\end{figure}

By comparing the dimensions of the symmetry-resolved block subspaces $\{e_j\}_{\alpha\beta}^{p,\Pi}$ --- provided in Fig.~\ref{fig:psi_npp_mapped} --- with $\dim\mathcal{N}_2^{(+,+)}$ listed in the figure caption, we see that the mapping of the subspace $\mathcal{N}_2^{(+,+)}$ to $\spn\{e_j\}_{\alpha\beta}^{p,\Pi}$ is bijective for $\alpha=\beta=(\pm,\pm)$.
Let us set $\alpha=\beta=(+,+)$ and apply Algorithm~\ref{alg:tsvd} to $\{e_j\}_{\alpha\beta}^{p,\Pi}$.
Note that this step is essentially the RMP over the span of 34 matrices of size $16\times 16$. \
When the maximal rank parameter $t=1$, the algorithm consistently converges to the same $\rho^*$ [the first block in Fig.~\ref{fig:phi1_img}] corresponding to the known scar state $\ket{\Phi_1}$ under the inverse mapping.
By incrementally increasing $t$, we hope to discover more complex states with higher bond dimensions \footnote{While this intuition is often correct, the relationship between a state's bond dimension and the ranks of its individual symmetry-resolved block images is not consistently predictable.}.
Indeed, with $t=2$, the algorithm consistently converges on four distinct $\rho^*$ corresponding to states whose symmetry-resolved block images are shown in Figs.~\ref{fig:phi1_img}--\ref{fig:theta2_img}.
We identify the result in Fig.~\ref{fig:phi2_img} with the known scar state $\ket{\Phi_2}$.
The results in Figs.~\ref{fig:theta1_img} and \ref{fig:theta2_img}, however, correspond to previously unknown eigenstates.
Upon further inspection, we note that these two newly discovered candidate scar states are related to each other by translation operator $T_x$ in the same way as $\ket{\Phi_1}$ and $\ket{\Phi_2}$ are related.
Per this structural analogy, we label them as $\ket{\Theta_1}$ and $\ket{\Theta_2}$.

For the subspace $\mathcal{N}_2^{(-,-)}$, mappings to all symmetry-resolved block images are bijective, see the subspace dimensions provided in Fig.~\ref{fig:psi_nmm_mapped}.
Let us choose the smallest block with $\alpha=(+,-)$ and $\beta=(-,+)$ (right above the main diagonal).
This gives an RMP over the span of 21 matrices of size $11\times 14$.
Applying Algorithm~\ref{alg:tsvd} to $\{e_j\}_{\alpha\beta}^{p,\Pi}$ yields the first successful outcome with $t=3$.
The image of the corresponding (and previously unknown) state --- which we will label as $\ket{\Im\Omega}$ anticipating its relation to the state $\ket{\Omega}$ --- is shown in Fig.~\ref{fig:imomega_img}.
Upon further analysis, we note that $\ket{\Im\Omega}$ is TI in the spin-$1/2$ basis.

The three interesting wavefunctions we have discovered --- $\ket{\Theta_{1,2}}$ and $\ket{\Im\Omega}$ --- can be expressed in a form with integer coefficients.
This further confirms their anomalous nature and also simplifies the process of obtaining exact parent Hamiltonians.
In App.~\ref{app:ph} we provide the basis of the nullspace of the smallest less-than-full-rank RDM for each of these three states.
The RDMs of $\ket{\Theta_{1,2}}$ become less-than-full-rank when $r = 3$ ($p=2$-blocked sites).
In the case of $\ket{\Im\Omega}$, using its natural spin-1/2 basis, the RDM becomes less-than-full-rank when $r = 7$.

Using the construction of Eq.~(\ref{eq:tiph}) with the basis $\{\ket{v_i}\}$ for $\ket{\Theta_1}$ in Eqs.~(\ref{eq:v1_theta1})--(\ref{eq:v5_theta1}) and setting all $q_i = 1$, we obtain a parent Hamiltonian with doubly-degenerate ground state for any $L_b \geq 5$.
However, Eq.~(\ref{eq:dscar}) --- evaluated numerically --- gives $d_\text{scar}^{\ket{\Theta_1}} = 1$.
This indicates that we have indeed found a new scar state, and also that $r=3$ projectors are insufficient for generating its complete bond algebra.
The extra ground state in the nullspace of the $r=3$ parent Hamiltonian is the trivial $\ket{O}^{\otimes L_b}$ --- it is easy to see that $\ket{O}^{\otimes 3} \perp \spn\{v_i\}$; the $r=4$ parent Hamiltonian, which we will not construct here, has a unique ground state $\ket{\Theta_1}$.

Repeating the above steps using the basis $\{v_i\}$ for $\ket{\Theta_2}$ in Eqs.~(\ref{eq:v1_theta2})--(\ref{eq:v6_theta2}) (again with all $q_i = 1$), we find that the resultant $r=3$ parent Hamiltonian has a unique ground state for any $L_b \geq 6$; correspondingly, $d_\text{scar}^{\ket{\Theta_2}} = 1$.

The above results indicate that $\ket{\Theta_{1,2}}$ are injective MPSs.
We find that their Schmidt indices associated with a bipartition into two subsystems of the same size saturate at $\chi^2 = 16$, which for injective TI MPSs corresponds to the bond dimension of $\chi = 4$.
Thus, fitting these states to appropriate variational MPS ans\"atze  --- a process, which, unfortunately, involves a significant amount of trial and error --- we obtain the tensors given in Eqs.~(\ref{eq:theta1mps}) and (\ref{eq:theta2mps}).

Performing similar steps using the basis $\{v_i\}$ for $\ket{\Im\Omega}$ in Eqs.~(\ref{eq:v1_imomega})--(\ref{eq:v6_imomega}), we find that with $r=7$ (spin-1/2 sites) it is impossible to construct a parent Hamiltonian $V_r(\mathbf q)$ with a finite-dimensional ground state manifold.
However, Eq.~(\ref{eq:dscar}) gives $d_\text{scar}^{\ket{\Im\Omega}} = 2$, which indicates that $\ket{\Im\Omega}$ is not an injective MPS.
Using our constructed $V_r(\mathbf q)$ (again with all $q_i = 1$), we can now numerically obtain $\ker (H +  V_r(\mathbf q)) = \spn\{\ket{\Re\Omega}, \ket{\Im\Omega}\}$, where $\ket{\Re\Omega}$ is the non-injective MPS sibling of $\ket{\Im\Omega}$.
In Fig.~\ref{fig:reomega_img} we show the symmetry-resolved block images of $\ket{\Re\Omega}$.
Note that $\ket{\Re\Omega} \in \mathcal{N}_2^{(+,+)}$ and $\ket{\Im\Omega} \in \mathcal{N}_2^{(-,-)}$ are orthogonal and can be distinguished by their $\mathcal{C}$ and $\mathcal{I}$ symmetry quantum numbers.
This holds for all system sizes $L\geq 9$ (both even and odd).

We find that the Schmidt indices for the bipartitions of $\ket{\Re\Omega}$ and $\ket{\Im\Omega}$, as well as those of their generic linear combinations, saturate at $2\chi^2 = 32$, which suggests that these states are superpositions of two injective MPSs with bond dimension $\chi = 4$.
Specifically, this is consistent with two complex-valued injective MPSs related by complex conjugation.
The simplest complex-valued variational MPS ansatz which generates states in $\mathcal{H}^1$ whose real and imaginary parts have definite and opposite $\mathcal{C}$ symmetry quantum numbers is the one with real $M^0$ and purely imaginary nilpotent $M^1$. 
Through this line of reasoning, we obtain the MPS in  Eq.~(\ref{eq:omegamps}).

Note that while $\ket{\Re\Omega} \propto \ket{\Omega} + \ket{\Omega^*}$ can be written as a TI MPS consisting of two diagonal blocks corresponding to the $\ket{\Omega}$ MPS and its complex conjugate, $\ket{\Im\Omega} \propto \ket{\Omega} - \ket{\Omega^*}$ does not have a regular system-size independent TI MPS form.
Even though $\ket{\Im\Omega}$ has the same bond dimension as $\ket{\Re\Omega}$, its MPS representation requires a system-size-dependent complex prefactor.
The fact that our algorithm is unbiased with respect to the representation of the scars enables the discovery of states like $\ket{\Im\Omega}$ as well as the TTI MPS eigenstates of the PPXPP model in Eqs.~(\ref{eq:s1tti})--(\ref{eq:ttti}).

It is worth mentioning that the volume-entangled scar state $\ket{\Lambda}$ reported in Ref.~\cite{Ivanov_2025} was, in fact, discovered via similar techniques, albeit without applying Algorithm~\ref{alg:tsvd}.
It is easily verified that $T\left(\mathfrak{M}f_\Pi\ket{\Lambda}\right)T^{-1}$ is a diagonal matrix whose four symmetry-resolved blocks are each proportional to identity matrices.
Consequently, $\rho^* = \mathbf 1 \in \spn\{e_j\}_{\alpha\beta}^{p,\Pi}$ maps back to $\ket{\Lambda}$ for suitable values of $\alpha$ and $\beta$.
This indicates that an anomalous $\rho^*$ need not necessarily be low-rank.

\begin{figure}
  \subfloat{\label{fig:perft2}}
  \subfloat{\label{fig:perft3}}
  \subfloat{\label{fig:perft6}}
  \includegraphics[width=\columnwidth]{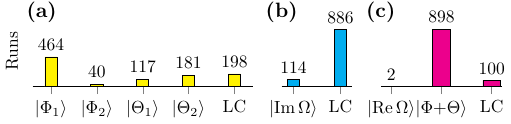}
  \caption{Performance of the scar distillation procedure over 1000 runs of Algorithm~\ref{alg:tsvd} for various input parameters. We show the number of runs that converged to a specific state as well as the number of runs that ended in a limit cycle (LC).
    (a)~$t = 2$, mapping of $\mathcal{N}_2^{(+,+)}$ to the block with $\alpha=\beta=(+,+)$.
    (b)~$t = 3$, mapping of $\mathcal{N}_2^{(-,-)}$ to the block with $\alpha=(+,-)$, $\beta=(-,+)$.
    (c)~$t = 6$, same mapping as in (a). $\ket{\Phi{+}\Theta}$ denotes any superposition of $\ket{\Phi_{1,2}}$ and $\ket{\Theta_{1,2}}$.
  }
  \label{fig:perf}
\end{figure}

To conclude this section, we make a few remarks on the performance of our algorithm and its limitations.
Figure~\ref{fig:perf} illustrates how outcomes of the scar distillation procedure vary with its input.
As should be expected, the discovery of more complex scars, whose symmetry-resolved block images have higher ranks, becomes increasingly harder.
This is exemplified by Fig.~\ref{fig:perft6}.
Although the algorithm is able to converge on $\ket{\Re\Omega}$, the probability of that is notably low.
The maximum rank parameter $t=6$ allows for a large manifold of linear combinations of the states $\ket{\Phi_{1,2}}$ and $\ket{\Theta_{1,2}}$.
The basin of attraction of this manifold is quite large, which is why the procedure becomes heavily biased against states like $\ket{\Re\Omega}$.
We can compare this with Fig.~\ref{fig:perft3}, where no such bias exists against $\ket{\Im\Omega}$; however, higher values of $t$ appear to make limit cycles more likely.
Unfortunately, since the exact scar states in our models of interest are often non-orthogonal in finite-size systems, this bias cannot be mitigated by removing the already known states from the search space.

For the PXP model, discovering more complex scars than those identified so far requires exploring larger system sizes.
This leads to a substantially larger search space due to the exponential growth of the zero-energy space dimension with system size.
In general, as the search space expands, limit cycles become more prevalent, and the likelihood of converging to increasingly complex states decreases significantly.
Combined with the bias discussed earlier, this renders our current approach ineffective in uncovering yet more intricate scars hosted by the PXP Hamiltonian.
Nevertheless, we hope that this initial investigation --- by providing a glimpse behind the seemingly impenetrable fa\c{c}ade of the exponentially degenerate zero-energy manifold --- inspires further explorations into the (partial) exact solvability and integrability of PXP-type models.

\subsection{Remarks on the PPXPP Hamiltonian}
Our algorithm similarly identifies all the scars of the PPXPP Hamiltonian listed in Eqs.~(\ref{eq:s1mps})--(\ref{eq:ttti}).
The scar distillation approach proves more effective here, which can --- at least in part --- be attributed to the smaller nullspaces compared to those of the PXP Hamiltonian.
Notably, by applying the algorithm to larger systems, we have identified multiple highly non-trivial TTI MPS-like scars, potentially of higher orders, which, at the time of writing, we have yet to express explicitly.
We leave finding the corresponding analytic expressions and proofs for future work.

\section{Dynamical signatures of exact  \texorpdfstring{$E=0$}{E=0} scars}
\label{sec:dynamics}
The existence of exact scar states has been associated with the presence of additional non-thermal low-entanglement states nearby \cite{Mori_2017,lin2019exact,Lin_2020}.
For instance, any TTI MPS (aka SMA) on top of an exact TI MPS has an upper-bounded energy variance in the thermodynamic limit, owing to the short-range correlations of local observables in MPSs (see App. F of Ref.~\cite{Lin_2020} for more details).

\begin{figure}
  \subfloat{\label{fig:z2overlaps}}
  \subfloat{\label{fig:z2dynamic}}
  \subfloat{\label{fig:v1overlaps}}
  \subfloat{\label{fig:z2dynamicphi}}
  \includegraphics[width=\columnwidth]{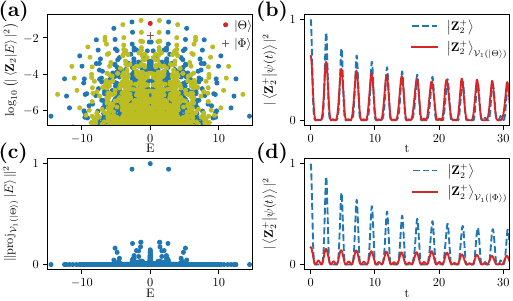}
  \caption{Relation between the $\mathbb{Z}_2$ CDW state revivals and $\ket{\Theta} = \ket{\Theta_1} + \ket{\Theta_2}$ (the TI variant of the states $\ket{\Theta_{1,2}}$).
    The system size is $L=24$.
    (a)~Overlaps of the eigenstates of $H_\text{PXP}$ with the CDW state $\ket{\mathbb{Z}_2}$.
    Blue symbols denote states in the sector $k = 0,~\langle I \rangle = 1$, while green symbols correspond to the the sector $k = \pi,~\langle I \rangle = -1$.
    Note that the CDW overlap with $\ket{\Theta}$ closely matches that of the neighboring members of the primary $\mathbb{Z}_2$ scar tower, whereas the CDW overlap with $\ket{\Phi} = \ket{\Phi_1} + \ket{\Phi_2}$ (the TI variant of the states $\ket{\Phi_{1,2}}$) similarly aligns with the secondary $\mathbb{Z}_2$ tower.
    This is observed across all numerically accessible system sizes.
    (b)~Dependence of the revival amplitudes of the $\ket{\mathbb{Z}_2^+}$ state on the initial state.
    See text for details.
    (c)~Overlaps of the eigenstates with the ten-dimensional $\smasub_1(\ket{\Theta})$ manifold.
    (d)~Same as (b), but with $\ket{\mathbb{Z}_2^+}$ projected onto the four-dimensional $\smasub_1(\ket{\Phi})$ manifold.
  }
\end{figure}

Consider the manifold $\smasub_1\left(\ket{\Theta}\right)$, which represents $\mathcal{H}^1$-projected TTI MPSs built upon the TI combination of the states $\ket{\Theta_{1,2}}$.
Specifically, $\smasub_1\left(\ket{\Theta}\right) = \spn\{\mathcal{P}_f\ket*{\widetilde\Theta^{(j)}}\}$, where
\begin{equation}
  \mps{\ket*{\widetilde\Theta^{(j)}}}{\left\{\sma{M}{M'_j}^s\right\}},
\end{equation}
$M^s$ is the representation of the state $\ket{\Theta}$ in App.~\ref{app:rydbti}, and $M'^s_j$ are arbitrary tensors with the same dimensions.
Operationally, using the linearity of the TTI MPS in $M'^s_j$, the states $\ket*{\widetilde\Theta^{(j)}}$ can be constructed by iterating over all choices for a single non-zero element in the $M'^s_j$ tensors.
We find numerically that $\dim\smasub_1\left(\ket{\Theta}\right) = 10$ for all system sizes $L > 8$ where $L$ is a multiple of 4.
In such systems, the high overlap between the TI state $\ket{\Theta}$ and the charge density wave (CDW) state $\ket{\mathbb{Z}_2} = \ket{0101\dots01}$ shown in Fig.~\ref{fig:z2overlaps} suggests a strong connection between $\ket{\Theta}$ and the nearby states within the $\mathbb{Z}_2$ scar tower.
We therefore expect $\smasub_1\left(\ket{\Theta}\right)$ to play an important role in the CDW state revivals.

This expectation is confirmed in Fig.~\ref{fig:z2dynamic}, which shows how the revivals (more precisely, oscillating presence) of the state $\ket{\mathbb{Z}_2^+} = (\ket{\mathbb{Z}_2} + T_x\ket{\mathbb{Z}_2})/\sqrt{2}$, quantified as $|\braket{\mathbb{Z}_2^+}{\psi(t)}|^2$, depend on the initial state $\ket{\psi(0)}$ for the two choices explained below.
Note that using $\ket{\mathbb{Z}_2^+}$ instead of $\ket{\mathbb{Z}_2}$ enables simulations to be performed within a sector where translation and inversion symmetries are fully resolved, while still yielding qualitatively similar results to those obtained with $\ket{\mathbb{Z}_2}$.
We compare the revival amplitudes when starting from the state $\ket{\psi(0)} = \ket{\mathbb{Z}_2^+}$ to those starting from the state $\ket{\psi(0)} = \ket{\mathbb{Z}_2^+}_{\smasub_1{(\ket{\Theta})}} \in \smasub_1{(\ket{\Theta})}$, which has the maximum possible $\ket{\mathbb{Z}_2^+}$ overlap among all states in $\smasub_1\left(\ket{\Theta}\right)$ --- this is effectively the normalized projection of $\ket{\mathbb{Z}_2^+}$ onto the subspace $\smasub_1(\ket{\Theta})$.
Although the initial $\ket{\mathbb{Z}_2^+}$ overlap is smaller in the latter case, its slower decay rate results in a slightly higher $\ket{\mathbb{Z}_2^+}$ overlap at later times compared to the former case.
The long-lived near-full revivals in the $\ket{\psi(0)} = \ket{\mathbb{Z}_2^+}_{\smasub_1{(\ket{\Theta})}}$ case suggest that this subspace contains states that are close to exact eigenstates of the PXP model with specific non-zero energies, as we will confirm directly below. 
It is also notable that the $\ket{\mathbb{Z}_2^+}$ contains more than $63\%$ of its weight in the $\smasub_1{(\ket{\Theta})}$ subspace for this system size, and the approximate agreement between the two cases at later time suggests that it is this weight that is mainly responsible for the observed revivals in the $\ket{\psi(0)} = \ket{\mathbb{Z}_2^+}$ case as well.

In Fig.~\ref{fig:v1overlaps} we show the squared norms of the projections of the eigenstates of $H_\text{PXP}$ onto $\smasub_1(\ket{\Theta})$.
Note the five approximately equally spaced in energy peaks aligned with the primary scars from the $\ev{\mathcal{I}} = 1$ sector in the middle of the spectrum.
Also note the remarkably high overlaps with the two primary scars from that sector closest to $E=0$, i.e., with energies $\approx \pm 2.67$.

Note that such projections of the exact eigenstates onto $\smasub_1(\ket{\Theta})$ can be viewed as best possible approximations to these eigenstates by states in the variational subspace $\smasub_1(\ket{\Theta})$.
In principle, we can perform ``unbiased'' searches for eigenstate approximations in this subspace, e.g., by minimizing the variance of the trial states orthogonal to $\ket{\Theta}$ or by diagonalizing the PXP Hamiltonian in this subspace, and this reproduces very closely the above projections for the closest ED $\mathbb{Z}_2$ scar states with non-zero energy.
Hence we can view these overlaps of the ED states with $\smasub_1(\ket{\Theta})$ as quantifying approximations to the ED scars by this variational subspace.
We note that these overlaps are significantly higher than any prior approximations to these ED $\mathbb{Z}_2$ scar states, see Figs.~S1-S3 in Ref.~\cite{lin2019exact} showing FSA approximation of Ref.~\cite{Turner_2018quantum} and SMA approximations on top of $\ket{\Phi_{1,2}}$ for a slightly larger system of size $L=26$.
This shows that (for the system size at hand) these ED $\mathbb{Z}_2$ scar states are well approximated by some TTI MPS on top of the state $\ket{\Theta}$.
Clearly, the states with lowest energy variance in $\smasub_1(\ket{\Theta})$ have energy spacing similar to that of the primary scars near $E=0$; hence, the observable oscillatory dynamics is expected to emerge within the $\smasub_1(\ket{\Theta})$ manifold.

For comparison, in Fig.~\ref{fig:z2dynamicphi}, we also show the $\ket{\mathbb{Z}_2^+}$ revivals when starting from the state $\ket{\psi(0)} = \ket{\mathbb{Z}_2^+}_{\smasub_1{(\ket{\Phi})}}$ defined analogously to $\ket{\mathbb{Z}_2^+}_{\smasub_1{(\ket{\Theta})}}$ but with respect to the TI combination of $\ket{\Phi_{1,2}}$.
Here, the amplitude of oscillations is significantly lower than in Fig.~\ref{fig:z2dynamicphi} due to both the lower $\ket{\mathbb{Z}_2^+}$ overlap of $\ket{\Phi}$ and the lower dimensionality of the manifold $\smasub_1{(\ket{\Phi})}$.
Our further exploration starting from the state $\ket{\psi(0)} = \ket{\mathbb{Z}_2^+}_{\smasub_1{(\ket{\Theta})} \cup \smasub_1{(\ket{\Phi})}}$ leads to a marginal, but noticeable, improvement in the overall amplitude of revivals.

It is important to emphasize that while the manifold $\smasub_1{(\ket{\Theta})}$ --- and, to a lesser extent, $\smasub_1{(\ket{\Phi})}$ --- plays a significant role in the $\ket{\mathbb{Z}_2^+}$ state revivals, it does not entirely capture the full dynamics of the phenomenon --- in particular, as the system sizes become larger.
States with more than one ``defect'' on top of the exact $E=0$ scars --- e.g., multi-mode approximations (MMA) ---  are also non-thermal, have finite energy variance and extended lifetimes as long as the density of defects is sufficiently small (see Sec.~V of Ref.~\cite{Mori_2017}).
Hence, while in smaller systems ($L\leq 16)$ the $\ket{\mathbb{Z}_2^+}$ revivals occur effectively fully within $\smasub_1{(\ket{\Theta})}$, in larger systems we need to consider generalized multi-defect manifolds $\smasub_n{(\ket{\Theta})}$.
For instance, by taking into account such generalized multi-defect manifolds, the higher initial amplitude of revivals when starting from the state $\ket{\mathbb{Z}_2^+}$ in Fig.~\ref{fig:z2dynamic}, as well as the amplitude's higher initial decay rate, can be explained.

Although $\ket{\Theta_{1,2}}$ do not produce exact eigenstates in systems with OBCs, similar reasoning applies. Specifically, by considering all possible terminations of the $\ket{\Theta_{1,2}}$ MPSs, 16-dimensional non-thermal manifolds can be constructed. These manifolds play a role in the CDW state revivals in systems with OBCs analogous to that of $\smasub_1{(\ket{\Theta})}$ in systems with PBCs.

\section{Summary and discussion}
In this paper, we extended the boundaries of exact solvability for several kinetically constrained models that are of experimental relevance.
Specifically, we uncovered multiple highly non-trivial new exact zero-energy scar states in the PXP and PPXPP models, employing a novel numerical approach that is also applicable to a broader class of Hamiltonians with exponentially degenerate subspaces.
To prove our new exact eigenstates, we formulated general sufficient conditions --- expressed as nonlinear tensor equations --- for an MPS to constitute an eigenstate of a given kinetically constrained Hamiltonian.
Furthermore, we showed that these conditions are satisfied not only by the new scars presented in this work but also by all previously reported exact eigenstates of the PXP/PPXPP and various other kinetically constrained models.

We note that the task of identifying exact $E=0$ eigenstates in the Hamiltonians of interest --- whether approached through our numerical method or via directly solving the nonlinear tensor equations representing the appropriate sufficient conditions for zero energy MPS eigenstates --- inevitably constitutes an instance of an NP-hard problem.
This leads us to conjecture that the exponentially degenerate nullspaces of the PXP/PPXPP and other related Hamiltonians may admit (possibly limited) NP-hard solvability.
In other words, there is no clear reason to assume that the increasingly complex hierarchy of exact $E=0$ scars in these Hamiltonians terminates at the states we were able to discover using the limited tools currently at our disposal.
In fact, there is no compelling reason to assume that this hierarchy terminates at all.
Thus we are left with an intriguing open question: Do exponentially degenerate nullspaces of PXP-type models harbor a finite or infinite number of in-principle exact, yet exceedingly challenging to uncover, zero energy eigenstates?

Since all exact $E=0$ scars have potentially non-thermal manifolds associated with them [obtained by using local defect (SMA/MMA) construction discussed in Sec.~\ref{sec:detection}] where examples of unusual dynamics are anticipated, we believe that addressing the above question and either uncovering or ruling out new scars beyond those introduced in this work is of considerable theoretical and experimental importance.
A potentially fruitful future direction is to apply our numerical approach to other models with symmetry-protected exponentially degenerate nullspaces to uncover new exact area-law scar states.

We demonstrated that the new scar states $\ket{\Theta_{1,2}}$ play a significant --- potentially, crucial --- role in the paradigmatic revivals of the $\mathbb{Z}_2$ CDW state in the PXP chain.
This offers a fresh perspective on the underlying ergodicity-breaking dynamics and potentially paves the way for new attempts to address questions regarding whether these revivals persist in the thermodynamic limit.
We defer such explorations to future dedicated studies.

While we focused on demonstrating the role of the states $\ket{\Theta_{1,2}}$ in ergodicity-breaking phenomena, other new scars are also expected to have non-thermal signatures.
In particular, the state $\ket{\Omega}$ stands out as a promising candidate for further investigation into possible observable non-ergodic phenomena.

Finally, we would like to highlight the recently discovered examples of non-ergodicity beyond the CDW state revivals discussed in Ref.~\cite{kerschbaumer2024quantummanybodyscarspxp}.
It remains unclear whether these phenomena can also be attributed to the presence of specific exact $E=0$ eigenstates in the respective models analogoulsy to how the $\ket{\mathbb{Z}_2}$ state revivals in the PXP chain were linked to the scar states $\ket{\Theta_{1,2}}$.

\begin{acknowledgments}
We thank Manuel Endres, Cheng-Ju Lin, Daniel Mark, Pablo Sala, Federica Surace, Sara Vanovac, and Leo Zhou for useful discussions and previous collaborations on related topics.
We are particularly grateful to Sanjay Moudgalya for participation in the initial stages of the project and for many discussions and encouragements throughout.
This work was supported by the National Science Foundation through grant DMR-2001186.
A.N.I.~acknowledges support also from the Eddleman Quantum Graduate Fellowship at Caltech.
\end{acknowledgments}

\appendix

\section{Structure of the ground (ceiling) manifold of  \texorpdfstring{$H_\text{PXP}$}{PXP} and  \texorpdfstring{$H_\text{PPXPP}$}{PPXPP}}
\label{app:ground}
In Ref.~\cite{Omiya_2023quantum} the $\mathbb{Z}_2$ scars of the PXP Hamiltonian were approximated as projections of certain exact eigenstates of the Hamiltonian $H_1$ in Eq.~(\ref{eq:h1pxp}) into the Rydberg-blockaded subspace (by eliminating all instances of $\ket{RL}_{j,j+1}$).
This approximation was shown to be quite good for the ground and first excited states and, by extension, their spectral reflections --- the ceiling and penultimate excited states; its quality, however, sharply declined for eigenstates closer to the middle of the spectrum.
In this section, we will obtain a TI MPS representation of the approximate ground (ceiling) state and, on top of it, construct a TTI MPS representing the approximate first (penultimate) excited state.
Both representations, initially in the blocked basis natural for $H_1$, will generate the corresponding states from Ref.~\cite{Omiya_2023quantum} for systems with even number of sites.
We will then lift both states to the single-site basis, thus making the approximations well-defined for systems with any number of sites.
In addition to showcasing the usefulness of the TTI MPS representation introduced in Sec.~\ref{sec:mps}, our results will provide new insights into the general structure of the low energy manifold of the PXP model, and, in particular, the poorly understood system size dependence of the bipartite entanglement entropy of the $\mathbb{Z}_2$ scars, 

Since $\frac{1}{2}(\ket{L}+\ket{R}\mp\sqrt{2}\ket{O})$ are eigenstates of individual terms of $H_1$ that minimize (maximize) local energy, the ground and ceiling states of $H_1$ can be written in the blocked basis as follows:
\begin{equation}
  \ket{\Psi}_\mp = \frac{1}{2^{L_b}}\left(\ket{L}+\ket{R}\mp\sqrt{2}\ket{O}\right)^{\otimes L_b}
\end{equation}

TI MPS representation of $\ket{\Psi}_\mp$ has bond dimension 1 and, up to
normalization, can be written as
\begin{equation}
  M_\mp^O = (\mp\sqrt{2}),\quad M^L = M^R = (1).
\end{equation}

Let us rewrite $M^O$, $M^L$, and $M^R$ as products of matrices $B^{0,1}$ and $C^{0,1}$ satisfying the following relations:
\begin{equation}
  \label{eq:bcgs}
  \begin{aligned}
    &B^0C^0 = M^O,\quad
    B^1C^0 = M^L,\quad\\
    &B^0C^1 = M^R,\quad
    B^1C^1 = \mathbf{0}.
  \end{aligned}
\end{equation}
Simplest solutions of Eqs.~(\ref{eq:bcgs}) involve $1 \times 2$ and $2 \times 1$ matrices for $B^s$ and $C^s$, and any such solution is gauge-equivalent to
\begin{equation}
  \begin{aligned}
    &B_\mp^0 = \begin{pmatrix}1 \mp \sqrt{2} & \mp\sqrt{2} \end{pmatrix},\quad
      C^0 = \begin{pmatrix}0 \\ 1 \end{pmatrix},\\
    &B^1 = \begin{pmatrix} 1 & 1\end{pmatrix},\quad
      C^1 = \begin{pmatrix} 1 \\ -1 \end{pmatrix}.
  \end{aligned}
\end{equation}
Switching to blocked basis $C^{s_1}B^{s_2}$ (translated by one spin-1/2 site relative to the one we started with) and setting $C^1B^1$ to zero to enforce Rydberg blockade, we obtain the approximation for the ground (ceiling) states
\begin{equation}  
  \mps{\ket{G}_\mp}{
    \underbrace{
      \begin{pmatrix}
        0 & 0 \\
        1 \mp \sqrt{2} & \mp\sqrt{2}
      \end{pmatrix}}_{M_\mp^O},
    \underbrace{
      \begin{pmatrix}
        1\mp\sqrt{2} & \mp\sqrt{2} \\
        -1\pm\sqrt{2} & \pm\sqrt{2}
      \end{pmatrix}}_{M_\mp^L},
    \underbrace{
      \begin{pmatrix}
        0 & 0 \\
        1 & 1
      \end{pmatrix}}_{M^R}},
\end{equation}
which, using a more symmetric gauge, can also be written as
\begin{equation}
  \label{eq:gsmpsblocked}
  \mps{\ket{G}_\mp}{
    \underbrace{
      \begin{pmatrix}
        0 & 0 \\
        0 & \mp\sqrt{2}
      \end{pmatrix}}_{M_\mp^O},
    \underbrace{
      \begin{pmatrix}
        0 & -1 + \sqrt{2} \\
        0 & 1
      \end{pmatrix}}_{M^L},
    \underbrace{
      \begin{pmatrix}
        0 & 0 \\
        -1-\sqrt{2} & 1
      \end{pmatrix}}_{M^R}}.
\end{equation}

We can go a step further and express $\ket{G}_\mp$ in the single-site basis. Clearly, up to an irrelevant normalization constants, the MPS in Eq.~(\ref{eq:gsmpsblocked}) is reproduced by the single-site TI MPS
\begin{equation}
  \label{eq:gspxp}
  \mps{\ket{G}_\mp}{
    \underbrace{
      \begin{pmatrix}
        0 & 0 \\
        0 & \mp\sqrt{2}
      \end{pmatrix}}_{M_\mp^0},
    \underbrace{
      \begin{pmatrix}
        -1 & -1 + \sqrt{2} \\
        -1 -\sqrt{2} & 1
      \end{pmatrix}}_{M^1}},
\end{equation}
which, in turn, is gauge-equivalent to
\begin{equation}
  \label{eq:gsmpssingle}
  \mps{\ket{G}_\mp}{
    \underbrace{
      \begin{pmatrix}
        1 & 0 \\
        1 & 0
      \end{pmatrix}}_{M^0},
    \underbrace{
      \mp\frac{1}{\sqrt{2}}\begin{pmatrix}
        0 & 1 \\
        0 & 0
      \end{pmatrix}}_{M_\mp^1}}.
\end{equation}

Note that even thought $H_1$, whose ground (ceiling) eigenstate was projected onto the Rydberg-blockaded subspace to obtain $\ket{G}_\mp$, is not well-defined for systems of odd size, the fidelity of the single-site TI MPS in Eq.~(\ref{eq:gsmpssingle}) relative to the true ground states of such systems is completely on par with that in systems of even size.

Since $\ket{L}-\ket{R}$ is annihilated by individual terms of $H_1$, the following gives TI versions of its first and penultimate excited states:
\begin{equation}
  \label{eq:feh1}
  \sqrt\frac{2}{L_b}\left(\sum_{i=1}^{L_b}\left(\ketbra{L}_i-\ketbra{R}_i\right)\right)\ket{\Psi}_\mp.
\end{equation}
Clearly, the operator acting on $\ket{\Psi}_\mp$ in Eq.~(\ref{eq:feh1}) commutes with the projection onto the Rydberg-blockaded subspace; therefore, disregarding the normalization constant, we can apply it directly to the TI MPS representation of $G_\mp$ in Eq.~(\ref{eq:gsmpsblocked}) to obtain the following TTI MPS form of the approximate first and penultimate excited states:
\begin{equation}
  \label{eq:fempsblockedtti}
  \mps{\ket{F}_\mp}{
    \sma{M}{\mathbf{0}}_\mp^O, \sma{M}{M}^L, \sma{\hphantom{-}M}{-M}^R.
  }
\end{equation}

Taking into account the Rydberg constraint, the operator acting on $\ket{\Psi}_\mp$ in Eq.~(\ref{eq:feh1}) can also be expressed in the manifestly TI form in the single-site basis as
\begin{equation}
  \label{eq:singleop}
  -\sqrt{\frac{2}{L_b}}\left(\sum_{r=1}^{L}e^{ikr}\ketbra{1}_r\right),
\end{equation}
where $L = 2L_b$ and $k = \pi$.
Applying the operator in Eq.~(\ref{eq:singleop}) to the single-site TI MPS in Eq.~(\ref{eq:gsmpssingle}) gives the inhomogeneous TTI MPS
\begin{equation}
  \label{eq:fempssingletti}
  \mps{\ket{F(k)}_\mp}{
    \sma{M}{\mathbf{0}}^0, \sma{\hphantom{e^{ikr}}M}{e^{ikr}M}_\mp^1.
  }
\end{equation}
When $L$ is even, it follows from the identity in Eq.~(\ref{eq:smaidentity}) that, up to normalization, the TTI MPS $\ket{F(\pi)}_\mp$ in Eq.~(\ref{eq:fempssingletti}) is equivalent to the blocked basis TTI MPS representation in Eq.~(\ref{eq:fempsblockedtti}) generating the approximate first (penultimate) excited state introduced in Ref.~\cite{Omiya_2023quantum}.
On the other hand, the case of odd $L$ is not addressed by Ref.~\cite{Omiya_2023quantum}.
If Eqs.~(\ref{eq:gspxp}) and (\ref{eq:fempssingletti}) are structurally valid, we should expect them to also generate approximations of the ground (ceiling) and first (penultimate) excited states for systems of odd sizes.

\begin{figure}
  \includegraphics[width=\columnwidth]{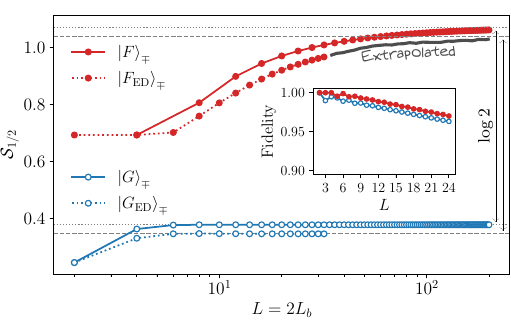}
  \caption{
    New insights into the structure of the low energy manifold of the PXP model.
    The bipartite entanglement entropies of the approximate states $\ket{G}_\mp$ and $\ket{F}_\mp$ appear to only have an approximately constant systematic error with respect to the corresponding true values (calculated for systems sizes up to $L=32$).
    This suggests that $\ket{F}_\mp$ likely provides a qualitatively accurate representation of the growth of the entanglement entropy with system size for the true first (penultimate) excited state.
    Therefore, we expect that the entanglement entropy of the first (penultimate) excited state should saturate at a finite value exactly $\log 2$ above that of the ground state.
    The structural validity of the single-site basis TTI MPS picture of Eq.~(\ref{eq:fempssingletti}) is further corroborated by the correct prediction of system-size dependent degeneracy of the first (penultimate) excited state and the independence of the fidelities (i.e., squared overlaps between $\ket{G}_\mp$ and $\ket{F}_\mp$ with the corresponding true states obtained via ED) on the parity of the system size (inset).
    To compute the entanglement entropy of the state $\ket{F}_\mp$ we used the blocked basis version Eq.~(\ref{eq:fempssingletti}), which for even $L$ is a homogeneous TTI MPS.
    For odd-sized systems, the fidelity of representation of the members of the doubly-degenerate first (penultimate) excited manifold was independent of the choice of $\ket{F_\text{ED}}$.
  }
\label{fig:gsfepxp}
\end{figure}

We find numerically that in systems of odd sizes, where the $k=\pi$ momentum sector is missing, the first (penultimate) excited state is doubly degenerate with momenta $k=\pm\frac{2\pi}{L}\left\lfloor\frac{L}{2}\right\rfloor$.
Hence, we expect $\ket{F}_\mp = \ket{F\left(\frac{2\pi}{L}\left\lfloor\frac{L}{2}\right\rfloor\right)}_\mp$ together with its complex conjugate to generate approximations of these degenerate states in systems of odd size.
Indeed, the fidelities of $\ket{G}_\mp$ and $\ket{F}_\mp$ show no significant dependence on the parity of the system size [see Fig.~\ref{fig:gsfepxp}, inset].
This confirms the structural validity of the model as well as indicates that the characteristics of the low energy manifold of $H_\text{PXP}$ do not require an effective spin-1 model like, for instance, the one discussed in Ref.~\cite{Omiya_2023quantum}.

It is worthwhile noting that the approximations in Eqs.~(\ref{eq:gspxp}) and (\ref{eq:fempssingletti}) can be substantially improved (in particular, for any specific system size) by introducing additional variational parameters on top of them.
In fact, somewhat similar variational wavefunctions were explored in Ref.~\cite{Pan_2022}, where auxiliary fermionic degrees of freedom introduced into the chain served effectively the same function as the virtual bonds in our MPS representations.
Their variational wavefunctions were shown to give remarkably high-fidelity approximations for both the ground (ceiling) and first (penultimate) excited states.
Since our primary interest is the structural relationship between the ground and first excited states, we will deliberately refrain from further variational optimizations here.

Using the MPS technique developed in App.~\ref{app:entspect} we can calculate the bipartite entanglement entropies of $\ket{G}_\mp$ and $\ket{F}_\mp$ for system sizes far exceeding the reach of exact diagonalization (ED).
The result of such a calculation is compared to the actual bipartite entanglement entropies of the corresponding states for numerically accessible systems in Fig.~\ref{fig:gsfepxp}, where we again see strong evidence of the structural correctness of Eq.~(\ref{eq:fempssingletti}).
Thus, we conclude that the first (penultimate) excited state of the PXP model is highly likely to be of the TTI MPS form on top of the ground (ceiling) state, and its bipartite entanglement entropy is expected to saturate at a finite value exactly $\log 2$ above that of the ground state~\cite{Haegeman_2013}.

Let us examine Eqs.~(\ref{eq:gspxp}) and (\ref{eq:fempssingletti}) to gain intuition about which features of the corresponding exact states are faithfully captured by them and which are not.
Recognizing that $M^0$ and $M_\mp^1$ are, respectively, idempotent and nilpotent matrices, it is easy to see that the amplitudes generated by the MPS representation in Eq.~(\ref{eq:gspxp}) depend only on the Hamming weight of the bitstring $s_1s_2\dots s_{L_b}$ in Eq.~(\ref{eq:timps}).
Specifically, these amplitudes are $(\mp 1/\sqrt{2})^{|s_1 s_2\dots s_{L_b}|}$, which is exact only when $L_b\leq 3$ [see Fig.~\ref{fig:gsfepxp}, inset]; in reality, for any $L_b>3$, the amplitudes also depend on the pattern of the bitstring.

The state $\ket{F(k)}_\mp$ in Eq.~(\ref{eq:fempssingletti}) was derived by acting with the specific operator in Eq.~(\ref{eq:singleop}) on the state $\ket{G}_\mp$.
Let us demonstrate that an entire family of extensive single-site diagonal operators produces exactly the same result.
This follows from the two properties of the TTI MPS form stated below.
\begin{prop}
  Suppose states
  \begin{equation}
    \mps{\ket{\psi_1}}{\left\{\sma{M}{M_1(r)}^s\right\}}_r,\quad \mps{\ket{\psi_2}}{\left\{\sma{M}{M_2(r)}^s\right\}}_r,
  \end{equation}
  where $r=1,2,\dots, L$ is the site number are (possibly inhomogeneous) TTI MPSs defined in some local basis $\{\ket{s}\}$ on top of the same TI MPS generated by tensor $M^s$.
Then any linear combination of $\ket{\psi_1}$ and $\ket{\psi_2}$ has the TTI MPS representation
\begin{equation}
  \mps{\alpha\ket{\psi_1} + \beta\ket{\psi_2}}{\left\{\sma{M}{\alpha M_1(r) + \beta M_2(r)}^s\right\}}.
\end{equation}
\end{prop}
\begin{prop}
  For any tensor $M^s$, the vanishing (zero) state vector $\ket{\varnothing(\alpha, k)}$ can be written in the inhomogeneous TTI MPS form as follows:
\begin{equation}
  \mps{\ket{\varnothing(\alpha, k)}}{\left\{\sma{\hphantom{\alpha e^{ikr}}M}{\alpha e^{ikr}M}^s\right\}}_r,
\end{equation}
where  $\alpha$ is an arbitrary constant and $k=\frac{2\pi}{L}p$ for $p \in \mathbb{Z}: \mod(p, L) \neq 0$.
\end{prop}
Thus,
\begin{equation}
  \label{eq:fesimple}
  \begin{aligned}
    &\ket{F(k)}_\mp = \ket{F(k)}_\mp + \ket{\varnothing(\alpha, k)} \\
    &\implies \mps{\ket{F(k)}_\mp}{\sma{\hphantom{\alpha e^{ikr}}M}{\alpha e^{ikr}M}^0,\sma{\hphantom{\beta e^{ikr}}M}{\beta e^{ikr}M}^1}\text{ for any } \alpha \neq \beta \\
    &\implies \ket{F(k)}_\mp \propto \left(\sum_{r=1}^Le^{ikr}\left(\alpha{\ketbra{0} + \beta\ketbra{1}}\right)\right) \ket{G}_\mp.
  \end{aligned}
\end{equation}

As an alternative to the more complex reasoning of Ref.~\cite{Omiya_2023quantum}, we can conclude that $\ket{G}_\mp$ is nothing but a minimally viable crude structural approximation of the ground (ceiling) state.
It captures the overall pattern of decreasing weight of the components with a higher number of excitations and assigns the correct parity-dependent signs while completely disregarding the finer details.
Similarly, the approximate first (penultimate) exited state $\ket{F}_\mp$ is merely the result of applying a modulated (with $k$ the closest wavevector to $\pi$) sum of completely generic diagonal operators --- like that in the last line of Eq.~(\ref{eq:fesimple}) --- to $\ket{G}_\mp$.

We can produce a similar crude MPS approximation for the ground (ceiling) state of the PPXPP model by stipulating that the amplitudes depend only on the Hamming weight of the bitstrings in Eq.~(\ref{eq:timps}) and that an exact state is generated for the system of size $L=5$ [in analogy with Eq.~(\ref{eq:gspxp}), which is exact for $L=3$] where the terms of $H_\text{PPXPP}$ become well-defined.
The MPS representation
\begin{figure}
  \includegraphics[width=\columnwidth]{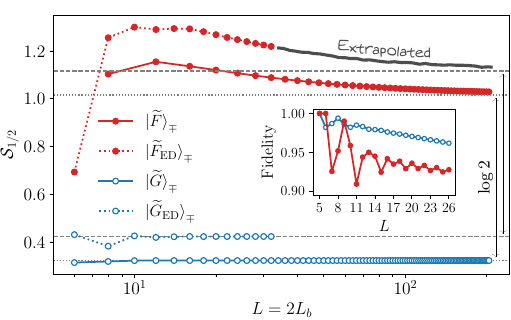}
  \caption{
    New insights into the structure of the low energy manifold of the PPXPP model.
    Similarly to the case of the PXP model in Fig.~\ref{fig:gsfepxp}, the bipartite entanglement entropies of the states $\ket*{\widetilde G}_\mp$ and $\ket*{\widetilde F}_\mp$ have an approximately constant systematic error with respect to the corresponding true values (calculated for system sizes up to $L=32$).
    The entanglement entropy of $\ket*{\widetilde F}_\mp$ correctly captures the overshoot and subsequent exponential decay towards a finite limiting value of the true entanglement entropy.
    Hence we expect the entanglement entropy of $\ket*{\widetilde F_\text{ED}}_\mp$ to saturate at a value exactly $\log 2$ above that of the ground (ceiling) state.
    Note that in the PPXPP model $\ket*{\widetilde F_\text{ED}}_\mp$ is not the first (penultimate) excited state; in fact, it is separated from the ground state by a linear in the system size number of doubly degenerate eigenstates (all of which can also be understood as inhomogeneous TTI MPSs with various momenta on top of $\ket*{\widetilde G}$).
    This means that $\ket*{\widetilde F_\text{ED}}_\mp$ is a somewhat highly excited state with finite entanglement entropy, which is a consequence of the state's direct structural relationship with the ground state.
  }
  \label{fig:gsfeppxpp}
\end{figure}

\begin{equation}
  \label{eq:gsppxpp}
  \mps{\ket*{\widetilde G}_\mp}{
    \underbrace{
      \begin{pmatrix}
        1 & -1 & 1 \\
        0 & 0 & 1 \\
        1 & -1 & 0
      \end{pmatrix}}_{M^0},
    \underbrace{
      \mp\frac{1}{\sqrt{5}}\begin{pmatrix}
        0 & 1 & 0 \\
        0 & 0 & 0 \\
        0 & 0 & 0
      \end{pmatrix}}_{M_\mp^1}}
\end{equation}
meets both criteria and gives an approximate ground (ceiling) state of $H_\text{PPXPP}$ with effectively the same fidelity as that of Eq.~(\ref{eq:gspxp}) for $H_\text{PXP}$ [see Fig.~\ref{fig:gsfeppxpp}, inset].
We can also define the state $\ket*{\widetilde F}_\mp$ to be formally the same as the state $\ket{F}_\mp$ using the tensor $M^s$ from Eq.~(\ref{eq:gsppxpp}).

We find that in the PPXPP model, for all numerically accessible systems,  $\ket*{\widetilde F}_\mp$ has a high overlap with a particular eigenstate $\ket*{\widetilde F_\text{ED}}_\mp$ that is not at the bottom of of the quasi-particle excitation band (it is separated from the ground (ceiling) state by energy difference of approximately $2.43$ for even $L$ and $2.40$ for odd $L$ with a linear in the system size number of doubly degenerate eigenstates with unique momenta different from those of both $\ket*{\widetilde G}_\mp$ and $\ket*{\widetilde F}_\mp$).
Comparing system size dependence of bipartite entanglement entropies of $\ket*{\widetilde G}_\mp$ and $\ket*{\widetilde F}_\mp$ with that of their ED counterparts in Fig.~\ref{fig:gsfeppxpp}, we again find striking qualitative agreement between our crude trial wavefunction approach and exact numerical results.

\section{Finite-size and asymptotic bipartite entanglement spectra of TI and TTI MPSs}
\label{app:entspect}
The approach we will take when calculating the entanglement spectrum
will be similar in spirit to that used in \cite{cirac2011entanglement,
  moudgalya2018entanglement, lin2019exact}, with the differences
due to the fact that the states we are considering are TI MPSs with no
simple OBC counterparts.

Let $\ket{\Psi}$ be a normalized $L$-site state generated from TI MPS
defined by $\chi\times \chi$ site-independent matrices $M^{s}$:
\begin{equation}
  \ket{\Psi} = \sum_{\{s\}}\Tr{M^{s_1}M^{s_2}\cdots
    M^{s_L}}\ket{s_1s_2\cdots s_L}.
\end{equation}
We want to split this TI state with PBC into two subsystems $A$ and
$B$, each containing $h = L/2$ sites, and determine the entanglement
between them. In order for such bipartition to be meaningful, we need
to assume that $L$ is even.

By resolving the trace in Eq.~(\ref{eq:timps}) and inserting a
resolution of identity $\mathbb{I} = \sum_{j=1}^\chi\ketbra{j}{j}$ (both
in the standard basis), we can rewrite $\ket{\Psi}$ as
\begin{equation}
  \begin{aligned}
    \label{eq:psisplit}
    \ket{\Psi} &= \sum_{i,j = 1}^\chi\sum_{\{s\}}
    \bigg(\mel{i}{M^{s_1}\cdots M^{s_h}}{j}\ket{s_1\cdots s_h}\\
    &\otimes\mel{j}{M^{s_{h + 1}}\cdots M^{s_L}}{i}\ket{s_{h+1}\cdots s_L}\bigg) \\
    &= \sum_{i,j = 1}^\chi\ket{\Psi^{A}_{ij}}\otimes\ket{\Psi^{B}_{ij}},
  \end{aligned}
\end{equation}
where
\begin{equation}
  \begin{aligned}
    &\ket{\Psi^A_{ij}} = \sum_{\{s\}}\mel{i}{M^{s_1}\cdots M^{s_h}}{j}\ket{s_1\cdots s_h}, \\
    &\ket{\Psi^B_{ij}} = \sum_{\{s\}}\mel{j}{M^{s_{h+1}}\cdots M^{s_L}}{i}\ket{s_{h+1}\cdots s_L}.
  \end{aligned}
\end{equation}
$\ket{\Psi^A_{ij}}$ and $\ket{\Psi^B_{ij}}$ form a complete basis for
their respective subsystems, but they are not in general linearly
independent.  We now define two $\chi^2 \times \chi^2$ Gram matrices
\begin{equation}
  \begin{aligned}
    G^A_{ij,i'j'} =\braket{\Psi^A_{ij}}{\Psi^A_{i'j'}}, \quad
    G^B_{ij,i'j'} =\braket{\Psi^B_{ij}}{\Psi^B_{i'j'}},
  \end{aligned}
\end{equation}
where double indices $ij$ and $i'j'$ are treated as a single index
running from $1$ to $\chi^2$.

Consider the Schmidt basis in which
\begin{equation}
  \ket{\Psi} = \sum_{i,j=1}^\chi \ket{\tilde\Psi_{ij}^A} \otimes
  \ket{\tilde\Psi_{ij}^B},
\end{equation}
with all unnormalized states $\ket{\tilde\Psi_{ij}^{A,B}}$ mutually
orthogonal (and some possibly corresponding to zero state vectors in
case the Schmidt rank isn't saturated). In this basis the reduced
density matrix for either subsystem is just
\begin{equation}
  \rho_{A, B} = \tilde G^A \tilde G^B,
\end{equation}
where $\tilde G_{A,B}$ are diagonal Gram matrices corresponding to the
mutually orthogonal states in the Schmidt decomposition. 

Based on the properties of bilinear forms and Gram matrices, it can be
argued that the eigenvalues of $G^AG^B$ are unaffected by the choice
of basis in Eq.~(\ref{eq:psisplit}), which means
\begin{equation}
  \rho_{A, B} = \diag\left\{p_1, p_2, \ldots, p_{\chi^2}\right\},
\end{equation}
where $p_i$ are the eigenvalues of $G^AG^B$.

For the case of a bipartition into two subsystems of equal size the
Gram matrices are obtained via simple reindexing of the transfer
matrix raised to a power:
\begin{equation}
  \label{eq:reindexing}
  \begin{aligned}
    G^A_{ij,i'j'} &= [E^{L/2}]_{ii',jj'} \\
    G^B_{ij,i'j'} &= G^A_{ji,j'i'},
  \end{aligned}
\end{equation}
where the double indices $ij$ map to a single index (e.g., $ij \to
i\chi + j$ in case the indices are zero-based).

To obtain the entanglement spectrum in the thermodynamics limit it is
helpful to first compute the limit of the transfer matrix:
\begin{equation}
  E^{\infty} = P\lim_{k\to\infty}\frac{J^k}{\braket{\Psi}{\Psi}(k)}P^{-1},
\end{equation}
where $P$ and $J$ are the basis transformation and Jordan canonical
form of the transfer matrix.

Similar techniques can easily be extended to TTI MPSs by incorporating the twist matrix into Eq.~(\ref{eq:psisplit}) and associating it with either of the half-systems.

\section{Basic properties of the \texorpdfstring{$\ket{\Theta_{1,2}}$}{Theta 1,2} eigenstates of the PXP model}
\label{app:propspxp}
In this section, we provide examples of typical MPS calculations for the MPSs $\ket{\Theta_1}$ and $\ket{\Theta_2}$.
Similar calculations can be performed for any other MPS.

\subsection{\texorpdfstring{$\ket{\Theta_1}$}{Theta1} and \texorpdfstring{$\ket{\Theta_2}$}{Theta2} are translations of each other}
Consider the following 4 matrices:
\begin{equation}
  \Scale[0.88]{\setlength{\arraycolsep}{2.5pt}
  \begin{aligned}
    &B_0 = \begin{pmatrix}
      0 & 3 & -1 & 0 \\
      -1 & 0 & 0 & 1 \\
      2 & 0 & 0 & -6 \\
      0 & 0 & 0 & 0
    \end{pmatrix},\quad
    B_1 = \begin{pmatrix}
      0 & 0 & 0 & 0 \\
      0 & 0 & 0 & 0 \\
      0 & 3 & 0 & 0 \\
      1 & 0 & 0 & -3 \\
    \end{pmatrix}, \\
    &C_0 = \begin{pmatrix}
      1 & 0 & 0 & 0 \\
      0 & 1 & -\frac{1}{3} & 0 \\
      0 & 0 & 0 & 0 \\
      0 & 0 & 0 & 1 \\
    \end{pmatrix},\quad
    C_1 = \begin{pmatrix}
      0 & -\frac{9}{2} & 0 & 0 \\
      0 & 0 & 0 & 0 \\
      -1 & 0 & 0 & 0 \\
      0 & -\frac{3}{2} & 0 & 0
    \end{pmatrix}.
  \end{aligned}
  }
\end{equation}
It is easy to verify that the MPS for $\ket{\Theta_1}$ can be
expressed as $M^{ij} = B_i C_j$, whereas that for $\ket{\Theta_2}$ is
obtained by setting $M^{ij} = C_i B_j$, where we identify $O$,
$L$, and $R$ with, respectively, $00$, $10$, and $01$;
also $B_1 C_1 = C_1 B_1 = 0$ as desired to satisfy the
nearest-neighbor Rydberg blockade constraint. Thus $T_x\ket{\Theta_1}
= \ket{\Theta_2}$, where $T_x$ is the operator that performs
translation by one spin-1/2 site.

\subsection{\texorpdfstring{$\ket{\Theta_1}$}{Theta1} and \texorpdfstring{$\ket{\Theta_2}$}{Theta2} are eigenstates of the particle-hole and inversion symmetry operators}
To make $C_\mathrm{ph}$ and $\mathcal{I}$ symmetries of
$\ket{\Theta_1}$ more apparent, it is helpful first make the $M^L$ and
$M^R$ matrices in Eqs.~(\ref{eq:theta1mps}) more symmetric. This can
be accomplished by computing an invertible matrix $P$ such that
$P^{-1}(M^L + M^R)P = J$, where $J$ is a diagonal matrix, and using it
to perform an MPS gauge transformation $M^s \to P^{-1}M^sP$. With
\begin{equation}
  \Scale[0.88] {
    \setlength{\arraycolsep}{2.5pt}
    P = \begin{pmatrix}
      0 & 0 & -4 & 0 \\
      0 & 0 & 0 & 4 \\
      0 & -1 & 0 & 3 \\
      1 & 0 & -1 & 0 \\
    \end{pmatrix},
  }
\end{equation}
we obtain the following transformed MPS:
\begin{equation}
  \label{eq:theta1mpstr}
  \Scale[0.88] {
    \setlength{\arraycolsep}{2.5pt}
    \underbrace{\frac{1}{4}\begin{pmatrix}
      0 & -1 & 0 & -9 \\ 27 & 0 & 17 & 0 \\ 0 & -1 & 0 & -9 \\ 1 & 0 & 3 & 0
    \end{pmatrix}}_{M^O},
  \underbrace{\begin{pmatrix}
    -3 & 0 & -1 & 0 \\ 0 & -1 & 0 & -9 \\ 0 & 0 & 0 & 0 \\ 0 & 0 & 0 & 0
  \end{pmatrix}}_{M^L},
\underbrace{\begin{pmatrix}
  0 & 0 & 1 & 0 \\ 0 & 0 & 0 & 9 \\ 0 & 0 & 1 & 0 \\ 0 & 0 & 0 & 3
\end{pmatrix}}_{M^R}.
  }
\end{equation}

Consider gauge transformation generated by $U = \diag(-1, 1, -1,
1)$. Such transformation leaves $M^L$ and $M^R$ invariant, whereas
$U^{-1}M^OU = -M^O$. This means only basis vectors with an even number
of $O$ blocks will be present in the state generated by the MPS in
Eq. (\ref{eq:theta1mpstr}). Hence it immediately follows that each
non-zero basis vector component of $\ket{\Theta_1}$ will have the same
parity (with respect to the number of excitations) equal to $L_b \mod
2$. Thus,
\begin{equation}
  \mathcal{C}_\mathrm{ph}\ket{\Theta_2} = (-1)^{L_b}\ket{\Theta_2}.
\end{equation}

Now consider gauge transformation generated by
\begin{equation}
  \Scale[0.88] {
    \setlength{\arraycolsep}{2.5pt}
    Q = \begin{pmatrix}
      0 & 0 & 0 & 1 \\
      0 & 0 & 9 & 0 \\
      0 & 9 & 0 & 0 \\
      1 & 0 & 0 & 0 \\
    \end{pmatrix}.
  }
\end{equation}
Its effect on the MPS in Eqs. (\ref{eq:theta1mpstr}) is as follows:
$Q^{-1}M^OQ = [M^O]^T$, $Q^{-1}M^LQ = -[M^R]^T$ and $Q^{-1}M^LQ =
-[M^R]^T$. In terms of spin-$\tfrac{1}{2}$ degrees of freedom
corresponding to each blocked site,
\begin{equation}
  \begin{aligned}
    &\tr\left\{M^{\sigma_1\sigma_2}M^{\sigma_3\sigma_4}\cdots M^{\sigma_{N-1}\sigma_N}\right\} \\
    &=(-1)^{L_b}\tr\left\{[M^{\sigma_2\sigma_1}]^T[M^{\sigma_4\sigma_3}]^T\cdots [M^{\sigma_N\sigma_{N-1}}]^T\right\} \\
    &=(-1)^{L_b}\tr\left\{M^{\sigma_N\sigma_{N-1}} \cdots M^{\sigma_4\sigma_3}M^{\sigma_2\sigma_1}\right\},
  \end{aligned}
\end{equation}
where the sign follows from the definite $C_\mathrm{ph}$ quantum
number of $\ket{\Theta_1}$ determined earlier. Thus,
\begin{equation}
  \mathcal{I}\ket{\Theta_1} = (-1)^{L_b}\ket{\Theta_1}.
\end{equation}
$\mathcal{C}_\mathrm{ph}$ and $\mathcal{I}$ symmetries of
$\ket{\Theta_2} = \hat T_x \ket{\Theta_1}$ match those of
$\ket{\Theta_1}$ because $[T_x, C_\mathrm{ph}] = 0$, and
$\mathcal{I}T_x = T_x^{-1}\mathcal{I} $ so that
$\mathcal{I}T_x\ket{\Theta_1} = T_x^{-1}\mathcal{I}\ket{\Theta_1} =
T_x^{-1}\mathcal{I}T_x^{-2}\ket{\Theta_1} =
T_x\mathcal{I}\ket{\Theta_1}$.

\subsection{Transfer matrices and norms of \texorpdfstring{$\ket{\Theta_1}$}{Theta1} and \texorpdfstring{$\ket{\Theta_2}$}{Theta2}}
\label{app:mpsnorm}
Many TI MPS calculations make use of the transfer matrix defined as
\begin{equation}
  \label{eq:tm}
  E = \sum_s  (M^s)^* \otimes M^s,
\end{equation}
where $M^{s}$ are site-independent matrices, $*$ denotes complex
conjugation, and the tensor product is taken over the auxiliary space
of the MPS.

The list of nonzero eigenvalues of the transfer matrices corresponding
to $\ket{\Theta_1}$ or $\ket{\Theta_2}$ in the blocked basis is
\begin{equation}
  \begin{aligned}
    \Lambda_E = \frac{1}{2}
    \bigg\{
    &3(3+\sqrt{21}),11+\sqrt{37}, \\
    &11+\sqrt{37}, 11-\sqrt{37}, \\
    &11-\sqrt{37}, 3(3-\sqrt{21}), \\
    &1+\sqrt{13}, 1-\sqrt{13}
    \bigg\},
  \end{aligned}
\end{equation}
and the norms are
\begin{align}
\label{eq:norm2}
\begin{split}
  &\braket{\Theta_1}{\Theta_1} = \braket{\Theta_2}{\Theta_2} = \tr
  {E^{L_b}} = \sum_{\lambda \in \Lambda_E} \lambda^{L_b}.
\end{split}
\end{align}

\subsection{MPS correlation length of \texorpdfstring{$\ket{\Theta_1}$}{Theta1} and \texorpdfstring{$\ket{\Theta_2}$}{Theta2}}
It can be shown~\cite{Orus_2014} that the correlation length $\xi$ of
an MPS is given by
\begin{equation}
  \xi = -\frac{1}{\log|\lambda_2/\lambda_1|},
\end{equation}
where $\lambda_1$ and $\lambda_2$ are the first and second largest
eigenvalues of the transfer matrix. In the case of
$\ket{\Theta_{1,2}}$, $\xi \approx 3.49$ blocked sites (cf.~$\xi
\approx 0.91$ for $\ket{\Phi_{1,2}}$).

\subsection{Overlaps of \texorpdfstring{$\ket{\Theta_1}$}{Theta1} and \texorpdfstring{$\ket{\Theta_2}$}{Theta2} with the \texorpdfstring{$\ket{\mathbb{Z}_2}$}{Z2} product states}
It is easy to calculate these overlaps (and they are the same for the states $\ket{\mathbb{Z}_2}$ and $T_x \ket{\mathbb{Z}_2}$):
\begin{equation}
  \begin{aligned}
    &\vert\braket{\mathbb{Z}_2}{\Theta_1}\vert / \sqrt{\braket{\Theta_1}} = \vert\braket{\mathbb{Z}_2}{\Theta_2}\vert / \sqrt{\braket{\Theta_2}}\\
    & = \Tr{[M^{R/L}]^{L_b}} / \sqrt{\braket{\Theta_1}} \\
    & = (3^{L_b} + 1) / \sqrt{\braket{\Theta_1}},
  \end{aligned}
\end{equation}
where $3$ and $1$ are the non-zero eigenvalues of the matrix $M^R$ in Eq.~(\ref{eq:theta1mps}).

Asymptotically, $\vert\braket{\mathbb{Z}_2}{\Theta_1}\vert / \sqrt{\braket{\Theta_1}} \simeq (6 / (3 + \sqrt{21}))^{L_b/2} \approx 0.8895^{L_b}$.
This becomes accurate for $L_b > 10$.
Note that this is substantially larger than the overlap of the $\mathbb{Z}_2$ CDW with the $\ket{\Phi_{1,2}}$ states, $\vert\braket{\mathbb{Z}_2}{\Phi_1}\vert / \sqrt{\braket{\Phi_1}{\Phi_1}} \simeq (2/3)^{L_b/2} \approx 0.8165^{L_b}$.

\section{Entanglement spectra of \texorpdfstring{$\ket{\Phi_{1,2}}$}{Phi 1,2} and \texorpdfstring{$\ket{\Theta_{1,2}}$}{Theta 1,2}}
\label{app:entpxp}
In this section, utilizing the techniques from App.~\ref{app:entspect}, we derive analytical asymptotic entanglement spectra for the states $\ket{\Phi_{1,2}}$ and $\ket{\Theta_{1,2}}$.

For example, using the transfer matrix and norm corresponding to $\ket{\Phi_1}$ one gets
\begin{equation}
  \Scale[0.88] {
    \setlength{\arraycolsep}{2.5pt}
    E^\infty = \frac{1}{2}
    \begin{pmatrix}
      1 & 0 & 0 & 1 \\
      0 & 0 & 0 & 0 \\
      0 & 0 & 0 & 0 \\
      1 & 0 & 0 & 1
    \end{pmatrix},
  }
\end{equation}
which upon reindexing prescribed by Eq.~(\ref{eq:reindexing}) yields
\begin{equation}
  G^A = G^B = \frac{1}{2}\cdot\mathbb{I}_{4\times 4}.
\end{equation}
Thus we immediately read off that 
\begin{equation}
  \rho_{A, B} = \frac{1}{4}\cdot\mathbb{I}_{4\times 4}
\end{equation}
and hence
\begin{equation}
  S_{1/2} = -\tr{\rho\log\rho} = \log 4.
\end{equation}
Note that this differs from the OBC case analyzed in~\cite{lin2019exact} due to the fact that in PBC the bipartition induces two entanglement cuts.
When $L$ is even these cuts are symmetric and the entanglement spectrum is expected to be a direct product of the identical spectra corresponding to each individual cut.
In the case of $\ket{\Phi_1}$ the decomposition is
\begin{equation}
  \begin{pmatrix}\frac{1}{4}, \frac{1}{4}, \frac{1}{4}, \frac{1}{4}\end{pmatrix} =
  \begin{pmatrix}\frac{1}{2}, \frac{1}{2}\end{pmatrix}^{\times 2}.
\end{equation}

Repeating the same analysis for $\ket{\Phi_2}$ and $\ket{\Theta_{1,2}}$ we get the following entanglement spectra (presented as direct products):
\begin{itemize}
\item{$\ket{\Phi_2}$:}
\begin{equation}
  \begin{pmatrix}\frac{2}{3}, \frac{1}{6}, \frac{1}{6}\end{pmatrix}^{\times 2};
\end{equation}
$S_{1/2} \approx 1.7351$.

\item{$\ket{\Theta_1}$:}
\begin{equation}
  \begin{aligned}
    \Bigg\{
    &\frac{1}{4}+\frac{\sqrt{7 \left(445+1574 \sqrt{21}\right)}}{1036},\\
    &\frac{1}{4}+\frac{\sqrt{7 \left(445+1574 \sqrt{21}\right)}}{1036},\\
    &\frac{1}{4}-\frac{\sqrt{7 \left(445+1574 \sqrt{21}\right)}}{1036},\\
    &\frac{1}{4}-\frac{\sqrt{7 \left(445+1574 \sqrt{21}\right)}}{1036}
    \Bigg\}^{\times 2};
  \end{aligned}
\end{equation}
$S_{1/2} \approx 1.8010$.

Note the degeneracy in the single-cut entanglement spectrum.
This indicates that $\ket{\Theta_1}$ (similarly to $\ket{\Phi_1}$) has symmetry-protected topological order~\cite{Pollmann_2010}.
\item{$\ket{\Theta_2}$:}
\begin{equation}
  \begin{aligned}
    \Bigg\{
    &\frac{1}{4} + \frac{13 + \sqrt{21}}{111} + \frac{\sqrt{7718\sqrt{21} + 22597}}{444\sqrt{7}}, \\    
    &\frac{1}{4} + \frac{13 + \sqrt{21}}{111} -  \frac{\sqrt{7718\sqrt{21} + 22597}}{444\sqrt{7}}, \\
    &\frac{1}{4} - \frac{13 + \sqrt{21}}{111} + \frac{\sqrt{906\sqrt{21} - 2541}}{444}, \\
    &\frac{1}{4} - \frac{13 + \sqrt{21}}{111} - \frac{\sqrt{906\sqrt{21} - 2541}}{444}
    \Bigg\}^{\times 2};
  \end{aligned}
\end{equation}
$S_{1/2} \approx 1.8839$.
\end{itemize}

\section{Alternative proof of Theorem \ref{thrm:h1eig}}
\label{app:proofh1eig}
\begin{proof}
Let $\ket{\psi} \in \mathcal{H}^1$ be a state whose TI MPS representation in the blocked basis is given by matrices $M^s$, $s = O, L, R$.
Suppose there exists a matrix $X$ satisfying Eqs.~(\ref{eq:h1eigcond_fo})--(\ref{eq:h1eigcond_fr}).
Combining Eqs.~(\ref{eq:h1pxp}) and (\ref{eq:timps}) and taking an inner product with an arbitrary product state $\ket{s_1 s_2 \cdots s_{L_b}} \in \mathcal{H}^1\oplus\mathcal{\overline H}^1$ we get
\begin{equation}
  \label{eq:mpsol}
  \begin{aligned}
    &\mel{s_1 s_2 \cdots s_{L_b}}{H_1}{\psi} = \Tr{F^{s_1}M^{s_2} \cdots M^{s_{L_b}}} \\ 
    &+ \Tr{M^{s_1} F^{s_2} \cdots M^{s_{L_b}}} + \cdots \\
    &+ \Tr{M^{s_1} M^{s_2} \cdots F^{s_{L_b}}},
  \end{aligned}
\end{equation}
where $F^s$ (the images of $M^s$ under individual terms of $H_1$) are given in Eq.~(\ref{eq:fs}).
Plugging in left sides of Eqs.~(\ref{eq:h1eigcond_fo})--(\ref{eq:h1eigcond_fr}) in place of every $F^s$ in Eq.~(\ref{eq:mpsol}) we obtain telescoping series that collapse to zero.
Thus, for arbitrary $\ket{s_1 s_2 \cdots s_{L_b}}$, $\mel{s_1 s_2 \cdots s_{L_b}}{H_1}{\psi} = 0$, which means $H_1\ket{\psi} = 0$.
Hence, the conditions of Lemma~\ref{lemma:fibh1} are satisfied and $\ket{\psi}$ is a zero energy eigenstate of $H_\text{PXP}$ with PBC.

Now consider an OBC state $\ket{\psi_{\alpha,\beta}}$ defined by the same bulk MPS matrices as follows:
\begin{equation}
  \ket{\psi_{\alpha,\beta}} = \sum_{\{s\}} v_\alpha^T M^{s_1} M^{s_2} \cdots M^{s_{L_b}}w_\beta \ket{s_1 s_2 \cdots s_{L_b}},
\end{equation}
where $v_\alpha^T$ and $w_\beta$ are some terminations.
The equivalent of Eq.~(\ref{eq:mpsol}) for OBC is
\begin{equation}
  \label{eq:mpsolobc}
  \begin{aligned}
    &\mel{s_1 s_2 \cdots s_{L_b}}{H_1}{\psi_{\alpha, \beta}} = v_\alpha^T F^{s_1}M^{s_2} \cdots M^{s_{L_b}} w_\beta \\
    &+ v_\alpha^T M^{s_1} F^{s_2} \cdots M^{s_{L_b}} w_\beta + \cdots \\
    &+ v_\alpha^T M^{s_1} M^{s_2} \cdots F^{s_{L_b}} w_\beta.
  \end{aligned}
\end{equation}
With the left sides of Eqs.~(\ref{eq:h1eigcond_fo})--(\ref{eq:h1eigcond_fr}) replacing every $F^s$ in Eq.~(\ref{eq:mpsolobc}), the telescoping series collapse to
\begin{equation}
  \label{eq:tsobc}
  \begin{aligned}
    &\mel{s_1 s_2 \cdots s_{L_b}}{H_1}{\psi_{\alpha, \beta}} = v_\alpha^T X M^{s_1}M^{s_2} \cdots M^{s_{L_b}} w_\beta \\
    &- v_\alpha^T M^{s_1}M^{s_2} \cdots M^{s_{L_b}} X w_\beta,
  \end{aligned}
\end{equation}
which in the case when $v_\alpha$ and $w_\beta$ are, respectively, left and right eigenvectors of $X$ with eigenvalues $\lambda_\alpha$ and $\lambda_\beta$ gives
\begin{equation}
  \begin{aligned}
    &\mel{s_1 s_2 \cdots s_{L_b}}{H_1}{\psi_{\alpha, \beta}} \\
    &=  (\lambda_\alpha - \lambda_\beta) \braket{s_1 s_2 \cdots s_{L_b}}{\psi_{\alpha, \beta}}.
  \end{aligned}
\end{equation}
Since $\ket{s_1 s_2 \cdots s_{L_b}}$ was arbitrary, we conclude that $\ket{\psi}$ satisfies the conditions of Lemma~\ref{lemma:fibh1} and hence is an eigenstate of $H_\text{PXP}$ with energy $\lambda_\alpha - \lambda_\beta$.
\end{proof}

We can therefore view the result of \cite{lin2019exact} as a particular solution of Eqs.~(\ref{eq:fibconstr}) and (\ref{eq:h1eigcond_fo})--(\ref{eq:h1eigcond_fr}) in terms of smallest possible matrices. For example, the following choice, identical to the MPS representation of state $\ket{\Phi_1}$ given in Eq.~(\ref{eq:phi1mps}), satisfies the requirements of Theorem~\ref{thrm:h1eig} for PBC:
\begin{equation}
  \label{eq:mpsj}
  \begin{aligned}
    &&M^O = -i\sigma_y, \quad M^L = \frac{1}{\sqrt{2}}(\sigma_z - I), \\
    &&M^R = \frac{1}{\sqrt{2}}(\sigma_z + I), \quad X = \frac{1}{\sqrt{2}}\sigma_x,
  \end{aligned}
\end{equation}
where $\sigma_{x,y,z}$ are the Pauli matrices.
Further, the two identical left and right eigenvectors of matrix $X$, $v_\pm = \begin{pmatrix}1 & \pm 1\end{pmatrix}^T$, whose corresponding eigenvalues are $\lambda_\pm = \pm 1/\sqrt{2}$, give four distinct ways to satisfy the requirements of Theorem~\ref{thrm:h1eig} for OBC.
Thus, we recover the PBC state $\ket{\Phi_1}$ with zero energy, and four OBC states $\ket{\Gamma_{-,+}}$, $\ket{\Gamma_{+,+}}$, $\ket{\Gamma_{-,-}}$, and $\ket{\Gamma_{+,-}}$ with the same bulk MPS as $\ket{\Phi_1}$ and energies, respectively, $-\sqrt{2}$, $0$, $0$, and $\sqrt{2}$.

Note that Theorem~\ref{thrm:h1eig} gives us no direct means for recovering the state $\ket{\Phi_2} = T_x\ket{\Phi_1}$ given in Eq.~(\ref{eq:phi2mps}) nor for proving that it is an eigenstates of $H_\text{PXP}$. This is not surprising because $\ket{\Phi_2}$ is not an eigenstates of $H_1$ and Lemma~\ref{lemma:fibh1} does not apply.

\section{Proof of Theorem \ref{thrm:fibh1dims}}
\label{app:fibh1dims}
\begin{proof}
  If $\ket{\phi} \in \mathcal{H}^1$ satisfies the conditions of Lemma~\ref{lemma:fibh1}, then, since $\ket{\phi}$ is an exact eigenstate of $H_1$, we also have $[H_1, \ketbra{v}]\ket{\psi}= 0 $ for any $\ket{v}$ satisfying $\ketbra{v}\ket{\phi} = 0$.
  By construction, $\ketbra{v}\ket{\phi} = 0$ for $\ket{v} = \ket{RL}$.
  Note that $[H_1, \ketbra{v}] = \ketbra{v'}{v} - \ketbra{v}{v'}$ has the same support as $\ketbra{v}{v}$, since $H_1 = \sum_j h^{(1)}_j$ is the sum of onsite terms; explicitly, $\ket{v'}_{j,j+1} = (h^{(1)}_j + h^{(1)}_{j+1}) \ket{v}_{j,j+1}$.
  Futhermore, it is easily seen that $\ketbra{v'}\ket{\phi} = 0$.
  This way, new local operators $\ketbra{v_i}$ that annihilate $\ket{\phi}$ can be produced.
  It is easy to check that we can obtain such operators with the following $\ket{v_i}$'s:
\begin{subequations}
  \begin{align}
    \label{eq:rl}
    & \ket{v_1} = \ket{RL}, \\
    \label{eq:lr}
    & \ket{v_2} = \ket{LR}, \\
    \label{eq:ol}
    & \ket{v_3} = \ket{OL} +\ket{RO}, \\
    \label{eq:lo}
    & \ket{v_4} = \ket{OR} +\ket{LO}, \\
    \label{eq:oo}
    & \ket{v_5} = \ket{RR} +\ket{LL} + 2\ket{OO}.
  \end{align}
\end{subequations}

Thus our task reduces to finding the dimensions of the subspaces spanned by states annihilated by all range-2 projectors generated from the state vectors in Eqs.~(\ref{eq:rl})--(\ref{eq:oo}) for systems with PBC and OBC.
It is worth noting that these projectors have previously appeared in the literature \cite{lin2019exact, Shiraishi_2019}.
They were introduced as terms in the parent Hamiltonian of the exact MPS $\ket{\Phi_1}$ identified in \cite{lin2019exact}, which in turn can be mapped to the unique ground state of the AKLT model with periodic boundary conditions (PBC).
Here, on the other hand, we arrived at the five projectors without assuming the knowledge of any exact state; instead, we argued that if a particular family of states satisfies the conditions of Lemma~\ref{lemma:fibh1}, it must be locally annihilated by all such projectors.
In what follows, we will set upper bounds on the dimensions the subspaces satisfying the conditions of Lemma~\ref{lemma:fibh1} in systems with PBC and OBC; let us denote these subspaces by, respectively, $V_\mathrm{PBC}$ and $V_\mathrm{OBC}$.

We start by observing that the algebra of projectors generated from the five range-2 state vectors in Eqs.~(\ref{eq:rl})--(\ref{eq:oo}) is invariant under the global unitary
\begin{equation}
  U_{L\leftrightarrow R}=\bigotimes_b (\ketbra{L}{R} + \ketbra{R}{L} + \ketbra{O}{O})_b,
\end{equation} 
and hence we can use eigenstates of $U_{L\leftrightarrow R}$ as the basis for $V_\mathrm{PBC}$ and $V_\mathrm{OBC}$.
The same is true about the unitary $\mathcal{C}_\text{ph} = \otimes_b (\ketbra{O} - \ketbra{R} - \ketbra{L})_b$ (in the blocked labels).
Further, $[U_{L\leftrightarrow R}, \mathcal{C}_\text{ph}] = 0$.
Therefore, we can use states $\ket{p, s}$, where $p = \pm 1$ and $s = \pm 1$ are two quantum
numbers corresponding to the eigenvalues of, respectively,
$\mathcal{C}_\text{ph}$ and $U_{L\leftrightarrow R}$, as our basis. Next, we argue that in systems with OBC all $\ket{p, s}$ are non-degenerate.

From Eqs. (\ref{eq:ol}) and (\ref{eq:lo}), we can deduce the following
relations between the amplitudes of basis vectors:
\begin{equation}
\label{eq:rlmove}
\begin{aligned}
&\braket{\ \circ\ OL\ \bullet\ }{p, s} =
-\braket{\ \circ\ RO\ \bullet\ }{p, s}, \\
&\braket{\ \circ\ OR\ \bullet\ }{p, s} =
-\braket{\ \circ\ LO\ \bullet\ }{p, s},
\end{aligned}
\end{equation}
where $\circ, \bullet$ are arbitrary sequences of $O, L, R$.
In other words, one can always shift $R$ or $L$ to the left or to the right into an adjacent $O$ as, respectively, $L$ or $R$, and flip the sign to obtain the new amplitude. 
In particular, the amplitude of any product state $\ket{\chi}$ for which $\mathcal{N}(\ket{L}) + \mathcal{N}(\ket{R}) = k$ ($\mathcal{N}$ denotes the number of specified blocks in the product state) can be reduced to
\begin{equation}
  \label{eq:lateral}
  \braket{\chi}{p, s} = \pm\braket*{\underbrace{OO\cdots
      O}_{L_b-k}W_1W_2\cdots W_k}{p, s},
\end{equation}
where $W_i = L, R$.
Next, using Eq.~(\ref{eq:rl}) and (\ref{eq:lr}), the RHS is nonzero only if all $W_i$ are equal to $L$ or are equal to $R$.
This, together with the choice of the quantum number $s = \pm 1$ gives exactly two possibilities for relating the amplitudes of the components with the same number of excitations.

For any fixed values of $p$ (allowing either even or odd number of excitations) and $s$ (defining the sign in the relations of the type $\braket{O^{\otimes L_b-k}R^{\otimes k}}{p, s} = s\braket{O^{\otimes L_b-k}L^{\otimes k}}{p, s}$), one can use Eq.~(\ref{eq:oo}) to unambiguously relate the amplitudes of all the components with different $k$ and that way uniquely define each such state.
Thus, as stated in the Theorem, there are four orthogonal and non-degenerate states $\ket{p, s}$, where $p$ and $s$ take values $\pm 1$, that span $V_\mathrm{OBC}$.

In systems with PBC and $L_b > 2$ only basis vectors with an even number of $O$'s can contribute to $\ket{p, s}$ ($L_b = 2$ is a special case, where $V_\mathrm{PBC} = V_\mathrm{OBC}$).
The amplitudes of basis vectors with an odd number of $O$'s and at least two non-$O$'s can be related to the amplitudes of basis vectors with an $RL$ or $LR$ violation by moving one of the edge excitations across the boundary and to the opposite side of the excitations block in Eq.~(\ref{eq:lateral}).
Further, Eq.~(\ref{eq:oo}) precludes components with an odd number of $O$'s and a single $R$ or $L$ because for $L_b > 2$ that would imply the existence of non-zero components with an odd number of $O$'s and three non-$O$'s, contradicting the previous statement in this paragraph.
Thus we conclude $p = (-1)^{L_b}$.

The even number of $O$'s and the fact that excitations can be moved around the chain when relating amplitudes of different components also gives
\begin{equation}
  \braket{O^{\otimes (L_b-k)}R^{\otimes k}}{p, s} =
  (-1)^{L_b}\braket{O^{\otimes (L_b-k)}L^{\otimes k}}{p, s},
\end{equation}
which fixes the $s$ quantum number to $s = (-1)^{L_b}$.
Thus we conclude that $\ket{(-1)^{L_b}, (-1)^{L_b}}$ is the only state residing in $V_\mathrm{PBC}$.
This state, due to having a definite $p$ quantum number, can only have zero energy.

Further, for PBC systems, by relating the amplitudes of appropriate components, it is not difficult to show that $\ket{(-1)^{L_b},(-1)^{L_b}}$ must be translationally invariant (in the blocked basis) and have definite inversion quantum number equal to $(-1)^{L_b}$ (inversion is defined as $I: j \to N - j + 1$ in the original spin-$\frac{1}{2}$ basis, where $N = 2L_b$).
\end{proof}

To make a few final remarks we prove the following:
\begin{prop}
  \label{prop:h1unitary}
  Any eigenstate of $H_1$ with eigenvalue $E = \sqrt{2}n$, where $n
  \in [-L_b,\dots,L_b]$, is also an eigenstate of $U_{L\leftrightarrow R}$
  with eigenvalue $(-1)^{L_b + n}$.
  \begin{proof}
    Per \cite{lin2019exact}, in the basis defined by
    \begin{equation}
      \begin{aligned}
        &\ket{+}_b = \frac{1}{2}\left(\ket{R} + \ket{L} +
        \sqrt{2}\ket{O}\right),\\ &\ket{-}_b = \frac{1}{2}\left(\ket{R} +
        \ket{L} - \sqrt{2}\ket{O}\right),\\ &\ket{\mathbf{0}}_b =
        \frac{1}{\sqrt{2}}\left(\ket{R} - \ket{L}\right),
      \end{aligned}
    \end{equation}
    $H_1$ is diagonal and can be written as
    \begin{equation}
      H_1 = \sqrt{2} \sum_{j = 1}^{L_b}\left(\ketbra{+}{+} -
      \ketbra{-}{-}\right).
    \end{equation}
    This means that every eigenstate is a superposition of product    states with constant $\mathcal{N}(\ket{+}_b) -    \mathcal{N}(\ket{-}_b) =n$.
    Thus $\mathcal{N}(\ket{\mathbf{0}}_b)$ always has definite parity in all contributing product states equal to that of $L_b - \mathcal{N}(\ket{+}_b) -\mathcal{N}(\ket{-}_b) = L_b - n - 2\mathcal{N}(\ket{-}_b)$, and exchanging $\ket{R}$ and $\ket{L}$ will only produce a global phase $(-1)^{L_b+n}$.
  \end{proof}
\end{prop}

If $\ket{(-1)^{L_b}, (-1)^{L_b}}$ is a zero energy eigenstate of $H_1$ in a system with PBCs, then it must also be a zero energy eigenstate of $H_1$ in a system with OBC (since $H_1$ is a sum of on-site terms and the boundary conditions only enter via the Rydberg blockade constraint).
This implies that $\ket{-(-1)^{L_b}, (-1)^{L_b}}$ is another zero-energy OBC candidate state with opposite ``particle-hole symmetry.''
The reason is that non-zero energy eigenstates of $H_1$ must consist of two components with different $p$ and same $s$ quantum numbers, and the only counterpart that $\ket{-(-1)^{L_b}, (-1)^{L_b}}$ has in $V_\mathrm{OBC}$ is another zero-energy state, so it itself can only have zero energy.
Finally, per Proposition \ref{prop:h1unitary}, states
\begin{equation}
\ket{\psi_\pm} = \ket{(-1)^{L_b}, -(-1)^{L_b}} \pm \ket{-(-1)^{L_b},
  -(-1)^{L_b}}
\end{equation}
are candidate finite energy states with $E$ being an odd multiple of
$\pm\sqrt{2}$.

\section{ \texorpdfstring{$X$}{X} matrices for Corollary~\ref{corr:proofx} and Corollary~\ref{corr:rydbproof} proving states  \texorpdfstring{$\ket{\Phi_{1,2}}$}{Phi 1,2},  \texorpdfstring{$\ket{\Theta_{1,2}}$}{Theta 1,2}, and  \texorpdfstring{$\ket{\Omega}$}{Omega}}
\label{app:proofxpxp}
To complete the proof of Corollary~\ref{corr:proofx}, for each state $\ket{\Phi_1}, \ket{\Phi_2}, \ket{\Theta_1}, \ket{\Theta_2}$, and $\ket{\Omega}$, we list the corresponding matrix $X$ such that the conditions of Theorem~\ref{thrm:genzm} in the blocked spin basis are satisfied:
  \begin{equation}
    % \Scale[0.88]
    {\setlength{\arraycolsep}{2pt}
      X_{\Phi_1} = \frac{1}{\sqrt{2}}
      \begin{pmatrix}
        0 & 1\\
        1 & 0
      \end{pmatrix}; ~ 
      X_{\Phi_2} = \frac{1}{\sqrt{2}}
      \begin{pmatrix}
        0 & 1 & 0 \\
        -1 & 0 & -2 \\
        0 & -2 & 0
      \end{pmatrix};
    }
  \end{equation}
  \begin{equation}
    %\Scale[0.88]
    {\setlength{\arraycolsep}{2pt}
      X_{\Theta_1} = \begin{pmatrix}
        1 & \frac{3}{2} & 1 & 0 \\ \frac{1}{2} & 1 & 0 & 1 \\ 2 & 0 & 1 & \frac{15}{2} \\ 0 & 0 & \frac{1}{2} & 1
      \end{pmatrix};\
      X_{\Theta_2} = \begin{pmatrix}
        0 & -3 & \frac{5}{4} & 0 \\ -\frac{1}{6} & 0 & 0 & -\frac{3}{2} \\ -1 & 0 & 0 & 0 \\ 0 & -\frac{3}{2} & \frac{3}{4} & 0
      \end{pmatrix};\hspace{-1.5em}
      }
    \end{equation}
    \begin{equation}
      %\Scale[0.88]
      {\setlength{\arraycolsep}{2pt}
        X_{\Omega} = \frac{i}{\sqrt{3}}\begin{pmatrix}
          0 & \frac{11}{32} & -\frac{1}{2} & \frac{7}{16} \\
          -6 & -\frac{9}{4} & 0 & \frac{1}{2} \\
          \frac{3}{8} & -\frac{3}{32} & -\frac{3}{4} & -\frac{5}{32} \\
          -3 & -\frac{9}{8} & 6 & -\frac{3}{2}
        \end{pmatrix}.
      }
    \end{equation}

In the case of $\ket{\Phi_1}$, stronger conditions of Theorem~\ref{thrm:h1eig} are satisfied and $X_{\Phi_1}$ is the same as in Ref.~\cite{lin2019exact}, while Theorem~\ref{thrm:genzm} and all other $X$ (including direct proof for the previously known state $\ket{\Phi_2}$) are new.
The above $X_\Omega$ also satisfies conditions of Corollary~\ref{corr:rydbproof} in the Rydberg atom basis, proving the state $\ket{\Omega}$ for any system size $L > 3$.

\section{TI variants of \texorpdfstring{$\ket{\Phi}$}{Phi} and \texorpdfstring{$\ket{\Theta}$}{Theta} MPS states in the Rydberg basis}
\label{app:rydbti}
For the sake of completeness and further applications, we also found non-injective MPS representations corresponding to the TI (in the spin-1/2 basis) versions of the $\ket{\Phi_{1,2}}$ and $\ket{\Theta_{1,2}}$ states.
For $\ket{\Phi} \equiv \ket{\Phi_1} + \ket{\Phi_2}$ (with $\ket{\Phi_2} = T_x \ket{\Phi_1}$) and $\ket{\Theta} \equiv \ket{\Theta_1} + \ket{\Theta_2}$ (with  $\ket{\Theta_2} = T_x \ket{\Theta_1}$) we have, respectively,
\begin{subequations}
  \begin{flalign}
    &\mps{\ket{\Phi}}{
      \underbrace{
      \sqrt{2}\begin{pmatrix}
        0 & 0 & 0 & \frac{1}{4} & 0 \\
        1 & 0 & 0 & 0 & -\frac{1}{4} \\
        0 & -\frac{1}{2} & 0 & 0 & 0 \\
        0 & 0 & 1 & 0 & 0 \\
        0 & 0 & 0 & -1 & 0 \\
      \end{pmatrix}}_{M^0}, ~
     \underbrace{
      \begin{pmatrix}
        0 & 1 & 0 & 0 & 0 \\
        0 & 0 & 0 & 0 & 0 \\
        0 & 0 & 0 & 0 & 0 \\
        0 & 0 & 0 & 0 & 1 \\
        0 & 0 & 0 & 0 & 0 \\
      \end{pmatrix}}_{M^1}}; \\
    &\mps{\ket{\Theta}}{
      \underbrace{
      \begin{pmatrix}
        0 & 0 & 0 & -\frac{17}{48} & 0 & 0 & -\frac{1}{16} & 0 \\
        3 & 0 & 0 & 0 & 0 & 0 & 0 & \frac{1}{16} \\
        0 & \frac{3}{4} & 0 & 0 & -\frac{1}{8} & 0 & 0 & 0 \\
        0 & 0 & 1 & 0 & 0 & \frac{1}{8} & 0 & 0 \\
        -6 & 0 & 0 & 0 & 0 & 0 & 0 & -\frac{17}{8} \\
        0 & 6 & 0 & 0 & -1 & 0 & 0 & 0 \\
        0 & 0 & 1 & 0 & 0 & \frac{1}{8} & 0 & 0 \\
        0 & 0 & 0 & -1 & 0 & 0 & -3 & 0 \\
      \end{pmatrix}}_{M^0}, ~
      \underbrace{
      \bigoplus\limits_{i=1}^4\begin{pmatrix}
        0 & 1\\
        0 & 0
      \end{pmatrix}}_{M^1}}.
  \end{flalign}
\end{subequations}
Note that in each case we chose a gauge and normalization such that $M^1$ matrix is a canonical nilpotent matrix.
Due to non-injectivity of the $\ket{\Phi}$ and $\ket{\Theta}$ MPS representations, these states are not defined for systems of odd sizes (i.e., they generate a trivial zero state vector).
Both states can easily be proven directly using Corollary~\ref{corr:rydbproof} by finding corresponding $X$ matrices (not provided here).

The states $\ket{\Phi}$ and $\ket{\Theta}$ were obtained by producing a block-diagonal combination of the corresponding translational-symmetry-breaking MPSs (which generates a manifestly TI state by construction), and then factoring this block-diagonal combination into the spin-1/2 basis.
One possible way to verify that these states are indeed TI variants of the respective MPSs is to repeat the above process.

One potential application of these TI versions is the construction of TTI MPS trial states using single-Rydberg-site ``defect'' matrices. This approach appears simpler than the two-site defect construction described in Ref.~\cite{lin2019exact}, and the capability to leverage MPS tools within our TTI formalism is especially powerful and appealing for systematic studies.

\section{\texorpdfstring{$X$}{X} matrices for Corollary~\ref{corr:ppxpp} and its extension  to TTI MPS}
We first list the $X$ matrices for proving the TI (in the blocked basis) scar states $\ket{S_1}$, $\ket{S_2}$, and $\ket{T}$ in the PPXPP model using Corollary~\ref{corr:ppxpp}:
\label{app:proofxppxpp}
\begin{equation}
  % \Scale[0.88]
  {\setlength{\arraycolsep}{2pt}
    X_{S_1} = 
    \begin{pmatrix}
      1 & 0\\
      0 & 1
    \end{pmatrix}; ~ 
    X_{S_2} =
    \begin{pmatrix}
      0 & 0 & 0 \\
      0 & 0 & 0 \\
      1 & 0 & 0
    \end{pmatrix};
  }
\end{equation}
\begin{equation}
  \Scale[0.88]{\setlength{\arraycolsep}{2pt}
    X_T = 
    \begin{pmatrix}
      0 & \gamma^{-1}-\gamma & 0 \\
      \gamma & 0 & \gamma^2 - \gamma\\
      0 & -\gamma & 0
    \end{pmatrix}.
  }
\end{equation}

To prove the TTI MPS states $\ket{S'_1}$, $\ket{S'_2}$, and $\ket{T'}$, we list the corresponding matrices $X$ such that the conditions of Corollary~\ref{corr:ppxpp} extended to TTI MPS per Theorem~\ref{thrm:gengenzmtti} are satisfied:
\begin{widetext}
\begin{equation}
  \Scale[0.88]{\setlength{\arraycolsep}{2pt}
    X_{S'_1} = 
    \begin{pmatrix}
      0 & 0 & 0 & 0 \\
      0 & 0 & 0 & 0\\
      0 & 1 & 0 & 0\\
      0 & 0 & 0 & 0
    \end{pmatrix}; ~ 
    X_{S'_2} = 
    \begin{pmatrix}
      0 & 0 & 0 & 0 & 0 & 0\\
      0 & 0 & 0 & 0 & 0 & 0\\
      1 & 0 & 0 & 0 & 0 & 0\\
      0 & 1 & 0 & 0 & 0 & 0\\
      0 & 0 & 1 & 0 & 0 & 0\\
      -1 & 0 & 0 & 1 & 0 & 0
    \end{pmatrix}; ~
    X_{T'} =
    \begin{pmatrix}
      0 & 2^{\frac{1}{3}}+ \frac{1}{2^{\frac{1}{3}}} - 2 \gamma & 0 & \frac{1}{2^{\frac{1}{3}}}-\gamma  & 0 & \gamma -\frac{1}{2^{\frac{1}{3}}} \\
      \gamma  & 0 & i \left(2^{\frac{1}{3}}-\gamma ^*\right) & 0 & 0 & 0 \\
      0 & -2^{2/3} \gamma ^2 & 0 & -\gamma  & 0 & \gamma  \\
      -2^{2/3} \gamma ^2 & 0 & \frac{i}{2^{\frac{1}{3}}} & 0 & 2^{\frac{1}{3}} \gamma ^*-\gamma  & 0 \\
      0 & 0 & 0 & 2^{2/3} \gamma ^2 & 0 & i \left(\frac{1}{2^{\frac{1}{3}}}-2^{\frac{1}{3}}\right) \\
      0 & 0 & -2^{2/3} \gamma ^2 & 0 & -\gamma  & 0
    \end{pmatrix}.
  }
\end{equation}
\end{widetext}
\section{OBC counterparts of PBC states satisfying the conditions of Theorem \ref{thrm:genzm}}
\label{app:obctheta}
Choosing left and right eigenvectors $v^T$ and $w$ of matrix $X$ as terminations to produce OBC states is guaranteed to work only for states satisfying the conditions of Theorem \ref{thrm:h1eig}.
For states that only satisfy the conditions of Theorem~\ref{thrm:genzm} --- in particular, $\ket{\Phi_2}$, $\ket{\Theta_{1,2}}$, and $\ket{\Omega}$ --- the following additional requirements need to be satisfied:
\begin{equation}
  \label{eq:condweakobc}
  \begin{aligned}    
    &v^T [X, M^L] = v^T F^L, \\
    &[X, M^R] w = F^R w,
  \end{aligned}
\end{equation}
where the first equation is the counterpart of Eqs.~(\ref{eq:padding_mofl})--(\ref{eq:padding_mlfl}) at the left boundary and the second equation is the counterpart of Eqs.~(\ref{eq:padding_frmo})--(\ref{eq:padding_frmr}) at the right boundary.

Equations~(\ref{eq:condweakobc}) are violated for all choices of eigenvectors of the $X$ matrices corresponding to the MPS representations of $\ket{\Phi_2}$, $\ket{\Theta_1}, \ket{\Theta_2}$, and $\ket{\Omega}$ given in the main text.
This means that, unlike $\ket{\Phi_1}$, these states do not have simple OBC counterparts.

\begin{widetext}
\section{Nullspaces of the smallest less-than-full-rank RDM for \texorpdfstring{$\ket{\Theta_{1,2}}$}{Theta 1,2} and components of \texorpdfstring{$\ket{\Omega}$}{Omega}}
\label{app:ph}
Here we list basis vectors $\{\ket{v_i}\}$ that produce minimal-range parent Hamiltonians, Eq.~(\ref{eq:tiph}), for the exact scars $\ket{\Theta_{1,2}}$ and $\ket{\Omega}$ (specificaly, its real and imaginary components) in the PXP chain.
\subsection{Basis for \texorpdfstring{$\ket{\Theta_1}$}{Theta1}}
In this case, the range is 3 in the 2-blocked sites and the basis vectors are: 
\begin{subequations}
  \begin{align}
    \label{eq:v1_theta1}
    \ket{v_1} = &6(\ket{OOR}+\ket{LOO})-7(\ket{ORO}+\ket{OLO})+2(\ket{RRR}+\ket{LLL})-6(\ket{LRR} + \ket{LLR}), \\
    \ket{v_2} = &9(\ket{ORR} + \ket{LLO})+3(\ket{OLR} + \ket{LRO})+11(\ket{OLL} + \ket{RRO})-24(\ket{ROR}+\ket{LOL}) + 22\ket{ROL}, \\
    \ket{v_3} = &3(\ket{OLR} - \ket{LRO}) + (\ket{OLL} - \ket{RRO} + 3(\ket{ROR} - \ket{LOL}), \\
    \ket{v_4} = &(\ket{ORR}-\ket{LLO}) + 4(\ket{OLR}-\ket{LRO}) + (\ket{ROR}-\ket{LOL}), \\
    \label{eq:v5_theta1}
    \ket{v_5} = &6(\ket{OOR}-\ket{LOO})-2(\ket{OOL}-\ket{ROO})-4(\ket{ORO}-\ket{OLO})+ \\
    &2(\ket{RRR}-\ket{LLL})-3(\ket{LRR}-\ket{LLR}).
  \end{align}
\end{subequations}

\subsection{Basis for \texorpdfstring{$\ket{\Theta_2}$}{Theta2}}
The range is 3 in the 2-blocked sites and the basis vectors are
\begin{subequations}
  \begin{align}
    \label{eq:v1_theta2}
    \ket{v_1} = &\ket{ORO} + \ket{OLO}, \\
    \ket{v_2} = &\ket{LRR} + \ket{LLR}, \\
    \ket{v_3} = &5(\ket{OLL} + \ket{RRO}) + \ket{OOO} + 15\ket{ROL}, \\
    \ket{v_4} = &(\ket{OOR}+\ket{LOO}) + 21(\ket{OOL}+\ket{ROO}) + 5(\ket{RRR} + \ket{LLL}), \\
    \ket{v_5} = &5(\ket{ORR} + \ket{LLO}) + 5(\ket{OLR} + \ket{LRO}) + 9\ket{OOO} - 15\ket{ROL}, \\
    \label{eq:v6_theta2}
    \ket{v_6} = &3(\ket{ORR} + \ket{LLO}) + 12(\ket{OLR} + \ket{LRO}) + 3(\ket{ROR} + \ket{LOL}) - 2\ket{LOR}.
  \end{align}
\end{subequations}

\subsection{Basis for \texorpdfstring{$\ket{\Re\Omega}$}{ReOmega} and \texorpdfstring{$\ket{\Im\Omega}$}{ImOmega}}
Here the range is 7 in the original Rydberg atom sites and the basis vectors are
\begin{subequations}
  \begin{align}
    \label{eq:v1_imomega}
    \ket{v_1} = &\ket{1000101} + \ket{1010001}, \\
    \ket{v_2} = &(\ket{0000010} + \ket{0100000}) + \ket{0001000} + \ket{0101010}, \\
    \ket{v_3} = &(\ket{0010101} + \ket{1010100})-(\ket{0101001}+\ket{1001010}), \\
    \ket{v_4} = &3(\ket{0000001} + \ket{1000000}) + 3(\ket{0000100} + \ket{0010000}) + 3(\ket{0101001}+\ket{1001010}) + 2\ket{1001001}, \\
    \ket{v_5} =&(\ket{0001001} + \ket{1001000}) + (\ket{0010010} + \ket{0100100}) + (\ket{0100001} + \ket{1000010}) + \\
                &3(\ket{0001010}+\ket{0101000}) + 3\ket{0100010} + 3\ket{0000000}, \\
    \label{eq:v6_imomega}
    \ket{v_6} =&(\ket{0001001} - \ket{1001000})-(\ket{0010010}-\ket{0100100}) + (\ket{0100001}- \ket{1000010}.
  \end{align}
\end{subequations}
\end{widetext}
\newpage

%\nocite{*}

\bibliography{main}

\end{document}